\documentclass[review]{elsarticle}
\usepackage{lineno}
\usepackage{hyperref}

\usepackage[utf8]{inputenc}
\usepackage{amsmath}
\usepackage{amsfonts}
\usepackage{amssymb}
\usepackage{graphicx}
\usepackage{accents}
\usepackage{color}
\usepackage{geometry}
\usepackage[dvipsnames]{xcolor}
\usepackage[normalem]{ulem}

\usepackage{hyperref}
\hypersetup{pdftitle={ },
            pdfsubject={  },
            pdfauthor={Anna Ask},
            pdfkeywords={pdftex,acrobat}
            bookmarks=false,
            pdfpagemode=UseNone,
            linkcolor=blue,
            citecolor=blue,
            colorlinks=true,
            bookmarksopen=false,
            bookmarksopenlevel=1,
            urlcolor=blue}
\usepackage{cleveref}

\biboptions{authoryear}

\def\vc  #1{\mbox{\boldmath $#1$}}
\def\virt #1{\mbox{$\dot{#1}$}}
\def\vx #1{\mbox{$\overset{\times}{\underline{\vc #1}}$}}
\def\vec #1{\mbox{$\underline{\vc #1}$}}

\def\ten #1{\mbox{$\undertilde{\vc #1}$}}
\def\TEN #1{\underset{^\approx}{\mbox{\boldmath $#1$}}} 
\newcommand{\leviciv}{\mbox{$\undertilde{\underline{\vc \epsilon}}$}}

\definecolor{mygray}{gray}{0.9}
\newcommand\cbox[1]{\colorbox{mygray}{$#1$}}

\let\oldequation\equation
\let\oldendequation\endequation

\renewenvironment{equation}
  {\linenomathNonumbers\oldequation}
  {\oldendequation\endlinenomath}
  
\let\oldalign\align
\let\oldendalign\endalign

\renewenvironment{align}
  {\linenomathNonumbers\oldalign}
  {\oldendalign\endlinenomath}

\begin{document}

\begin{frontmatter}
\title{A Cosserat crystal plasticity and phase field theory for grain boundary migration \footnote{$\copyright$ 2018. This manuscript version is made available under the CC-BY-NC-ND 4.0 license \url{http://creativecommons.org/licenses/by-nc-nd/4.0/}}}


\author[address1]{Anna Ask}

\author[address1]{Samuel Forest\corref{mycorrespondingauthor}}
\cortext[mycorrespondingauthor]{Corresponding author}
\ead{samuel.forest@ensmp.fr}

\author[address2,address3]{Benoit Appolaire}
\author[address1]{Kais Ammar}
\author[address4]{Oguz Umut Salman}

\address[address1]{MINES ParisTech, PSL Research University, MAT -- Centre des mat\'{e}riaux,\\ CNRS UMR 7633, BP 87 91003 Evry, France}
\address[address2]{Universit\'e  de Lorraine, CNRS, IJL, F--54000 Nancy, France}
\address[address3]{Laboratoire d'Etude des Microstructures, CNRS/Onera, BP72, 92322 Châtillon Cedex, France} 
\address[address4]{CNRS, LSPM UPR3407, Université Paris 13, Sorbonne Paris Cité, 93430, Villetaneuse, France}

\begin{abstract}
The microstructure evolution due to thermomechanical treatment of metals can largely be described by viscoplastic deformation, nucleation and grain growth. These processes take place over different length and time scales which present significant challenges when formulating simulation models. In particular, no overall unified field framework exists to model concurrent viscoplastic deformation and recrystallization and grain growth in metal polycrystals. In this work a thermodynamically consistent diffuse interface framework incorporating crystal viscoplasticity and grain boundary migration is elaborated. 
The Kobayashi--Warren--Carter (KWC) phase field model is extended to incorporate the full mechanical coupling with material and lattice rotations and evolution of dislocation densities. The Cosserat crystal plasticity theory is shown to be the appropriate framework to formulate the coupling between phase field and mechanics with proper
distinction between bulk and grain boundary behaviour.
\end{abstract}


\end{frontmatter}


\section{Introduction}

\subsection{Scope of the work}
The microstructure of a polycrystalline metallic material is characterized by the shape 
and distribution of differently oriented grains. Macroscopic material properties such as 
strength and ductility can be tuned by thermomechanical processing which significantly 
alters the microstructure of the metal at the grain scale through viscoplastic deformation 
and subsequent (sequential) or concurrent (dynamic) nucleation and growth of new grains. 
 
To date, the deformation, nucleation and grain 
growth tend to be treated using a combination of different methods, even in coupled 
approaches. Regardless of the choice of method, it is important to identify the variables which are 
sufficient to describe the microstructure and its evolution. Nucleation tends to be 
favored at sites of large misorientation or lattice curvature and the mobility of 
migrating grain boundaries is also closely related to the misorientation \citep{humphreys04,sutton07,gottstein10}. On the one hand, 
viscoplastic deformation may result in significant reorientation 
of the crystal lattice and the development of heterogeneous orientation distributions within previously 
uniformly oriented grains. On the other hand, a migrating grain boundary 
will also result in reorientation since the grain orientation of material points in the 
region it passes will change from that of the old to that of the growing grain. 
In order to obtain a strongly coupled scheme suitable to model concurrent viscoplastic 
deformation and grain growth, it is necessary to consider the evolution of the grain 
orientation during these respective processes. Plastic deformation also leads to build-up 
of dislocations inside the grains. The resulting dislocation densities tend not to be 
homogeneous and nucleation is favored at sites with a high dislocation content. 
Furthermore, stored energy due to dislocations is an important driving force 
for grain boundary migration. 

Crystal plasticity models provide the most complete tools to model microstructure 
evolution at the grain scale induced by viscoplastic deformation, including heterogeneous 
reorientation and dislocation density build-up \citep{juergen03,rotersreview10}. 
For the nucleation and growth of new grains, it is common to combine some criterion for the generation of new nuclei with a model for migrating grain boundaries. 
The tracking of the moving interface can be done explicitly or implicitly. 
Examples of the latter are level-set \citep{Chen1995,Bernacki2008} and phase-field methods. In the literature, there are two main phase field approaches used to model grain growth 
in solid polycrystals: multi-phase field methods as introduced by 
\cite{Steinbach1996, SteinbachPezzolla1999} or \cite{FanChen1997} 
and orientation field based models of the type introduced by \cite{KWC2000, KWC2003}. 
In particular, by using 
the formalism of generalized forces introduced by \cite{Gurtin1996,GurtinLusk1999}, 
the phase field methods can be derived from the same energy principles as the governing 
equations of mechanics, making them a strong candidate for a coupled theory. 

\subsection{Phase-field models for GB migration}
To model grain growth using a multi-phase field model, each of $N$ non-conserved phase 
field variables $\theta_i$, $i=1, \hdots, N;\,\sum_i^N \theta_i = 1$ is associated with a 
grain orientation. Note that in the multi-phase approach, the phase fields themselves 
are not the crystal orientations (even if they are referred to as orientation 
fields in e.g. \cite{FanChen1997}) but rather represent the fraction of phase $i$, 
of a given orientation, present at a material point. Inside a grain, only one phase field 
variable takes the value 1 and all the others are zero. \cite{Steinbach1996} suggested 
that all possible orientations be divided into $N=10$ classes according to a Potts model. 
More generally, any discrete number of possible orientations may be used but of course, 
the problem size will grow with $N$. \cite{FanChen1997} studied the influence of the 
number of orientations on the grain growth simulations, demonstrating that in their 
proposed model, with few allowed orientations the tendency was that grains of equal 
orientation coalesced whereas with a larger number of allowed orientations, 
grain boundary migration was favored. One attractive feature of the multi-phase field 
approach is that the model parameters can easily be associated with the measurable 
quantities mobility, grain boundary energy and interface thickness. The inclusion of 
misorientation dependency and anisotropy is likewise straightforward \citep{Steinbach1996}. 
However, the multi-phase field model is not suitable for a strongly coupled approach 
based on lattice orientation since it does not contain an equation that allows for 
evolution of the orientation within a grain. The orientation of a material point can 
only change due to grain boundary migration. This is not compatible with heterogeneous 
lattice rotation developing in a grain due to plastic deformation. Since each phase 
is associated with a single orientation, new phase fields must be added to the model 
to accommodate each incremental change in orientation due to e.g. viscoplastic 
deformation. Even then, only a discrete number of orientations can be modeled. 
In reality, all possible crystal orientations should be considered. 

The orientation phase field model due to \cite{KWC2000,KWC2003} (often and in what follows denoted the KWC model) requires only two fields. One is an order parameter $\eta \in [0,1]$ 
which can be regarded as a measure of crystalline order called {\it crystallinity}. 
In a solidification model $\eta=0$ would correspond to the liquid phase and $\eta=1$ 
to the solid phase. For grains in a polycrystal the variation of $\eta$ enables a 
distinction between the bulk of the grain and the grain boundary, $\eta$ being minimal
inside diffuse grain boundaries. The other phase field 
parameter is the orientation field $\theta$, which in the KWC two-dimensional formulation 
denotes the orientation of the crystal lattice with respect to the laboratory frame. 
It is less straightforward to associate the parameters of the KWC model with measurable 
quantities compared to the multi-phase model.   
\cite{LobkovskyWarren2001} found the sharp 
interface limit of the KWC model through formal asymptotics, providing expressions for 
the grain boundary energy and mobility which enables parameter calibration based on 
comparisons with experimental data. While originally formulated in two dimensions, 
extensions to three dimensional orientation field based formulations are discussed 
in e.g. \cite{KobayashiWarren2005} and \cite{PusztaiBortel2005}. In the orientation phase 
field model, the lattice orientation evolution is explicitly taken into consideration and the 
orientation is defined as a continuous field everywhere in the solid body. This enables 
a direct coupling with the evolution of lattice orientation provided by a crystal 
plasticity approach, although a strong coupling between lattice orientation 
as a field variable and lattice reorientation due to deformation was not considered 
by \cite{KWC2000,KWC2003} and does not exist today. Bridging this gap is the main purpose of the present
paper.

A particular feature of the KWC model is that it allows for overall grain rotation,
reminiscent of grain rotation in granular media like soils. 
While there is some evidence supporting rotation towards preferential misorientations 
in nanograins, from experimental observations as well as atomistic simulations 
(cf. \cite{CahnTaylor04,Upmanyu06,TrauttMishin12} and references therein), in larger
micron size 
grains such grain boundary diffusion-driven rotation of the crystal lattice is not 
observed. It can only be suppressed in the uncoupled KWC model by a differentiated 
mobility function distinguishing between behavior in the grain boundary and the bulk of 
the grain \citep{LobkovskyWarren2001}. 

The multi-phase field model of \cite{SteinbachPezzolla1999} was used by \cite{Takaki2010} 
to model static recrystallization in combination with crystal plasticity and 
in \cite{Takaki2009} to model dynamic recrystallization in combination with a 
Kocks-Mecking model to calculate energy stored in the grains during deformation. 
Multi-phase field models were likewise used together with crystal plasticity calculations 
to model sequential recrystallization by e.g. \cite{Guvenc2013, Guvenc2014, Vondrous2015} {and dynamic recrystallization together with a statistical model for nucleation by \cite{ZhaoNiezgoda2016}}. 
In these works, output from the crystal plasticity simulations, such as stored energy, 
was used as input for the nucleation process and/or subsequent grain growth, modeled 
by the phase-field method. \cite{Abrivard2012a,Abrivard2012b} used crystal plasticity 
together with the KWC model in a sequential scheme where the stored energy and calculated 
lattice reorientation served as input to the phase field model. In none of these 
approaches is there a direct coupling of the evolution of mechanical and phase field 
quantities. A trivial, but important, example is that a superposed rigid body motion would 
give rise to lattice reorientation in the mechanical model but no corresponding 
automatic update of the phase fields. 

\subsection{Cosserat crystal plasticity and higher order theories of GB}
Although sequential combinations of different methods have been used successfully 
to model aspects of static and dynamic recrystallization, a complete field theory 
that reconciles the lattice reorientation due to the deformation of a polycrystal 
with the change in orientation at a material point due to migrating grain boundaries 
is still missing. In this work such a theory is proposed based on Cosserat crystal 
plasticity together with an orientation field based approach for diffuse, migrating 
grain interfaces. A full account of the Cosserat theory, which goes back to the brothers 
Cosserat \citep{Cosserat1909}, is given in the seminal work on polar and non-local 
field theories by \cite{Eringen19761}. A Cosserat medium is enriched with additional 
degrees of freedom which describe the rotation of an underlying triad of rigid 
directors at each material point. The notion of oriented microelements characterized 
by some hidden directors were introduced into crystal viscoplasticity by \cite{mandel73}. 

The links between the Cosserat continuum and crystal containing dislocations have been
recognized very early \citep{guenther58}. In the recent Cosserat crystal plasticity theories
by \cite{archmech97,ijss2000,mayeur11,mayeur13}, the Cosserat directors are identified
with lattice crystal directions by means of suitable internal constraints. 
The Cosserat torsion-curvature tensor defined as the gradient of lattice rotation 
has been identified as an essential part of the dislocation density tensor \citep{nye53,kroener63,kroener01,svendsen02}.
The relative rotation between lattice and material directions results from plastic slip
processes with respect to definite slip systems depending on the material crystallographic
structure. It is computed by suitable plasticity flow rules according to standard
crystal plasticity. The lattice curvature follows from non-homogeneous plastic deformation
arising in single or polycrystals. Cosserat crystal plasticity equations were proposed
by \cite{ijss2000,mayeur11} based on the definition of Helmholtz free energy and
dissipation potentials. Finite element implementations of these models were used to study
continuous dislocation pile-up formation at grain-boundaries.

\cite{Blesgen2017} recently proposed a modeling approach for dynamic recrystallisation based on 
a Cosserat crystal plasticity calculation followed by separate recovery and 
softening using a stochastic model for the nucleation of new grains. In a final third step, a 
level set approach is used to model the migration of grain boundaries. 
The Cosserat rotations are restricted to coincide with the lattice rotations through 
a penalty term in the energy functional as in \cite{ijss2000,mayeur11,blesgen14}. However, the evolution of 
orientation through migrating grain boundaries is not directly coupled to the evolution 
of lattice orientation due to deformation and the approach cannot be regarded as a 
complete field theory reconciling the equations 
of crystal plasticity with grain boundary migration but rather as an algorithmic procedure
requiring the use of  several successive types of simulation steps for each time increment.

More elaborate continuum mechanical descriptions of grain boundaries have been recently
proposed in a series of contributions by \cite{fressengeas11,taupin13}.
They involve a detailed description of the GB structure by means of 
appropriate disclination density fields, and resort to higher order continuum theories 
involving couple stresses and incompatible plastic fields \citep{capo13pm}.
Direct comparison with the exact atomic structure of defects in grain boundaries was
possible, showing the accuracy and potential of the continuum approach
\citep{upa16,cordier16}.
The simulation of grain boundary motion under shear loading was performed by
\cite{berbenni13} in agreement with molecular dynamics computations \citep{mcdowell11}.
The Cosserat crystal plasticity coupled with phase field proposed in the present work
 can be seen as a simplified theory 
compared to these higher order theories. It does not account for the inner structure of 
grain boundaries and is applicable at larger scales than the atomic scale.
It is intended to simulate polycrystal plasticity and microstructure evolution at the
micron scale.

\subsection{Objective of the present work}
The objective of the present work is to reconcile the KWC grain boundary motion phase
field model and mechanics within a consistent field theory incorporating crystal elasto-visco-plasticity
and lattice curvature effects on static and dynamic microstructure evolution.
The approach is based on continuum thermodynamics theory of standard materials 
formulated by means of two potentials.
In the previous attempt by \cite{Abrivard2012a} and other previously mentioned 
contributions, the orientation phase field variable
and the lattice rotation induced by crystal plasticity were treated as separate
variables computed from distinct equations. In the present work, the Cosserat theory is used 
to combine both variables into a single physical one. For that purpose, 
the KWC orientation phase field evolution equation is first interpreted as the balance of
moment of momentum of a Cosserat continuum with suitable constitutive laws. 
The Cosserat framework enables a 3D generalization of the KWC models to the fully 
anistropic case. The application examples presented after the theory construction
aim at illustrating the consistency of the model with respect to lattice rotation 
effects induced by various physical phenomena: rigid body motion, elastic deformation and
grain boundary motion induced by non-homogeneous dislocation densities.

The proposed continuum theory is applicable between the
nano and micron scales relevant for grain boundary behavior.
The general setting including kinematics and balance equations is valid 
but the explicit constitutive laws depend on the chosen modeling scale.
Such a constitutive framework is given in the present work for plasticity
and grain boundary migration phenomena at the micron scale with a view
to dynamic recrystallisation simulations in polycrystalline aggregates.
At this scale, the mechanical initial state of grain boundaries is assumed to be
stress--free in contrast to atomic description of defects in grain boundaries
necessarily associated with elastic straining of the lattice in the 
grain boundary region, see the corresponding generalized continuum modeling
proposed at this scale by \cite{fressengeas11}. 
In the proposed theory, GBs possess a fictitious thickness as usual in phase field models.
Asymptotic analyses can be used to estimate some parameters of the KWC model from 
grain boundary energy values.

The outline of the paper is as follows. The Cosserat framework including diffuse
interfaces is presented in Section \ref{sec:seC} where the balance equations for 
momentum, moment of momentum and crystallinity are also derived. The skew-symmetric part of the 
Cosserat stress tensor is then related to the relative rotation between 
lattice directions and material lines by suitable elasto-visco-plastic equations
in Section \ref{sec:seC2}. The theory is presented in a small deformation framework for the
sake of simplicity. Skew-symmetric tensors therefore play an essential role since they
characterize linearized rotations in the model. Also in Section \ref{sec:seC2}, explicit
constitutive choices are proposed that identically fulfill the Clausius-Duhem inequality
and that encompass the previous KWC and Cosserat crystal plasticity models.
The new coupling terms arising in the theory are highlighted and commented especially in the
examples of Section \ref{sec:NumEx}. The first applications deal with the equilibrium profiles
of all variables in the initial grain boundary state and with the evolution of the lattice
orientation field when elastic straining is prescribed to a laminate polycrystal.
The role of the Cosserat coupling modulus on the control of lattice rotation is discussed.
The last example shows grain boundary migration induced by
a gradient of stored energy due to non-homogeneous field of dislocation densities.
The attention is drawn to the evolution of dislocation density due to the sweeping 
of the dislocated crystal by the GB.

\subsection*{Notation}
The following is a brief explanation of the notation used throughout the paper. 
Vectors $a_i$ are denoted by $\vec a$ and second order tensors $A_{ij}$ by $\ten A$. 
Fourth order tensors $C_{ijkl}$ are written as $\TEN C$. 
The third order Levi-Civita permutation tensor $\epsilon_{ijk}$ is denoted by $\leviciv$. 
A skew-symmetric tensor $\ten A^{{\rm skew}}$ can be represented by a pseudo-vector (denoted by a superposed cross) given by
\begin{equation}
\vx A = - \frac{1}{2} \leviciv : \ten A  \,, 
\end{equation}
and likewise the skew-symmetric tensor can be found from the pseudo-vector through
\begin{equation}
\ten A = - \leviciv \cdot \vx A \,.
\end{equation}
Colon indicates double contraction, i.e. $A_{ij}B_{ij}$ is written 
$\ten A : \ten B$. Simple contraction, $a_ib_i$ is written as the usual dot 
product $\vec a \cdot \vec b$. 
The tensor product $\ten A = \vec a \otimes \vec b$ indicates the construction 
$A_{ij} = a_ib_j$. The gradient $\partial a_i/\partial x_j$ 
and divergence $\partial a_i/\partial x_i$ of a vector $\vec a$ are written 
as $\vec a \otimes \nabla$ and $\nabla \cdot \vec a$, respectively. 
The divergence of a tensor is given by $\ten A \cdot \nabla$ with differentiation acting on the second index, i.e. $\partial A_{ij}/\partial x_j$.

\section{Cosserat framework with diffuse interfaces}
\label{sec:seC}
This section is dedicated to the description of the kinematics and 
order parameters, the derivation of the balance laws and the general formulation
of the constitutive laws. Detailed expressions for the latter are 
postponed to the next section. The presentation is given within the 
small strain, small rotation and small curvature framework.
\subsection{Cosserat kinematics, order parameters and deformation measures}
In a Cosserat continuum each material point is endowed with three 
translational degrees of freedom, the displacement vector $\vec u$, 
and three independent rotational 
degrees of freedom, the microrotation pseudo-vector $\vec \varTheta$. {Later in this work, $\vec \varTheta$ will be associated with the crystallographic lattice rotations by suitable constitutive relations.}
In the small deformation setting, the current state of a triad of orthonormal 
directors is related to the original state by the Cosserat microrotation 
tensor $\ten R$, given by
\begin{equation}
\ten R = \ten I - \leviciv \cdot \vec \varTheta \,,
\end{equation}
where $\ten I$ is the identity tensor and $\leviciv$ is the Levi-Civita permutation tensor. The objective deformation measures are the deformation tensor $\ten e$ and the curvature (or wryness) tensor $\ten \kappa$, given by
\begin{equation}
\ten e= \vec u \otimes \nabla + \leviciv \cdot \vec \varTheta \,, 
\qquad
\ten \kappa = \vec \varTheta \otimes \nabla \,.
\end{equation}
The deformation tensor can be decomposed into symmetric and skew-symmetric contributions given by
\begin{align}
\ten e^{{\rm sym}} = \, &  \frac{1}{2}\,\left[\,
\vec u \otimes \nabla +\nabla \otimes \vec u
 \,\right] = \ten \varepsilon \,, \\
 \ten e^{{\rm skew}} = \, &  \frac{1}{2}\,\left[\,
\vec u \otimes \nabla - \nabla \otimes \vec u
 \,\right] + \leviciv \cdot \vec \varTheta  \,,
\end{align}
and equivalently the rate of deformation can be decomposed so that
\begin{align}
\dot{\ten e}^{{\rm sym}} = \, &  \frac{1}{2}\,\left[\,
\dot{\vec u} \otimes \nabla +\nabla \otimes \dot{\vec u}
 \,\right] = \dot{\ten \varepsilon} \,, \\
 \dot{\ten e}^{{\rm skew}} = \, &  \frac{1}{2}\,\left[\,
\dot{\vec u} \otimes \nabla - \nabla \otimes \dot{\vec u}
 \,\right] + \leviciv \cdot \dot{\vec \varTheta} =
 \ten \omega + \leviciv \cdot \dot{\vec \varTheta} \,,
\end{align}
where $\dot{\vec u}$ is the material velocity vector.
The symmetric tensor $\ten \varepsilon$ is the usual small strain tensor 
and $\ten \omega$ is the so-called spin tensor. The skew-symmetric tensor 
$\ten e^{{\rm skew}}$ can be represented by a pseudo-vector whose 
rate can be written as
\begin{equation}
\dot{\vx e} =  \vx \omega - \dot{\vec \varTheta} \, ,
\end{equation}
where it is apparent that $\ten e^{{\rm skew}}$ represents the relative
rotation of the material with respect to the microstructural directors.
The small deformation setting allows for additive decomposition of the 
deformation measures into elastic and plastic contributions. Furthermore, 
a third contribution is introduced in the form of an eigen-deformation 
tensor $\ten e^\star$. It is analogous to an eigenstrain and is introduced 
to allow for cases where a Cosserat deformation exists but does not give 
rise to any stresses: 
\begin{equation}
\ten e = \ten e^{e}+ \ten e^{p} + \ten e^\star \,,
\end{equation}
It is sufficient for the purpose of this work to assume that the eigendeformation is skew-symmetric, and therefore the following holds
\begin{align}
\ten \varepsilon = \, & \ten \varepsilon^e + \ten \varepsilon^p \,, \\
\vx e = \, & \vx e\,^e + \vx e\,^p + \vx e\,^\star \,.
\end{align}
The pseudo-vector $\vx e\,^\star$ is now called {\it eigen-rotation} to highlight that it represents a relative rotation with respect to a fixed frame. {With the Cosserat micro-rotation related to the crystallographic 
lattice rotation, the eigen-rotation represents a reference crystallographic 
orientation. Such a change of reference is likely to occur in a moving grain boundary. }
Likewise, for the rates 
\begin{align}
\dot{\ten \varepsilon} = \, & \dot{\ten \varepsilon}^e + \dot{\ten \varepsilon}^p \,, \\
\dot{\vx e} = \, & \vx \omega - \dot{\vec \varTheta} = \dot{\vx e}\,^e + \dot{\vx e}\,^p + \dot{\vx e}\,^\star \,. 
\label{rate_easym}
\end{align}
The plastic and elastic and spin tensors are defined as 
$\vx \omega\,^p := \dot{\vx e}\,^p$ and $\vx \omega\,^e := \vx \omega -\vx \omega\,^p$. Equation (\ref{rate_easym}) can then be written as
\begin{equation}
\vx \omega\,^e - \dot{\vec \varTheta} = \dot{\vx e}\,^e + \dot{\vx e}\,^\star \,.
\label{eedot_omega}
\end{equation}
As usual in crystal plasticity, the elastic spin tensor {still} accounts for 
the rotation rate of the crystal lattice with respect to the initial orientation {inside of the grain}.
At this stage, the lattice directions and the Cosserat directors are 
distinct vectors.    
The previous relation provides a link between the lattice and microrotation
rates. In the absence of eigen-rotation, the lattice and Cosserat rotations coincide if the elastic deformation is symmetric, i.e. 
\begin{equation}
\vx e\,^e \equiv 0. 
\end{equation}
This internal constraint is imposed in Cosserat crystal plasticity so that
the Cosserat directors remain parallel to lattice vectors during the 
deformation process \citep{archmech97,ijss2000,mayeur11,mayeur14,blesgen14}.
In the latter references, the constraint is enforced by a penalty 
method involving high values of the Cosserat elastic plastic modulus,
as defined in Section \ref{sec:specons}.
As a result, the Cosserat curvature tensor coincides with the lattice 
curvature tensor. {It would also be possible to strictly enforce the constraint through a Lagrange multiplier method.}

In order to study moving grain boundaries, the Cosserat model is enhanced 
with a phase field variable $\eta \in [0,1]$ and its gradient $\nabla \eta$. 
The variable $\eta$ can be considered as a coarse-grained or macroscopic order parameter where $\eta=1$ 
in the bulk of a grain of perfect crystalline order and $\eta < 1$ 
in the diffuse grain boundary where crystal order is lower. 
In the presence of plastic deformation the requirement $\eta = 1$ in the bulk 
of the grain must be relaxed, allowing for the disordering of the crystal 
lattice through dislocation densities building up and associated stored energy. 

\subsection{Balance laws}
\label{sec:balancelaws}
The method of virtual power is used to introduce the generalized stresses of the
theory and derive the corresponding balance laws. The set of virtual field 
variables entering the principle of virtual power is given by 
\begin{equation}
\mathcal{V} = \lbrace 
\virt{\eta},
\nabla \virt{\eta},
\virt{\vec u},
\virt{\vec u}\otimes \nabla,
\virt{\vec \varTheta},
\virt{\vec \varTheta}\otimes \nabla
\rbrace
\,.
\end{equation}
The Cosserat deformation tensor $\ten e$ is work conjugate to the force stress tensor
$\ten \sigma$ of unit [Pa] and the curvature tensor $\ten \kappa$ is work 
conjugate to the couple stress tensor $\ten m$ of unit [Pa$\cdot$m]. 
In accordance with the formalism developed in \citep{Gurtin1996, GurtinLusk1999,ammar09},
 the phase field variable $\eta$ and its gradient are assumed to be associated 
with the generalized stresses $\pi_\eta$ and $\underline{\vc \xi}_\eta$, 
of units [Pa] and [Pa$\cdot$m] respectively. The virtual power density of 
internal forces is then given by
\begin{equation}
p^{(i)} = -\pi_\eta \virt \eta + \vec \xi_\eta \cdot \nabla \virt \eta
+ \ten \sigma : \virt{\ten e} +
 \ten m : \virt{\ten \kappa}
= -\pi_\eta \virt \eta + \vec \xi_\eta \cdot \nabla \virt \eta
+ 
\ten \sigma :
\virt{\vec u}\otimes \nabla - 2\,\vx \sigma \cdot \virt{\vec \varTheta}
+
\ten m : \virt{\vec \varTheta}\otimes \nabla
\,.
\label{pvp_internal_coupled}
\end{equation}
where  the pseudo-vector $\vx \sigma$ contains the skew-symmetric contributions 
to the stress tensor. It can be checked that the power density of internal forces
is invariant with respect to the superposition of rigid body motion, as it should
be, due to the fact that the relative deformation satisfies this invariance 
property. The external virtual power contribution due to body forces and couples is
\begin{equation}
p^{(e)} = \pi_\eta^{\rm{ext}} \virt \eta + \vec f^{{\rm ext}} \cdot \virt{\vec u} +
\vec c^{{\rm ext}} \cdot \virt{\vec \varTheta}
\label{pvp_external_coupled}
\end{equation}
and the external virtual power density due to contact forces and couples is given by
\begin{equation}
p^{(c)} = \pi_\eta^{c} \virt \eta + \vec f^{c} \cdot \virt{\vec u} +
\vec c^{c} \cdot \virt{\vec \varTheta} \,.
\label{pvp_contact_coupled}
\end{equation}
The principle of virtual power then states that, over any region $\cal{D}$ with boundary $\partial \cal{D}$ of $\Omega$,
\begin{equation}
\int_{\cal{D}} p^{(i)}  {\rm d} V = \int_{\cal{D}} p^{(e)}  {\rm d} V + \int_{\partial \cal{D}} p^{(c)}  {\rm d} V \,.
\end{equation}
for all virtual fields $\dot \eta, \virt{\vec u}, \virt{\vec \varTheta}$.
By integrating by parts
\begin{align}
& \vec \xi_\eta \cdot \nabla \virt \eta = 
\nabla \cdot \left[\vec \xi_\eta\,\virt \eta\right] 
-\nabla \cdot \vec \xi_\eta\,\virt \eta \,, \\
& \ten \sigma :
\virt{\vec u}\otimes \nabla = \left[
\virt{\vec u} \cdot \ten \sigma
\right] \cdot \nabla -\virt{\vec u} \cdot 
 \ten \sigma \cdot \nabla \,, \\
& \ten m:
\virt{\vec \varTheta}\otimes \nabla = \left[
\virt{\vec \varTheta} \cdot \ten m
\right] \cdot \nabla -\virt{\vec \varTheta} \cdot 
 \ten m \cdot \nabla \,,
\end{align}
and applying the divergence theorem, the principle of virtual power for all ${\cal{D}} \subset \Omega$ and $\forall \,( \virt \eta, \virt{\vec \varTheta}, \virt{\vec u})$ can be rewritten as
\begin{equation}
\begin{aligned}
&
\int_{\cal{D}} \virt{\eta} \,\left[
\nabla \cdot \underline{\vc \xi}_\eta + \pi_\eta + \pi_\eta^{{\rm ext}}
\right] \, {\rm d} V  
+  \int_{\partial \cal{D}} \virt{\eta} \,\left[
   \pi_\eta^{c} - \vec \xi_\eta \cdot \vec n 
\right] \, {\rm d} S 
\\
+ & \int_{\cal{D}} \virt{\vec u} \cdot \left[
 \ten \sigma\cdot \nabla + \vec f^{{\rm ext}}
\right] \, {\rm d} V  
+  \int_{\partial \cal{D}} \virt{\vec u} \cdot \left[
 \vec f^{c} - \ten \sigma\cdot \vec n 
\right] \, {\rm d} S 
\\
+ & 
\int_{\cal{D}} \virt{\vec \varTheta} \cdot \left[
\ten m \cdot \nabla + 2\,\vx \sigma  + \vec c^{{\rm ext}}
\right] \, {\rm d} V 
+
\int_{\partial \cal{D}} \virt{\vec \varTheta} \cdot \left[
\vec c^{c} - \ten m \cdot \vec n 
\right] \, {\rm d} S = 0 \,, 
\end{aligned}
\end{equation}
where $\vec n$ is the outward unit normal to $\partial \cal{D}$. The balance equations and boundary conditions follow directly and are given by
\begin{align}
& \nabla \cdot \underline{\vc \xi}_\eta + \pi_\eta + \pi_\eta^{{\rm ext}} = 0  \qquad 
& {\rm in} \,\, \Omega \,, \label{gurtinbal} \\ 
&  \ten \sigma \cdot \nabla + \vec f^{{\rm ext}} = \vec 0
\qquad & {\rm in} \,\, \Omega \,, \label{momentum} \\
& \ten m \cdot \nabla + 2\, \vx \sigma + \vec c^{{\rm ext}} = \vec 0 \qquad & {\rm in} \,\, \Omega \,,  \label{momomentum}\\
& \vec \xi_\eta \cdot \vec n = \pi_\eta^{c}
&\qquad {\rm on} \,\, \partial \Omega \,, \\
& \ten \sigma \cdot \vec n = \vec f^{c}
\qquad & {\rm on} \,\, \partial \Omega \,, \\
& \ten m \cdot \vec n = \vec c^{c}
\qquad & {\rm on} \,\, \partial \Omega \,.
\end{align}
Equation (\ref{momentum}) represents the balance of momentum for a generally
non-symmetric stress tensor,  whereas the balance of moment of momentum 
is given by equation (\ref{momomentum}). The latter equation is coupled to the former via
the skew-symmetric part of the stress tensor, as expected in a Cosserat
continuum \citep{withers01}.
Equation (\ref{gurtinbal}) is the balance of generalized stresses 
derived by \citep{Gurtin1996,ammar09}
for phase field models and by \cite{Abrivard2012a} in the context of grain boundary migration.

\subsection{Constitutive equations}
The Clausius-Duhem inequality states that (for isothermal conditions)
\begin{equation}
-\rho \, \dot{\Psi} + p^{(i)} \geq 0 \,,
\label{clausius-duhem}
\end{equation}
where $\Psi$ is the Helmholtz free energy density which is assumed to take $\eta$, $\nabla \eta$, $\ten e^e$ and $\ten \kappa$ as arguments together with additional internal variables $r^\alpha$ related to the inelastic behavior, so that
\begin{equation}
\rho \,\Psi = \psi(\eta,\nabla \eta, \ten e^e, \ten \kappa,r^\alpha) \,.
\end{equation}
With the density of internal power given by (\ref{pvp_internal_coupled}), the Clausius-Duhem inequality becomes
\begin{equation}
\begin{aligned}
& -\left[\pi_\eta + \frac{\partial \psi}{\partial \eta}  \right]\dot{\eta}
+
\left[ \vec \xi_\eta- \frac{\partial \psi}{\partial \nabla \eta}  \right]\cdot \nabla \dot{\eta}
+
\left[\ten \sigma-  \frac{\partial \psi}{\partial \ten e^{e}}
\right]:\dot{\ten e}^{e} +
\left[\ten m -  \frac{\partial \psi}{\partial \ten \kappa}
\right]:\dot{\ten \kappa} \\
& +
\ten \sigma:\dot{\ten e}^{\star}+
\ten \sigma:\dot{\ten e}^{p}
-\frac{\partial \psi}{\partial r^\alpha}
\dot{r}^\alpha \geq 0 \,.
\end{aligned}
\end{equation}
The construction of the model is based on the appropriate choice of dissipative
and non-dissipative contributions in the previous inequality, with a view
to modelling grain boundary migration in elasto-visco-plastic crystals.
In order to account for the relaxation behavior of a phase field model, the stresses $\pi_\eta$ and $\vx \sigma$ are assumed to contain dissipative contributions (in analogy with a rheological Kelvin-Voigt element) so that
\begin{align}
& \pi_\eta = -  \frac{\partial \psi}{\partial \eta} + \pi_\eta^{neq} \,, \\ 
& \vx \sigma = \vx \sigma\,^{eq} + \vx \sigma\,^{neq} \,,
\end{align}
with the non-dissipative, non-symmetric stress given by
\begin{equation}
\ten \sigma^{eq} =  \frac{\partial \psi}{\partial \ten e^e} \,.
\end{equation}
Note that the dissipative contribution of the stress is limited to its 
skew-symmetric part, for reasons that will be explained in the 
next section where specific explicit constitutive laws are presented. For the microstress $\vec \xi_\eta$ and the couple stress $\ten m$ such dissipative effects are neglected and the constitutive relations are given by
the following state laws:
\begin{align}
& \vec \xi_\eta =  \frac{\partial \psi}{\partial \nabla \eta} \,,  \\
& \ten m =  \frac{\partial \psi}{\partial \ten \kappa} \,.
\end{align}
The thermodynamic forces associated with the internal variable are called
\begin{equation}
R^\alpha =  \frac{\partial \psi}{\partial r^\alpha} \,.
\end{equation}
The remaining part of the dissipation inequality can now be written as
\begin{equation}
-\pi_\eta^{neq}\,\dot{\eta} +
2\,\vx \sigma\,^{neq} \cdot [\,\vx \omega\,^e - \dot{\vec \varTheta} \,] + 
2\,\vx \sigma\,^{eq} \cdot \dot{\vx e}\,^\star+
\ten \sigma:\dot{\ten e}^{p}
-R^\alpha \dot{r}^\alpha \geq 0 \,.
\label{diff_part}
\end{equation}
where the result in equation (\ref{eedot_omega}) was used. In the absence of displacements the format above corresponds to the format found by \citep{Abrivard2012a} for the KWC model. Note that the non-equilibrium contribution to the stress is only associated with the skew-symmetric part of the deformation.

\subsection{Evolution equations}
For the dissipative contributions, a dissipation potential is introduced 
and decomposed in five parts
\begin{equation}
\varOmega = \varOmega^\theta(\vx \sigma\,^{neq}) + \varOmega^\star(\vx \sigma\,^{eq}) 
+ \varOmega^p(\ten \sigma) + \varOmega^\alpha(R^\alpha) 
+ \varOmega^\eta(\pi_\eta^{neq}) \,,
\end{equation}
such that
\begin{equation}
\vx \omega\,^e - \dot{\vec \varTheta} =  \frac{\partial \varOmega^\theta}{\partial \vx \sigma\,^{neq}}\,, \quad
\dot{\vx e}\,^\star =  \frac{\partial \varOmega^\star}{\partial \vx \sigma\,^{eq}} \,, \quad
\dot{\ten e}^p  =  \frac{\partial \varOmega^p}{\partial \ten \sigma}, \quad
\dot r^\alpha = - \frac{\partial \varOmega^\alpha}{\partial R^\alpha}, \quad
\dot \eta  = -\frac{\partial \varOmega^\eta}{\partial \pi^{neq}_\eta}.
\label{dissip_pots}
\end{equation}
The skew-symmetric viscous and elastic parts of the stress tensor are therefore
taken as the driving forces for the evolution of the difference between
lattice and Cosserat rotation and for the eigen-rotation, respectively. These quantities are conjugate in the dissipation rate inequality (\ref{diff_part}). {The first and last contributions of equation (\ref{dissip_pots}) are necessary to recover the evolution equations corresponding to the orientation phase field model in \cite{KWC2000,KWC2003}. A similar format was used in the uncoupled approach of \cite{Abrivard2012a}.}
The plastic flow rule and evolution laws for internal variables are also derived
from the dissipation potential, as in standard generalized materials
\citep{germain83,maugin92book,bccf09}. 
Inserted into the dissipation equation (\ref{diff_part}), this gives
\begin{equation}
\pi_\eta^{neq}\, \frac{\partial \varOmega^\eta}{\partial \pi^{neq}_\eta}+
2\,\vx \sigma\,^{neq} \cdot \frac{\partial \varOmega^\theta}{\partial \vx \sigma\,^{neq}}+
2\,\vx \sigma\,^{eq} \cdot \frac{\partial \varOmega^\star}{\partial \vx \sigma\,^{eq}}+
\ten \sigma:\frac{\partial \varOmega^p}{\partial \ten \sigma}
+ R^\alpha \frac{\partial \varOmega}{\partial R^\alpha}
\geq 0 \,.
\label{diff_part_pot}
\end{equation}
The positivity of the dissipation rate for any process is ensured by 
suitable convexity properties of the dissipation potential $\varOmega$
with respect to its arguments.

\section{Free energy and dissipation potentials}
\label{sec:seC2}
In order to close the system of equations it is necessary to choose specific 
forms of the free energy function $\psi$ and of 
the dissipation potential $\varOmega$.
A general anisotropic setting is first proposed and then specialized to the
isotropic case in order
to recover the KWC equations and highlight their coupling with mechanical
contributions.

\subsection{General anisotropic behavior}
A general form of the free energy function which incorporates the phase field 
variables and the deformation measures is proposed:
\begin{equation}
\begin{aligned}
\psi(\eta,\nabla \eta, \ten e^e, \ten \kappa,r^\alpha) 
= \, &
f_0 
\left[ 
f(\eta) + \frac{1}{2} \, \nabla \eta \cdot \ten A \cdot \nabla \eta + 
s\,g(\eta)||\ten \kappa|| + 
\frac{\varepsilon^2}{2} \, \ten \kappa : \TEN H(\eta) : \ten \kappa 
\right ] 
\\
+ \, & \frac{1}{2}\ten e^e : \TEN E(\eta):
\ten e^e  + \psi_\rho(\eta,r^\alpha)
\,.
\label{psitot}
\end{aligned}
\end{equation}
Simple quadratic contributions are chosen for the gradient phase field as
usual in phase field models, and for Cosserat elasticity via the fourth
rank tensors $\TEN H$ and $\TEN E$. The latter are generally anisotropic.
They display major symmetry but not minor
symmetry due to the 
non-symmetric nature of the deformation and curvature tensors $\ten e^e$ and $\ten \kappa$. 
They may depend on the order parameter in order to distinguish 
the behavior in the bulk of the grain from that in the grain boundary. 
The second order tensor $\ten A$ is symmetric. 
The normalization parameter $f_0$ is taken with unit [Pa] or [J/m$^3$] 
while $\ten A$ has unit [m$^2$], $s$ and $\varepsilon$ have unit [m], and $\TEN H$ 
is dimensionless. {The function $g(\eta)$ is required to satisfy $g(0) = 0$ and should increase monotonically with $\eta$ \citep{KWC2000}.}
The term involving the norm of the elastic curvature tensor comes in addition
to the quadratic contribution. It is inspired 
by the curvature vector in the KWC model and from potentials involving 
the dislocation
density tensor proposed in \citep{ortizconti05,ohno07,inspection13,wulfinghoffForest14}. The lattice curvature tensor is often used as an approximation of the 
full dislocation density tensor following \citep{nye53}. {In the KWC model, the linear contribution (norm of the curvature tensor) 
acts to localize the grain boundaries whereas the quadratic terms acts as regularization by diffusing them \citep{KWC2000}. It may also be noted that for small misorientations, the format they suggest can be considered as a series expansion of the Read-Shockley energy to second order. In terms of the physical interpretation of the format (\ref{psitot}), the linear term in the norm of the curvature can interpreted as the self energy of the geometrically necessary dislocations (GND) whereas the quadratic term represents the elastic interaction between them \citep{ohno07,mesarovicforest2015}.} 
The term $\psi_\rho(\eta,r^\alpha)$ contains the contribution due to $N$ internal variables $r^\alpha$ which will be associated with {statistically
stored dislocation (SSD) density based 
hardening, as a complement to GND stored energy.}  

The resulting state laws for the generalized stresses are given by
\begin{align}
\ten \sigma^{eq} = \, & \TEN E : \ten e^e  \,, \\
\ten m = \, &  f_0\left[
\,s\,g(\eta)\frac{\ten \kappa}{||\ten \kappa||}
 +\varepsilon^2\,\TEN H(\eta) : \ten \kappa\,
\right] \,, \label{mstress_constitutive}\\
\pi_\eta = \,& - f_0 \left[\,f_{,\eta} + 
s\,g_{,\eta} ||\ten \kappa|| +
\frac{\varepsilon^2}{2} \, \ten \kappa : \TEN H\,_{,\eta} : \ten \kappa 
\,\right]  - \frac{1}{2} \, \ten e^e :\TEN E\,_{,\eta}:\ten e^e
- \psi_{\rho,\eta}  + \pi_\eta^{neq}   \,, \\
\vec \xi_\eta = \, &  f_0\, \ten A \cdot \nabla \eta \,.
\end{align}
where the possible dependence on $\eta$  of all elastic moduli has been 
taken into account, the notation $\bullet_{,\eta}$ standing for partial derivation 
w.r.t. $\eta$. {Due to the inclusion of the linear term $||\ten \kappa||$ in the energy density, equation (\ref{mstress_constitutive}) requires special attention. According to Eq. (\ref{mstress_constitutive}), the couple stress tensor is singular at zero curvature.
Taking the divergence of $\ten m$, such as in the balance equation (\ref{momomentum}), results in a singular diffusive type equation for which the mathematical treatment was given by \cite{KobayashiGiga1999}. In the numerical treatment, the singular term is regularized (c.f. \ref{app1}).}

Quadratic dissipation potentials $\varOmega^\theta$ and $\varOmega^\star$ are chosen. 
They ensure the positivity of the corresponding
contributions in the dissipation inequality. Such simple forms are appropriate for the 
description of the relaxation phenomena in grain boundaries: 
\begin{align}
\varOmega^\theta = \, & \frac{1}{2} \vx \sigma\,^{neq} \cdot \ten \tau_\theta^{-1} \cdot \vx \sigma\,^{neq} \,, \label{sigv}\\
\varOmega^\star = \, & \frac{1}{2} \vx \sigma\,^{eq} \cdot \ten \tau\,_\star^{-1} \cdot \vx \sigma\,^{eq} \,, \label{estar}
\end{align}
where $\ten \tau_\theta$ 
and $\ten \tau_\star$ are constitutive symmetric 
positive definite, thus invertible, second order tensors. 
These material parameters govern the dissipation, they are associated with the 
mobility of the grain boundaries in the phase field model and have unit [Pa$\cdot$s]. 
They may depend in general on the full set of state variables 
$\lbrace \eta,\nabla \eta, \ten \kappa, T \rbrace$.
The equation for the dissipative stress and the evolution law for the eigen-rotation, respectively, then become
\begin{align}
\vx \sigma\,^{neq} = \, & \ten \tau_\theta \cdot [\,\vx \omega\,^e - \dot{\vec \varTheta}\,] \,, \label{stress_v_evol} \\
 \vx \sigma\,^{eq}= \, &  \ten \tau_\star \cdot  \dot{\vx e}\,^\star \,. \label{estar_evol}
\end{align}
These constitutive equations therefore are viscosity laws for the skew-symmetric
parts of the stress tensors. 
Equation (\ref{stress_v_evol}) is used to compute the
viscous stress from the relative rotation rate, whereas (\ref{estar_evol}) 
is the evolution law for the eigen-rotation in the form of a
relaxation equation driven by the reversible part of the stress. \\
The evolution of the crystal order parameter is likewise provided by a 
quadratic potential:
\begin{equation}
\varOmega^\eta = \tau_\eta^{-1} \, \pi_\eta^2
\end{equation}
in the form of the usual phase field relaxation equation
as in \citep{Gurtin1996,KWC2000,Abrivard2012a}:
\begin{equation}
- \pi_\eta^{neq} =  \tau_\eta \, \dot{\eta} \label{eta_evol}\,,
\end{equation}
where $\tau_\eta(\eta,\nabla \eta, \ten \kappa, T)$ is a positive scalar 
function. From inspection of (\ref{diff_part_pot}) it is apparent that the 
formats (\ref{stress_v_evol}), (\ref{estar_evol}) and (\ref{eta_evol}) lead 
to non-negative contributions to the energy dissipation. 

\subsection{Isotropic grain boundary behavior}
\label{sec:specons}
The model equations can be simplified by specialization to the case of isotropic grain boundary behavior and a separation between symmetric and skew-symmetric contributions to the deformation. The free energy is then given by
\begin{equation}
\begin{aligned}
\psi(\eta,\nabla \eta, \ten e^e, \ten \kappa,r^\alpha) 
= \, &
f_0 
\left[ 
f(\eta) + \frac{a^2}{2}| \nabla \eta|^2 + 
s\,g(\eta)||\ten \kappa|| + 
\frac{\varepsilon^2}{2}h(\eta)||\ten \kappa||^2 
\right ] 
\\
+ \, & \frac{1}{2}\ten \varepsilon^e : \TEN E^s :
\ten \varepsilon^e + 2\,\mu_c(\eta)\,\vx e\,^e \cdot\vx e\,^e + \psi_\rho(\eta,r^\alpha)
\,.
\end{aligned}
\label{energy_isotropic}
\end{equation}
The term inside brackets is essentially the same\footnote{The only 
difference lies in the 3D formulation instead of the 2D KWC model.} as introduced by \citep{KWC2000,KWC2003}. 
The elasticity tensor $\TEN E^s$ now has both major and minor symmetry and 
coincides with the classical Hooke tensor. It may be anisotropic 
whereas the skew-symmetric contribution is taken to be isotropic.
The Cosserat elastic modulus $\mu_c$ is called the coupling modulus
\citep{lakes85,neff06}. 
It plays an essential role in Cosserat mechanics since it relates the 
relative rotation to the skew-symmetric part of the stress tensor. 
It acts as a penalty term limiting the magnitude of the relative 
rotation.
In the present model, it
may depend on the phase field $\eta$ and thus take different values
in the grain boundary and in the bulk.
The resulting state laws are given by
\begin{align}
\pi_\eta = \, & -  \frac{\partial \psi}{\partial \eta} + \pi_\eta^{neq} 
= - f_0 \left[\,f_{,\eta} + 
s\,g_{,\eta} ||\ten \kappa|| +
\frac{\varepsilon^2}{2}h_{,\eta}||\ten \kappa||^2  
\,\right]  - 2\,\mu_{c,\eta} \,\vx e\,^e \cdot\vx e\,^e  
- \psi_{\rho,\eta} + \pi_\eta^{neq}  \,, \\
\ten \sigma^{eq} = \, &  \frac{\partial \psi}{\partial \ten e^e} 
=
\TEN E^s : \ten \varepsilon^e - 2\,\mu_c(\eta) \leviciv \cdot \vx e\,^e  \,, \label{stress_isotrop} \\
\vec \xi_\eta = \, &  \frac{\partial \psi}{\partial \nabla \eta} = f_0\, a^2 \nabla \eta \,, \\
\ten m = \, &  \frac{\partial \psi}{\partial \ten \kappa} = f_0\left[
\,s\,g(\eta)\frac{\ten \kappa}{||\ten \kappa||}
 +\varepsilon^2\,h(\eta)\,\ten \kappa\,
\right] \,.
\end{align}
The skew stress pseudo-vector is given by
\begin{equation}
\vx \sigma = 2\, \mu_c(\eta) \, \vx e\,^e + \vx \sigma\,^{neq} \,.
\label{skewstress_pseudo}
\end{equation}
The isotropic forms of equations (\ref{stress_v_evol},\ref{estar_evol}) are
\begin{align}
\vx \sigma\,^{neq} = \, & \tau_\theta \, [\,\vx \omega\,^e - \dot{\vec \varTheta}\,] \,, \\
\vx \sigma\,^{eq} = \, & \tau_\star \, \dot{\vx e}\,^\star \,. \label{evol_estar_isotropic}
\end{align}
Inserting the above results into the balance laws for $\vec \xi_\eta$ and $\ten m$ (while assuming all external body forces and couples are zero) 
provides the evolution and partial differential equations for the phase field $\eta$ 
and the relative rotation in the form
\begin{align}
\, & \tau_\eta\,\dot{\eta} =  f_0\, a^2 \Delta \eta - f_0 \left[\,f_{,\eta} + 
s\,g_{,\eta} ||\ten \kappa|| +
\frac{\varepsilon^2}{2}h_{,\eta}||\ten \kappa||^2  
\,\right] 
\cbox{- 2\,\mu_{c,\eta} \,\vx e\,^e \cdot\vx e\,^e}
- \psi_{\rho,\eta} 
\,, \label{evol_eta} \\
- \, & \tau_\theta \cdot [\,\cbox{\vx \omega\,^e} - \dot{\vec \varTheta}\,] =   \frac{f_0}{2}\left[
\,s\,g(\eta)\frac{\ten \kappa}{||\ten \kappa||}
 +\varepsilon^2\,h(\eta)\,\ten \kappa\,
\right]  \cdot \nabla \cbox{+ 2\,\mu_c(\eta) \, \vx e\,^e} \label{evol_theta} \,.
\end{align}
The fundamentally new terms compared to the KWC and Abrivard models 
are highlighted in 
{gray}. They correspond to the contributions ensuring full coupling between
phase field and mechanics.
In the case of $\mu_c(\eta) = 0$, the format of the evolution equations 
closely resembles the model by \cite{KWC2000,KWC2003} with the additional 
SSD stored energy term introduced by \cite{Abrivard2012a}. 

In the present work the Cosserat pseudo-vector $\vec \varTheta$ is interpreted as the rotation of the lattice vectors with respect to a fixed reference. Typically, $\vec \varTheta$ might be associated with the lattice orientations of the grains in a polycrystal with respect to the laboratory frame. It follows that $\vec \varTheta \neq \vec 0$ in general even in the otherwise undeformed configuration, i.e. with $\vec u = \vec 0$. 
This highlights the need for introducing the eigen-rotation $\vx e\,^\star$. 
When $\vec u = \vec 0$, the skew-symmetric deformation is given by $\vx e\,^e = -\vec \varTheta - \vx e\,^\star$, according to Eq. (\ref{eedot_omega}). 
The choice of initial values $\vx e\,^\star = -\vec \varTheta$ ensures that the skew-symmetric 
equilibrium stress of equation (\ref{stress_isotrop}) vanishes. Without the introduction of $\vx e\,^\star$, it would not be possible to have a stress-free state where $\vec u=\vec 0$ at the same time as $\vec \varTheta \neq \vec 0$. The introduction of the eigen-rotation can therefore be seen as the introduction of a stress-free reference orientation of the grain.
This reference evolves according to Eq. (\ref{evol_estar_isotropic}) during grain boundary migration. 

In the original KWC model, the evolution law for the orientation is given by the terms not marked with red in equation (\ref{evol_theta}). This format allows for the orientation inside a grain, even far from the grain boundary, to evolve through rotation \citep{KWC2000}.  {In order to prevent this rotation it is necessary to separate the time scales of the evolution in the grain boundary and in the bulk of the grain.} In \citep{KWC2003}, this was done by considering a mobility function of the form \begin{equation}
\tau_\theta(\eta,\nabla \eta, \nabla \theta) = \frac{1}{2} P(||\ten \kappa||)\,\hat{\tau}_\theta\,\eta^2 \,,
\label{tau_theta_gen}
\end{equation}
where $\hat{\tau}_\theta$ is a constant and the mobility function $P(||\ten \kappa||)$ takes a high value for vanishing lattice curvature and a low value in the grain boundaries where the lattice curvature is pronounced.  {In the coupled model it is expected that this separation of time scales is not necessary since the rotation in the bulk can be prevented by the Cosserat coupling term and the penalty parameter $\mu_c(\eta)$. This represents an improvement
of the KWC model} and will be subject of the section \ref{sec:rotcontrol}.


\subsection{Crystal plasticity}

In a crystal, plastic deformation takes place by slip on preferred directions 
given by $N$ discrete slip systems. The driving force for the activation 
of slip on the crystallographic slip system number $\alpha$ is the resolved 
shear stress $\tau^\alpha$ calculated as
\begin{equation}
\tau^\alpha = \vec \ell^\alpha \cdot \ten \sigma \cdot  \vec n^\alpha
\end{equation}
where $ \vec \ell^\alpha$ and $ \vec n^\alpha$ are, respectively, 
the slip direction and normal to the slip plane. 
The kinematics of plastic flow is governed by 
\begin{equation}
\dot{\ten e}^p = \sum_{\alpha=1}^N \dot{\gamma}^\alpha \, \vec \ell^\alpha \otimes \vec n^\alpha \,.
\label{plastic_flow}
\end{equation}
The slip rate $\dot{\gamma}^\alpha$ is calculated according to 
a viscoplastic flow rule taken from \cite{cailletaud-92}:
\begin{equation}
\dot{\gamma}^\alpha = \left< \frac{| \tau^\alpha | - R^\alpha}{K_v} \right>^n
{\rm sign} \, \tau^\alpha
\end{equation}
where $\left< \bullet \right> = {\rm Max}(\bullet, 0)$.
The critical resolved shear stress for slip system $\alpha$ is $R^\alpha$ 
and $K_v$ and $n$ are viscosity parameters, with $n$ being the Norton power law exponent.

An essential driving force for grain boundary migration is the energy stored 
due dislocation accumulation. The SSD dislocation part of the free energy density
function in equation (\ref{psitot}) is now taken as
\begin{equation}
\psi_\rho(\eta,r^\alpha) = \eta \, \sum_{\alpha = 1}^N \frac \lambda 2
\mu r^{\alpha 2} \,,
\label{dislo_energy}
\end{equation}
where $\lambda$ is a parameter close to 0.3 \citep{hirthlothe}
and $\mu$ is the shear modulus.
The internal variables $r^\alpha$ are
now related to the SSD dislocation densities $\rho^\alpha$
in the following way:
\begin{equation}
R^\alpha = \frac{\partial \psi}{\partial r^\alpha} = \lambda \eta \mu r^\alpha
= \lambda \eta \mu b \sqrt{\sum_{\beta = 1}^N h^{\alpha \beta} \rho^\beta}
\label{rrho}
\end{equation}
where $b$ is the norm of the Burgers vector of the considered slip system
family. The relation (\ref{rrho}) defines the internal variable $r^\alpha$
as a function of the dislocation densities. 
The interaction between the dislocations of the various slip 
systems is represented by the interaction matrix $h^{\alpha \beta}$. 
The energy term is multiplied by the order parameter $\eta$ 
because the formula for stored energy by dislocation accumulation is 
valid in the bulk crystal but loses its meaning inside the grain 
boundary. 
During plastic deformation, dislocations are subjected to multiplication
and annihilation mechanisms recorded by the following Kocks-Mecking-Teodosiu
evolution equation:
\begin{equation}
\dot\rho^{\alpha} =
\left\{
\begin{array}{lr}
\frac{1}{b} \left( K\sqrt{\sum_{\beta}\rho^{\beta}} -2d\rho^{\alpha}\right) |\dot\gamma^{\alpha}| \cbox{-\rho^{\alpha}\,C_D\,A(||\ten  \kappa||)\,\dot{\eta}}
\, &   \text{if  } \dot{\eta} > 0
\\ \, \vspace{-1.5mm}\\
\frac{1}{b} \left( K\sqrt{\sum_{\beta}\rho^{\beta}} -2d\rho^{\alpha}\right) |\dot\gamma^{\alpha}|
\, &  \text{if  } \dot{\eta} \leq 0
\end{array}
\right. 
\label{KocksMecking}
\end{equation}
The evolution equation contains the term highlighted in {gray} which is added 
to account for the annihilation of dislocations behind a sweeping grain 
boundary. The other terms are, respectively, for the competing dislocation storage and dynamic recovery taking place during the deformation process. The parameter $K$ is a mobility constant and $d$ is the critical annihilation distance between opposite sign dislocations whereas the parameter $C_D$ determines the dynamics of the 
static recovery due to a migrating grain boundary. 
A lower value means only partial recovery whereas a higher value leads to full 
recovery, see \citep{humphreys04} for TEM images of such static recovery
processes. The function $A(||\ten  \kappa||)$ serves to localize the process 
in a thin region of the grain boundary domain. In order to ensure that 
recovery only takes place in the wake of the sweeping grain boundary, not in 
front of it, the corresponding term is active only if $\dot{\eta} > 0$.
Such a static recovery term coupled with the crystallinity phase field 
was proposed by \cite{Abrivard2012a} in the evolution equation for the total 
stored dislocation 
energy. As a further refinement, the static recovery is written here
at the level of each individual slip system. {Dislocation densities are not transferred between slip systems by (\ref{KocksMecking}). Instead, when the GB sweeps a material point, the lattice rotates to change its orientation. It is assumed that each slip system is rotated in the same way keeping track of each associated dislocation density. This represents a naive view of the
actual process of atomic motion and reordering process taking place in the actual
migration. To improve the physical relevance of equation (\ref{KocksMecking}), analysis of the crystallographic structures of dislocation debris after GB migration would be required \citep{humphreys04}.}

In the present paper, the decomposition of the curvature tensor into
elastic and plastic parts or of the couple stress tensor into equilibrium
and non-equilibrium parts was not considered for simplicity. 
The lattice curvature is therefore synonymous of energy storage. The reader is referred to \citep{archmech97,fleckhutchinson97,ijss2000,mayeur11,mayeur15} 
for the consideration of a dissipative contribution of lattice curvature.

\section{Application to lattice rotation and grain boundary migration in a laminate
microstructure}
\label{sec:NumEx}

In this section results from finite element simulations of several test examples are presented 
and discussed. The general geometry used for the calculations is that shown in figure \ref{equilibre}. It is two-dimensional and represents a laminated crystal with periodically replicated bicrystal layers, where each layer is 10 $\mu$m wide. 

The system of equations for isotropic grain boundaries reduced to two dimensions 
under plane strain conditions is detailed in \ref{app1}. Model parameters for pure copper are given in 
\ref{app2} along with guidelines 
on their calibration.  The simulations are done on a dimensionless system of characteristic length $\Lambda$ and time $\tau_0$, such that $\overline{x} = x/\Lambda$ and $\overline{t}=t/\tau_0$, according 
to \ref{app2}. The free energy in (\ref{energy_isotropic}) is non-dimensionalized by the parameter $f_0$. All model parameters are given in table \ref{material_parameters_table} 
unless otherwise indicated. In the simulations, a $15$ degree misorientation over the grain 
boundary is applied. The parameter $f_0$ in the free energy has been calibrated so that a GB at $15$ degree misorientation has an energy of 0.5 J/m$^2$.

\subsection{Finite element implementation}

The proposed phase field Cosserat theory with the constitutive equations summarized in \ref{app1}
has been implemented in the finite element code Z--set \citep{zset}. 
It is based on a variational formulation derived from the virtual power form of the
balance equations presented in Section \ref{sec:balancelaws}, as initially proposed
by \cite{ammar09}.
The detailed formulation is not presented here for the sake of brevity.
The essential features of the implementation combine elements given 
in \cite{ijss2000} for Cosserat crystal plasticity and in \cite{Abrivard2012a}
for the KWC model.
In the two-dimensional case, the \textit{monolithic} weak formulation involves 4 nodal degrees 
of freedom: 2 components of
displacement, Cosserat microrotation component $\theta_3$ and phase field $\eta$.
Quadratic elements with reduced integration are used.
An implicit iterative resolution scheme is used to solve the
balance equations based on the Newton-Raphson algorithm. A fourth order Runge-Kutta
method with automatic time-stepping is applied to solve the differential equations
driving the internal variables of the model.

The global orthogonal coordinate system $(x_1,x_2,x_3)$ is given in figure \ref{equilibre}, where direction 1 is along the thickness of the bicrystal layers and direction 2 is along the height. Due to periodicity, the system is invariant along direction 2 so that
only one finite element is used in that direction. The discretization along 1 is specified
in each case.  A specific feature of the model is that the Cosserat rotation is used to transform tensors between the global coordinate frame and the local frame of the crystal. {The local integration of the constitutive equations is carried out in the crystal frame.} The Cosserat pseudo-vector $\vec \varTheta$ is associated with the lattice orientations with respect to a fixed reference frame, which is here taken to be $(x_1,x_2,x_3)$. With the calculations restricted to the $(x_1,x_2)$-plane, the axis of rotation is the $x_3$-axis. This gives $\vec \varTheta =  [\begin{array}{ccc} 0 & 0 & \theta_3 \end{array}]^T$ with $\theta_3$ the angle of rotation used to construct the appropriate orthogonal transformation tensors between the global and the local frame. This is different from a model using a sharp interface description since $\vec \varTheta$ in the present model varies as a phase-field variable over grain boundaries, i.e. it is continuous everywhere in the computational domain. In what follows, the subscript will be dropped when referring to the rotation angle so that $\theta=\theta_3$.

\subsection{Grain boundary equilibrium profiles}
\begin{figure}[t!]
\centering
\includegraphics[height=5.1cm]{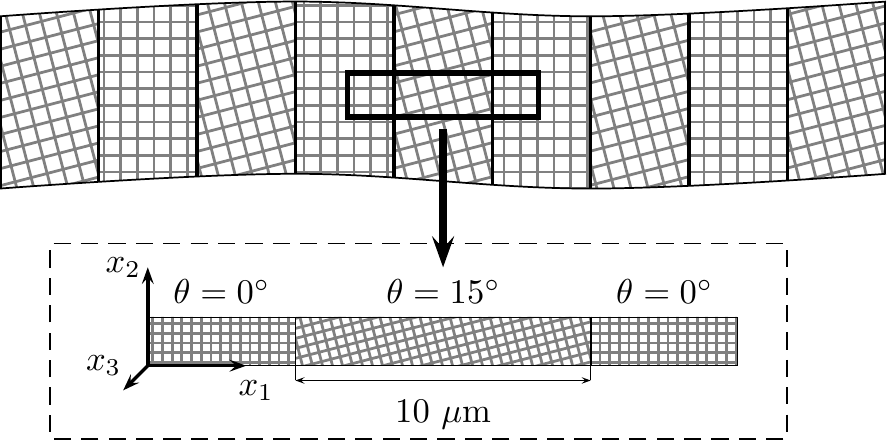}
\caption{\label{equilibre} A microstructure made of periodically replicated bicrystal layers is represented by a periodic geometry used for simulations. The misorientation between the layers is 15$^\circ$.}
\end{figure}
\begin{figure}[thb!]
\centering
\includegraphics[height=5.5cm]{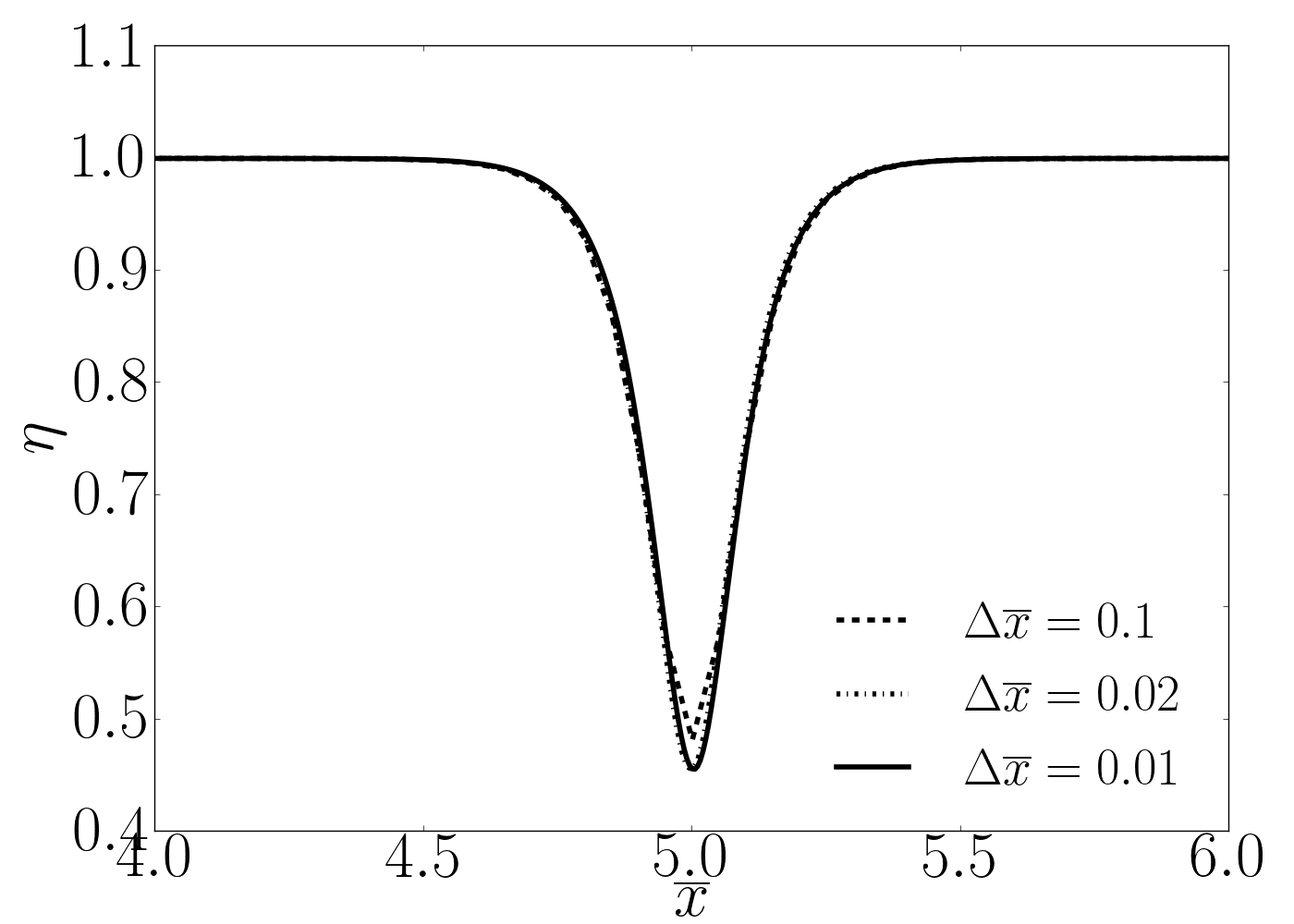}
\includegraphics[height=5.5cm]{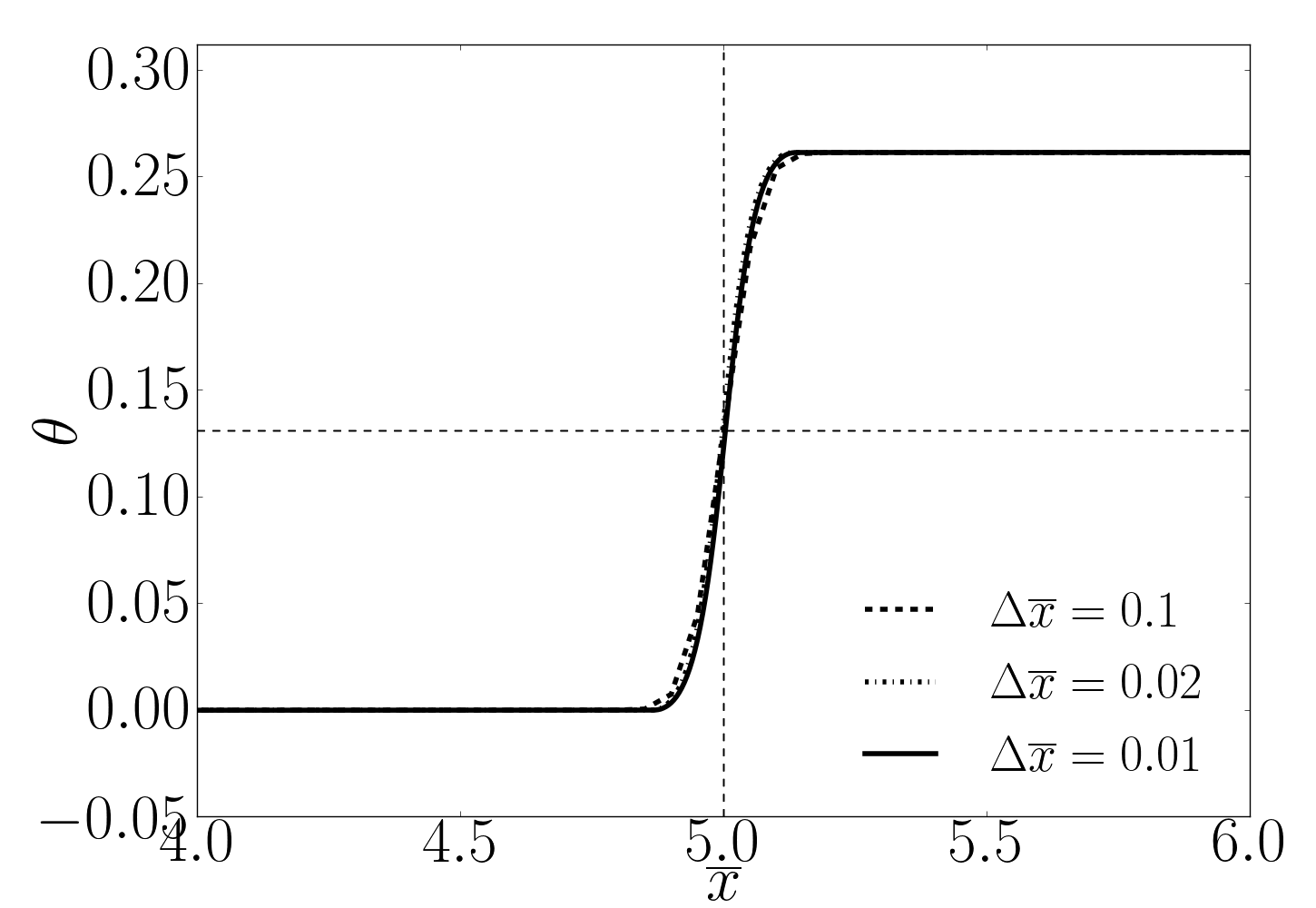}
\caption{\label{equilib_profiles} {Grain boundary} profiles of $\eta$ (left) and $\theta$ (right) at equilibrium for the KWC model. Solid lines: 2000 elements ($\Delta \overline{x}=0.01$); dashed-dotted lines: 1000 elements ($\Delta \overline{x}=0.02)$ and dashed lines: 200 elements ($\Delta \overline{x}=0.1$).}
\end{figure}

Prior to any microstructure evolution calculation, the field variables of the 
model must be determined for the initial static equilibrium state of the grain boundaries.
The initial crystallinity and orientation fields are obtained here from phase field
simulations based solely on the KWC model. 
According to the free energy function (\ref{energy_isotropic}), the KWC model is recovered 
for the choice $\mu_c = 0$ for the Cosserat parameter and in the absence of any displacements, 
i.e. $\vec u = \vec 0$. 

The geometry (see figure \ref{equilibre}) is discretized using 200, 1000 or 2000 elements such that each element size is $\Delta \overline{x} = 0.1$, $\Delta \overline{x} = 0.02$ or $\Delta \overline{x} = 0.01$, respectively. Initial values are constructed for $\vec \varTheta$ and $\eta$, in the form of 
$\tanh$ functions and $\cosh$ functions, respectively, and the system is then allowed to relax towards equilibrium. 
Spurious grain rotation inside the grains is suppressed by appropriate choices 
for the mobility functions $\tau_\eta$ and $\tau_\theta$. Specifically, $\tau_\eta$ is taken 
to be constant whereas $\tau_\theta$ is assumed to be of the form (\ref{tau_theta_gen}). The specific format of the mobility function is given by (\ref{Pmob_dim}). The (dimensionless) mobility parameters used are $\overline{\tau}_\eta=1$ and $\overline{\tau}_\theta=1000$, 
cf.  equations (\ref{tau_phi}) and (\ref{tau_nodim}), together with $\overline{\mu}_P=1000$ and $\overline{\beta}_P=1000$ in the (dimensionless) mobility function given by equation (\ref{Pmob_nodim}). {The remaining material parameters can be found in table \ref{material_parameters_table}.} {From the results presented in figure \ref{equilib_profiles}, it seems that around 10 elements are needed inside the GB region to resolve the profiles of the phase fields, corresponding to the discretization with 1000 elements along direction 1. This discretization will be used in all the examples that follow.}

The found static equilibrium 
orientation field and field of order parameter are used to initialize the field Cosserat micro-rotation $\vec \varTheta$ and the field $\eta$ in the next calculations. It is also necessary to initialize the field of eigen-rotation
$\vx e\,^\star_0$ for subsequent simulations involving the mechanical coupling. 
In the absence of plastic deformation and strain, the constitutive law (\ref{skewstress_pseudo}) 
providing the equilibrium skew-symmetric stress component becomes:
\begin{equation}
\vx \sigma\,^{eq} = 2\, \mu_c \, \vx e\,^e = 2\, \mu_c (\vx e - \vx e\,^\star_0)
= -2\,\mu_c (\vec \varTheta + \vx e\,^\star_0)  
\end{equation}
It is assumed that the initial state is stress-free. This assumption therefore requires
that initially $\vx e\,^\star_0 = -\vec \varTheta_0$.
The equilibrium solution can be computed again using the full Cosserat model 
including the previous initialized field values.
It can be checked in Fig. \ref{equilibre} that the KWC model and the full Cosserat
model lead to the same equilibrium 
profiles as it should be. The stress values also identically vanish.

This will allow for an initially undeformed state to be stress-free even when $\mu_c \neq 0$, as is the case in the following simulations.

\subsection{Control of grain rotation}
\label{sec:rotcontrol}

\begin{figure}[t!]
\centering
\includegraphics[height=5.5cm]{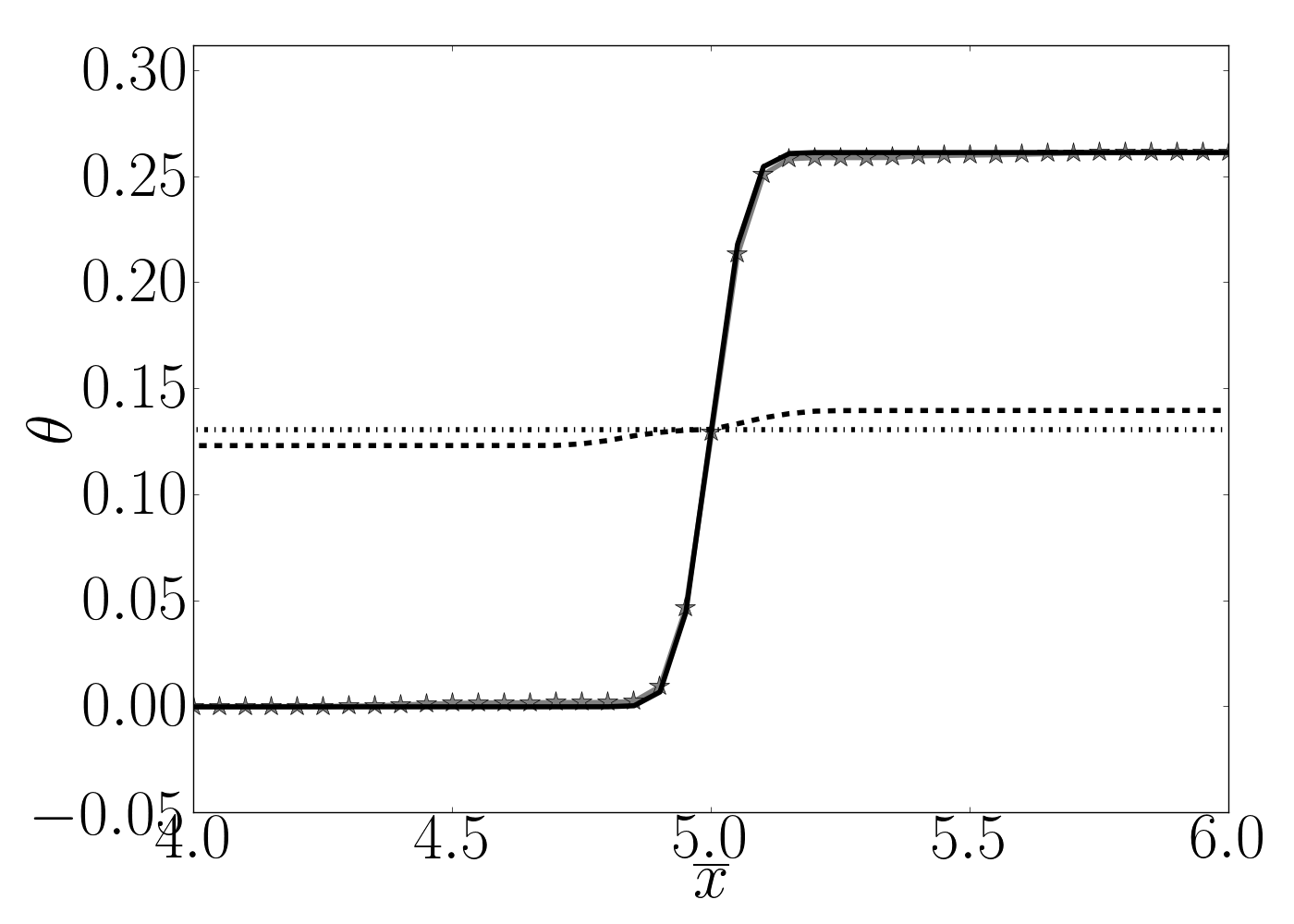}
\includegraphics[height=5.5cm]{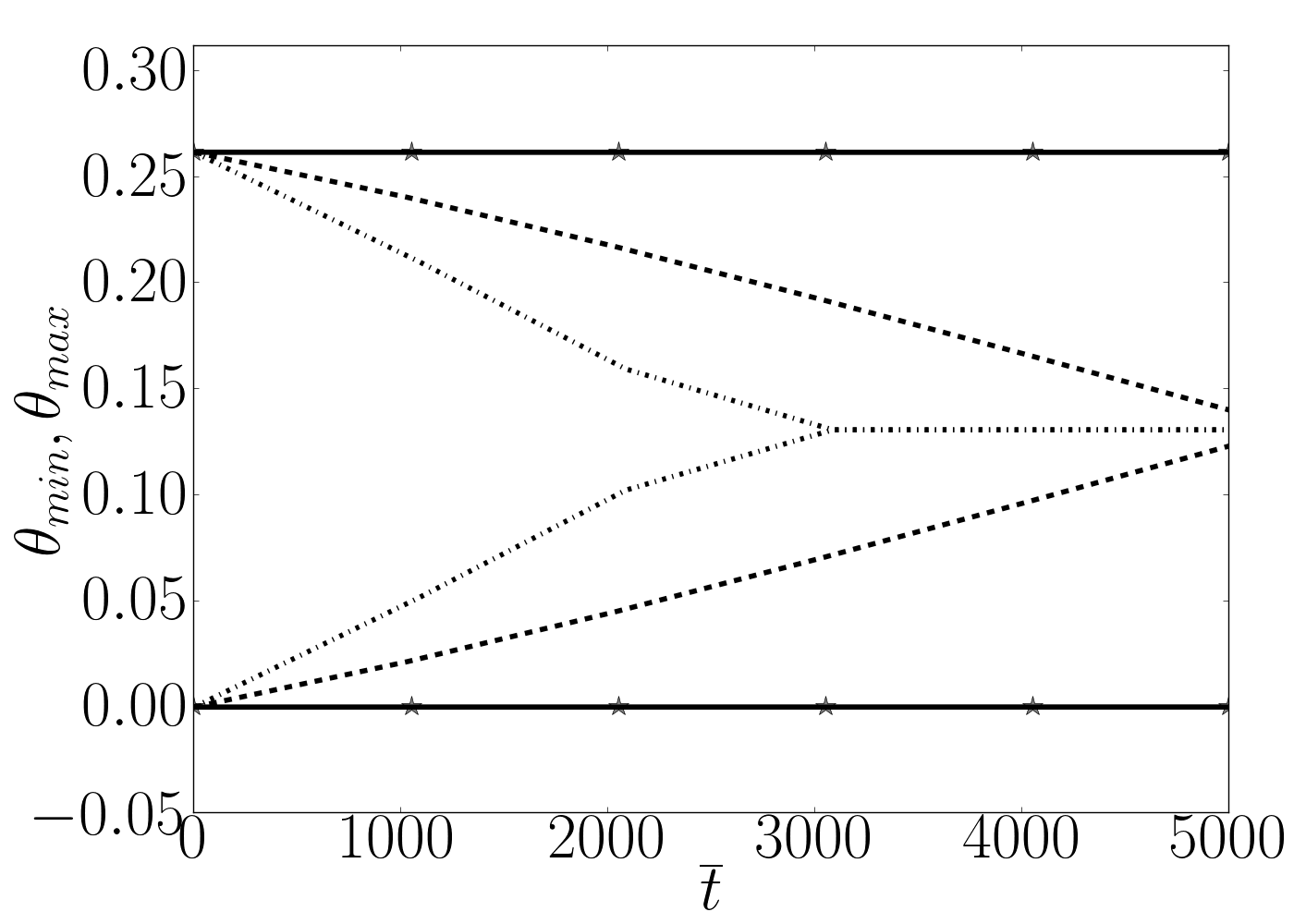}
\caption{\label{rotation_time} Profiles of $\theta$ at $\overline{t}=5000$ (left) and evolution of $\theta$ over time (right). Solid line: Cosserat coupling modulus is active (d) $\mu_c/f_0=1000$. In the other cases, it is not active. {Gray line with stars} (a): $\overline{\mu}_P=1000$, $\overline{\tau}_\theta=1000$. Dashed-dotted line (b): $\overline{\mu}_P=0.5$, $\overline{\tau}_\theta=1000$. Dashed line (c): $\overline{\mu}_P=1000$, $\overline{\tau}_\theta=1$. .}
\end{figure}

The KWC model does not restrict the 
change in orientation to the grain boundaries but also allows for rotation 
in the interior of a grain \citep{KWC2000}. While such overall grain rotation was discussed for certain cases, notably in nanograined structures \citep{Upmanyu06}, it is not observed 
on the scale of interest here. In the KWC phase field model, the rotation must 
be controlled by differentiating the mobility function between the bulk of the grain 
and the grain boundary \citep{KWC2003}. 
Even with such a distinction, rotation is not prohibited but merely delayed to 
a time scale larger than that of interest. 

\begin{table}[thb!]
\centering
\begin{tabular}{c|c|c|c}
Case  & $\mu_c/f_0$ & $\overline{\tau}_\theta$ & $\overline{\mu}_P$ \\
\hline
(a) & 0 & 1000 & 1000 \\
(b) & 0 & 1000 & 0.5 \\
(c) & 0 & 1 & 1000 \\
(d) & 1000 & 1 & 0.5
\end{tabular}
\caption{\label{material_parameters_s4_3} List of model parameters for the simulatons in section \ref{sec:rotcontrol}. All other parameters are given in table \ref{material_parameters_table}.}
\end{table}

In the present coupled approach, the Cosserat mechanics provides a way 
to prevent rotations inside the grains. This is due to the non-dissipative 
contribution to the skew-symmetric stress (\ref{skewstress_pseudo}) which penalizes deviations between the elastic lattice (re-)orientation and the Cosserat rotation plus the eigen-rotation, by means of the Cosserat coupling modulus,
$\mu_c$. The evolution of the eigen-rotation is taken to be confined to the grain 
boundary and it therefore cannot compensate for a changing Cosserat rotation 
in the bulk of the grain.

To illustrate the role of the Cosserat coupling modulus, the previous test 
example is studied once more. Initial values $\vec \varTheta$ and $\eta$ are given in the form of 
$\tanh$ functions and $\cosh$ functions, respectively, and the system is then allowed to relax towards equilibrium. On the one hand, the Cosserat coupling modulus is zero and the influence of the mobility parameters on the bulk grain rotation is studied. On the other hand, the coupling modulus is non-zero. In this latter case, in order to have a stress-free equilibrium state, the eigen-rotation is initiated with the values obtained from the calculations in the previous test example. Four different sets of parameters are considered; three (a)--(c) where $\mu_c=0$ and one (d) where $\mu_c/f_0=1000$. {The dimensionless mobility parameters associated with each test case are given in table \ref{material_parameters_s4_3} and all other material parameters in table \ref{material_parameters_table}. Note that the choice $\overline{\mu}_P=0.5$ gives a constant value $P=1$ of the mobility function, c.f. equation (\ref{Pmob_nodim}), i.e. there is no separation of the time scales between the grain boundary and the bulk for this choice.} 

Figure \ref{rotation_time} shows the 
profiles of $\theta$ (left) at time step $\overline{t} = 5000$ for the 
different combinations of parameters whereas figure \ref{rotation_time} 
(right) shows the evolution of $\theta_{min}$ and $\theta_{max}$ 
(the minimum and maximum values over the unit cell) w.r.t. time. {Gray line with stars} is for case (a), {where the mobility parameters are chosen sufficiently high to prevent grain rotation on the simulated time scale}. In both cases (b)--dashed-dotted line--and (c)--dashed line--grain rotation takes place. At the end the bicrystal unit cell has become a single grain with intermediate
orientation. Solid line is for the case (d) (non-zero Cosserat coupling modulus). No grain rotation takes place despite the choice of mobility parameters.

This test shows that the Cosserat coupling term effectively prevents the rotation in the bulk. It also demonstrates that if the bulk rotation can be prevented by the Cosserat modulus, it is not necessary any more to distinguish between the mobility of the bulk 
and the grain boundary.

\subsection{Superposed rigid body rotation}
\label{sec:rigidrot}
%
%
\begin{figure}[thb!]
\centering
\includegraphics[height=5.5cm]{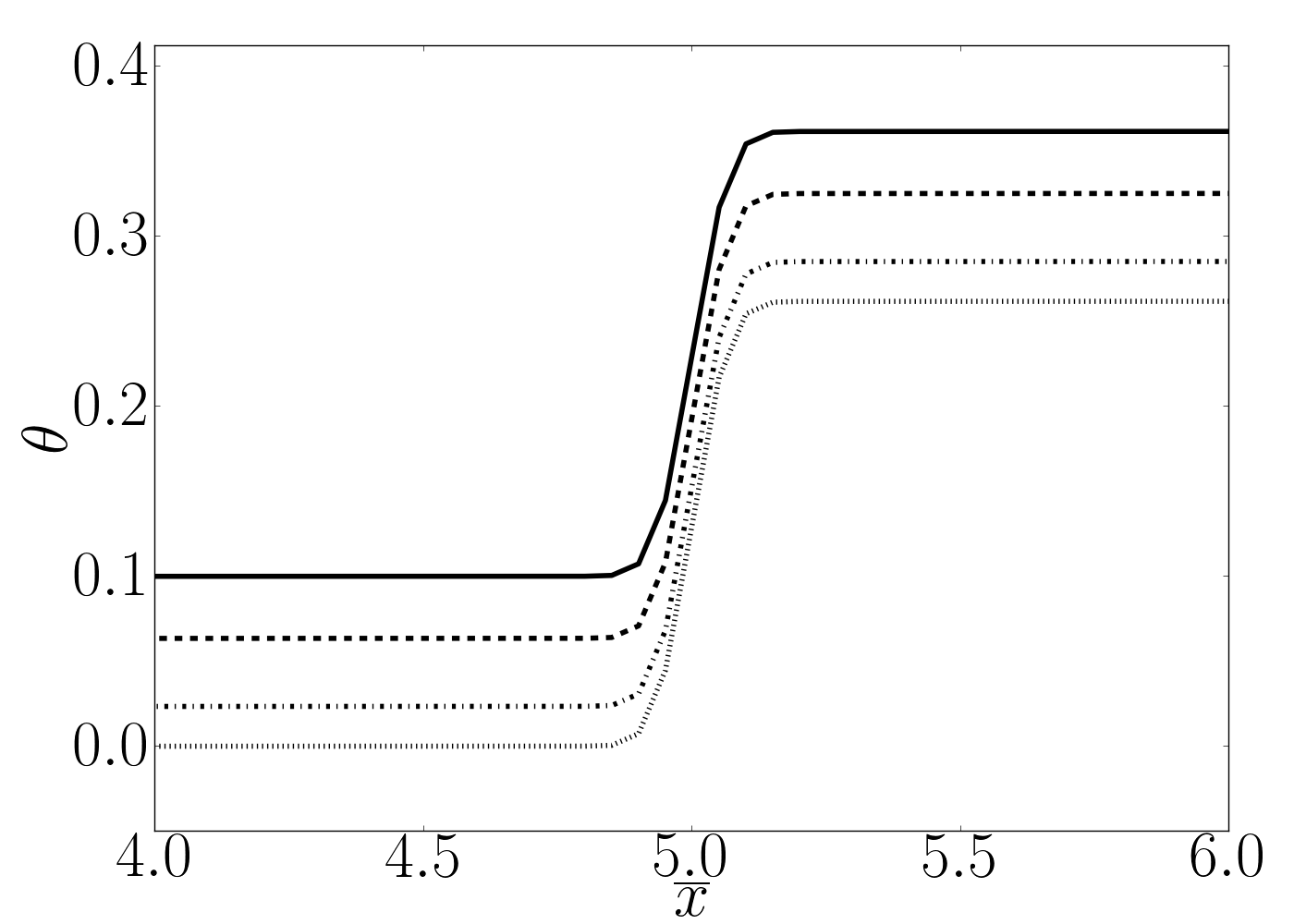}
\includegraphics[height=5.5cm]{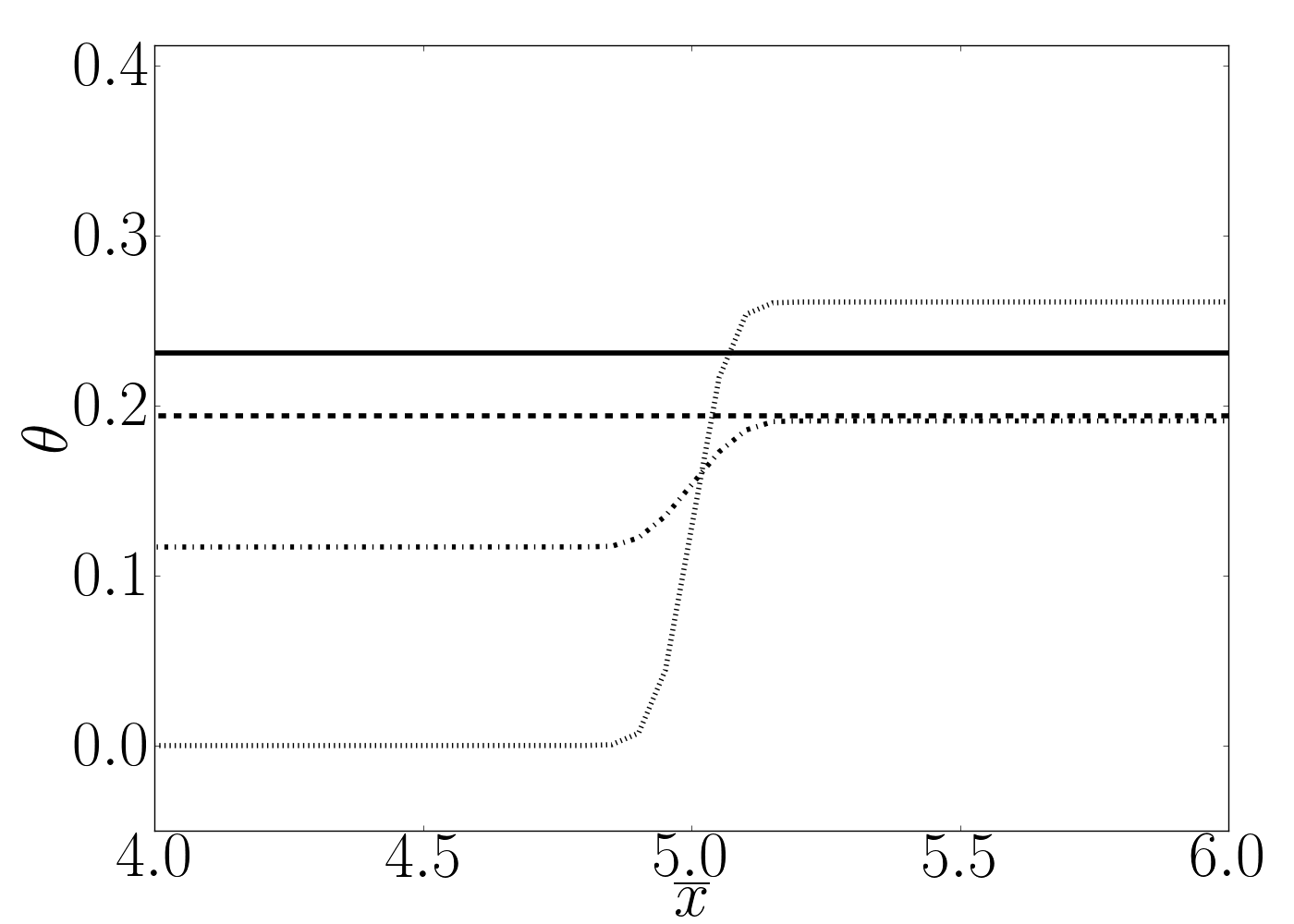}
\includegraphics[height=5.5cm]{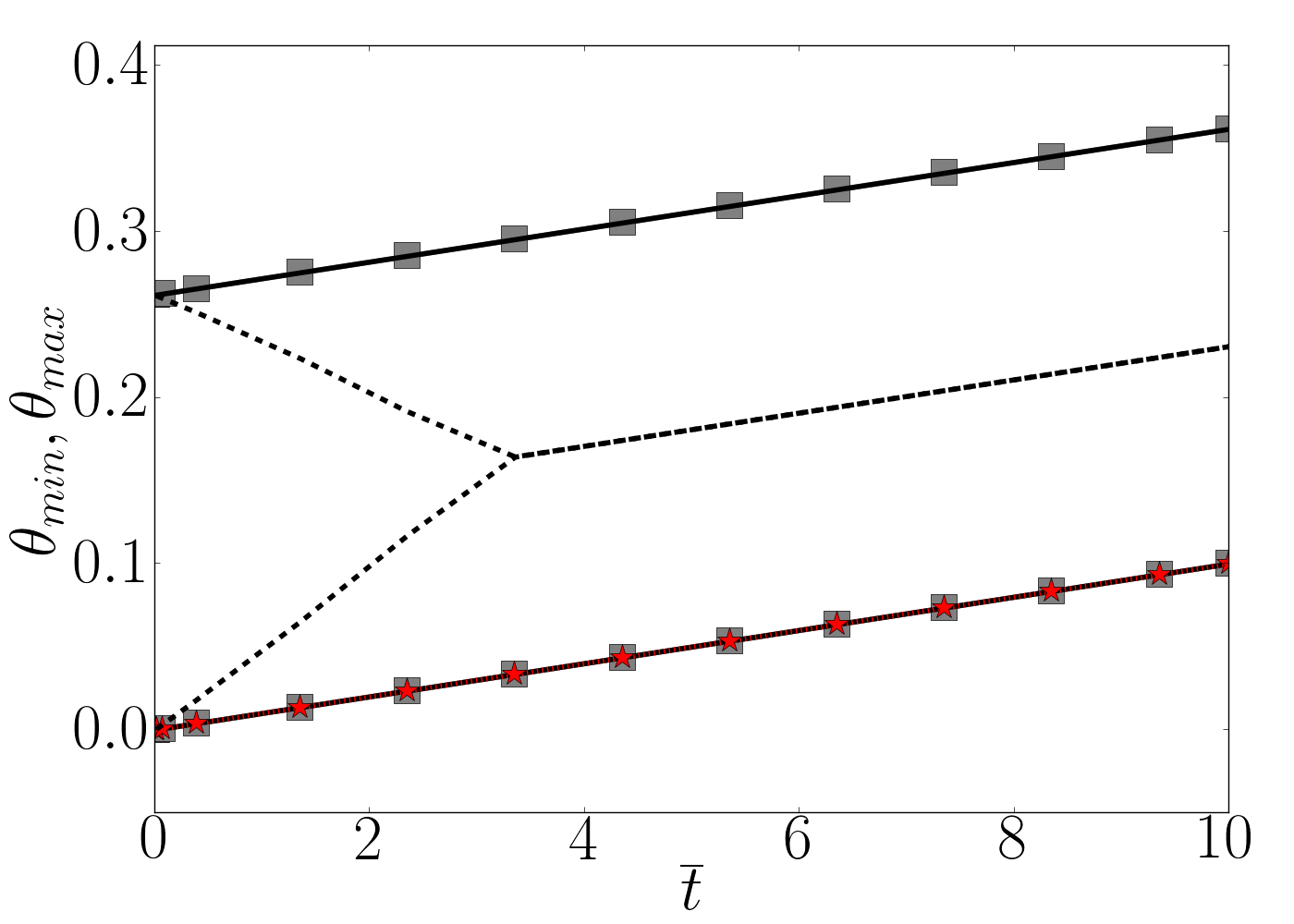}
\caption{\label{bigrain_rotate_profiles} Superposed rigid body motion: Profiles for $\theta$: top left with Cosserat coupling and top right without Cosserat coupling active. In both cases $\overline{\mu}_P=0.5$, $\overline{\tau}_\theta=1$. The profiles are shown at time steps $\overline{t} = 0.01, 2.35, 6.35, 10$ respectively (dotted, dashed-dotted, dashed, solid lines). The bottom figure shows the evolution over time of $\theta_{min}$ and $\theta_{max}$ together with the progress of imposed rotation (stars). Solid line is for for the case where $\mu_c/f_0=1000$. The other lines show $\mu_c =0$ with $\overline{\mu}_P=0.5$, $\overline{\tau}_\theta=1$ (dashed line) and $\overline{\mu}_P=1000$, $\overline{\tau}_\theta=1000$ {(gray line with square markers)}.}
\end{figure}

The superposition of a rigid body motion deserves some attention. It 
is a trivial check that must be 
performed in order to validate the objectivity of the model.
It must be noted first that the original KWC model is not objective with 
this respect. Looking at the equation (\ref{evol_theta}) without the {gray} terms, 
it appears
that the superposition of a rigid body motion does not induce any
lattice rotation $\dot {\vec \varTheta}$. This is not consistent with the fact
that the lattice follows the rotating material and that it should therefore
rotate. In the new model, the rigid body motion arises in the first red
term in equation (\ref{evol_theta}). If the right hand side vanishes (no microstructure
evolution), the microrotation rate $\dot {\vec \varTheta}$ will be equal to
the rigid body rotation $\vx \omega\,^e$. In the case of a rigid body 
motion, the elastic lattice rotation rate $\vx \omega\,^e$ coincides
with the material spin $\vx \omega$.
This is illustrated in the following example.

In this third example, a periodic two-grain structure is exposed to a 
rigid body rotation. According to the model, the crystal (Cosserat) 
orientation should follow the superposed rotation in order for the motion 
to be stress free.  
Three different sets of parameters (a,b,c) are considered. 
{On the one hand (a), the coupling term is active with $\mu_c/f_0=1000$ whereas for case (b) and (c) it is set to zero. The different combinations of parameters can be found in table \ref{material_parameters_s4_4} and all other material parameters are given in table \ref{material_parameters_table}.}
The values at static equilibrium from the first example above are used 
as initial values for the Cosserat rotation $\vec \varTheta$, 
the order parameter $\eta$ and the eigen-rotation $\vx e\,^\star$.
{Figure \ref{bigrain_rotate_profiles} shows the evolution of the crystal 
orientation during the rotation with the profiles over time for case (a) top left and case (b) top right. The bottom figure shows the evolution of the maximum and minimum orientations (corresponding to the bulk values of the respective grains) over time for cases (a)-(c).} When the Cosserat coupling modulus is active, 
the lattice (Cosserat) orientation follows the imposed rotation. This is not guaranteed if 
the coupling parameter is zero, as can be seen by the different behavior depending on the 
different choices of mobility parameters (b) and (c).
\begin{table}[t!]
\centering
\begin{tabular}{c|c|c|c}
Case  & $\mu_c/f_0$ & $\overline{\tau}_\theta$ & $\overline{\mu}_P$ \\
\hline
(a) & 1000 & 1 & 0.5 \\
(b) & 0 & 1 & 0.5 \\
(c) & 0 & 1000 & 1000 \\
\end{tabular}
\caption{\label{material_parameters_s4_4} List of model parameters for the simulatons in section \ref{sec:rigidrot}. All other parameters are given in table \ref{material_parameters_table}.}
\end{table}

\subsection{Simple shear deformation of the periodic bicrystal structure}

The same periodic structure as in the previous examples is now exposed to simple shear loading. The initial conditions for $\eta$, $\vec \varTheta$ and the eigen-rotation $\vx e\,^\star$ are taken from the equilibrium solution found in the first example. 
Isotropic elastic behavior as well as cubic anisotropy are considered. As described above, the transformation of strain and stress between the global (reference) frame and the local frame of the crystal lattice is done by means of a transformation tensor constructed using the rotation $\theta$ around the $x_3$-axis. The diffuse boundary description with a continuous change in lattice orientation over the grain boundaries differs from a sharp interface model where the lattice orientation varies abruptly between adjacent grains. For the periodic boundary conditions that are applied, it is possible to find analytic solutions for the stress components for a given rotation field. The details of the analytic solutions for simple shear loading are given in \ref{app3}. For the purpose of illustration, elasticity is assumed even though in reality the applied deformations result in stresses which exceed the yield limit for pure copper. Below, the stress components $\sigma^s_{ij}$, when given, always refer to the components of the symmetric stress tensor $\ten \sigma^{s} = \TEN E^s:\ten \varepsilon^e$. 

Periodic boundary conditions are assumed so that loading is applied by imposing
\begin{equation}
\begin{aligned}
\vec u = \ten B \cdot \vc x + \vec p 
\end{aligned}
\end{equation}
where $\vec p$ is the periodic displacement fluctuation  
taking the same value in opposite points of the boundary 
and the components of the tensor $\ten B$ are given by
\begin{equation}
\ten B = \left[\,
\begin{array}{ccc}
0 & B_{12} & 0 \\
B_{21} & 0 & 0 \\
0 & 0 & 0 
\end{array} \,
\right] \,,
\end{equation} 
for shear loading. Simple shear is applied by setting $B_{12}=0$ and $B_{21}=0.001$.
The simple shear contains overall straining and rotation imposed to the unit 
cell. {In the simulations, $\mu_c/f_0 = 1000$, $\overline{\tau}_\theta=1$, $\overline{\mu}_P=0.5$ and the elasticity parameters are specified for each case (isotropy or anisotropy). All other material parameters are given in table \ref{material_parameters_table}.}

\subsubsection{Isotropic elastic behavior}

The elastic parameters are taken to be $E=120$ GPa for the elastic modulus and $\nu = 0.3$ for the Poisson's ratio. 
The resulting stress computed using 200 or 1000 elements is $\sigma^s_{12}=46.15$ MPa (identical results) which corresponds to the analytic solution for the given material parameters. The imposed loading results in stretching as well as lattice reorientation. Since isotropic behavior is assumed, the lattice reorientation is uniform. Figure \ref{bigrain_shear_simple_wtheta} (left) shows the elastic reorientation (red stars) and the relative rotation $\theta$ - $\theta_0$ of the Cosserat directors (associated here with the crystal orientations with respect to a fixed frame). The values of the relative rotations have been magnified by a factor 100.
The rotation induced by elasticity is very small and in accordance with the
analytical solution.

\subsubsection{Anisotropic elastic behavior} 

\begin{figure}[t!]
\centering
\includegraphics[height=5.5cm]{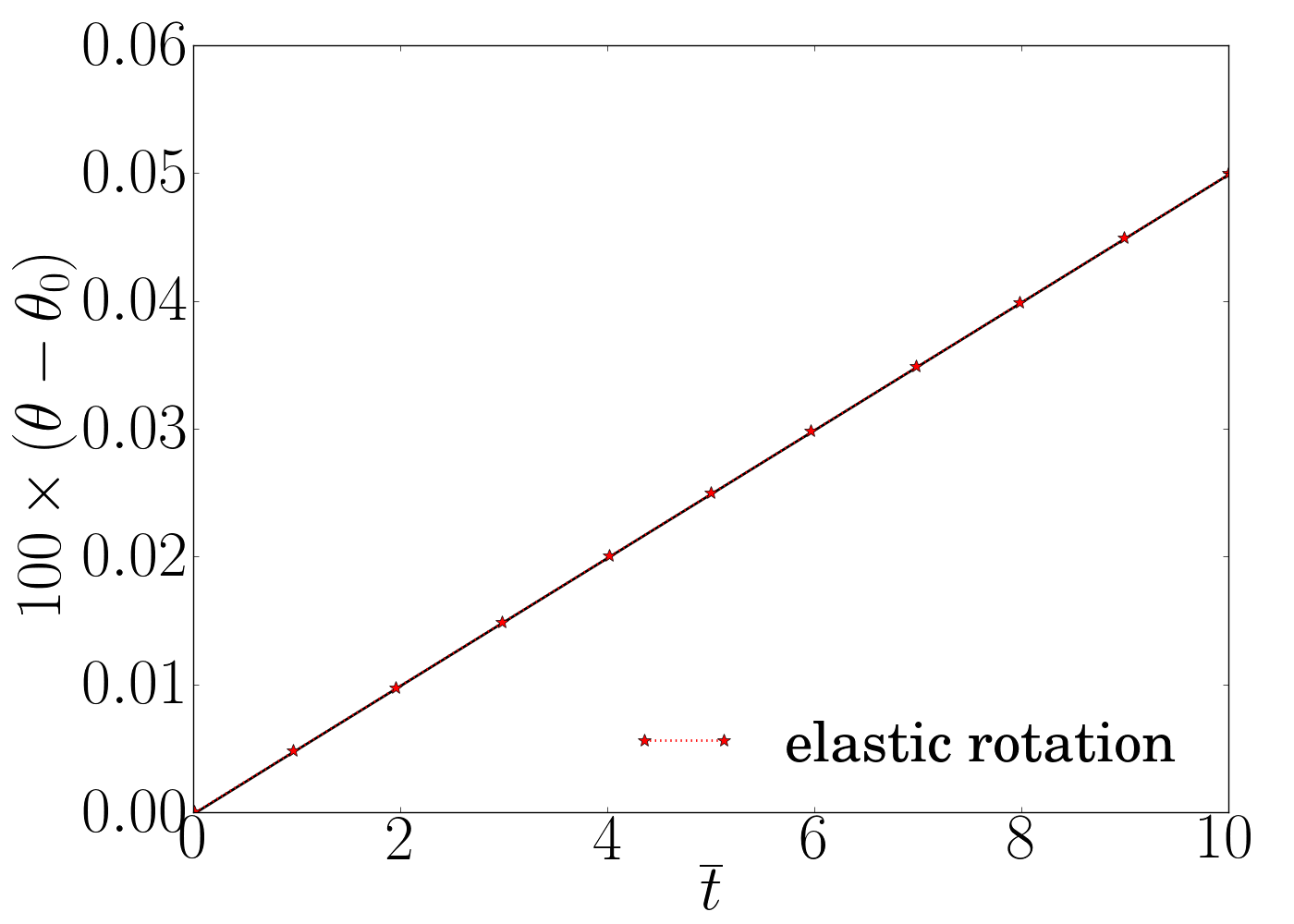}
\includegraphics[height=5.5cm]{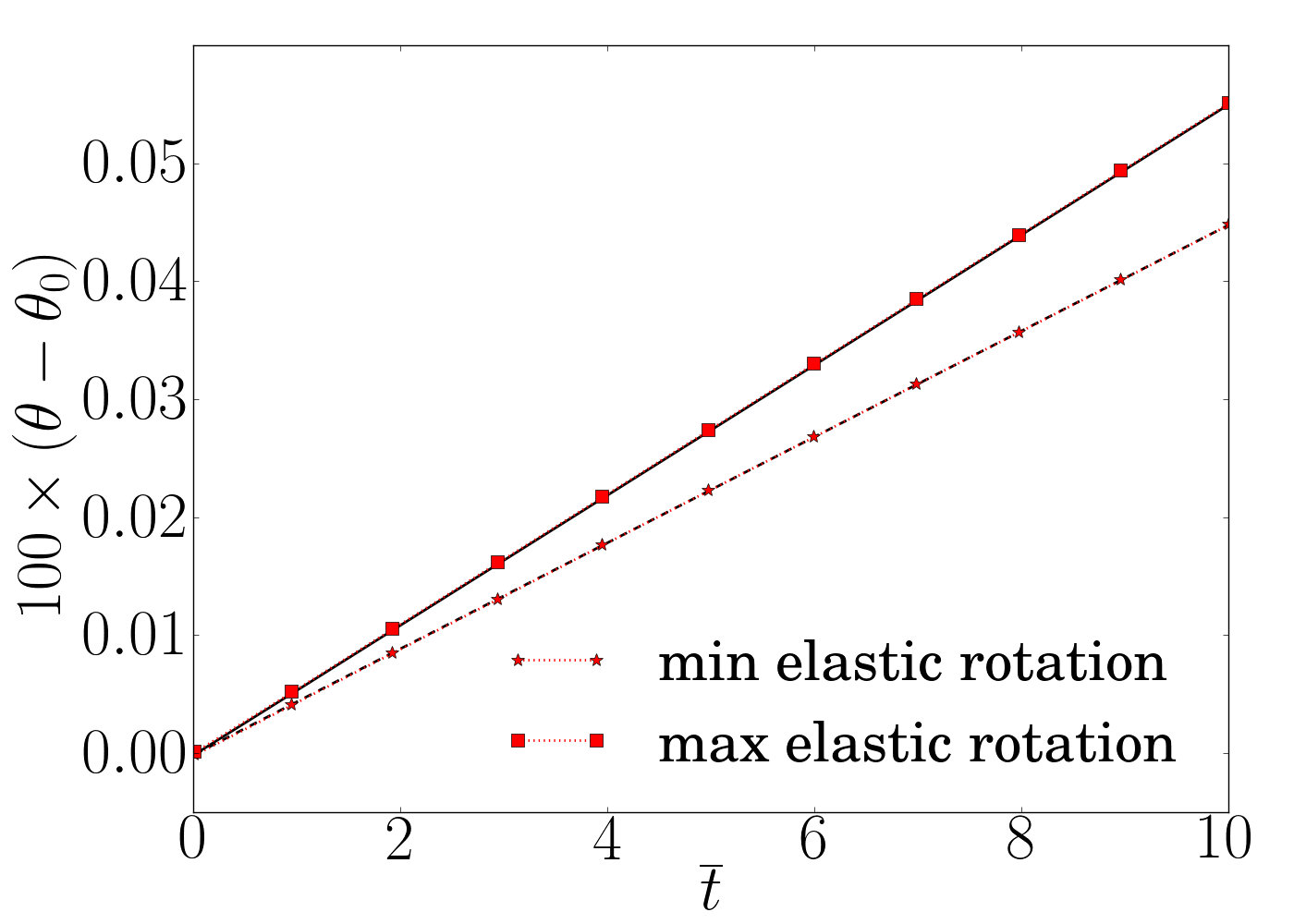}
\caption{\label{bigrain_shear_simple_wtheta} Relative elastic rotation due to simple shear loading and corresponding change of $\theta$ (magnified by 100).  Left: isotropic and right: anisotropic behavior with 1000 finite elements. For isotropic behavior (left) the relative rotation is the same everywhere. For anisotropic behavior it is different in the differently oriented grains, with solid line representing the relative rotation of the grain at initially $15^\circ$ orientation relative to the reference frame and dashed line the grain at $0^\circ$ initial orientation.}
\end{figure}
\begin{figure}[t!]
\centering
\includegraphics[height=5.5cm]{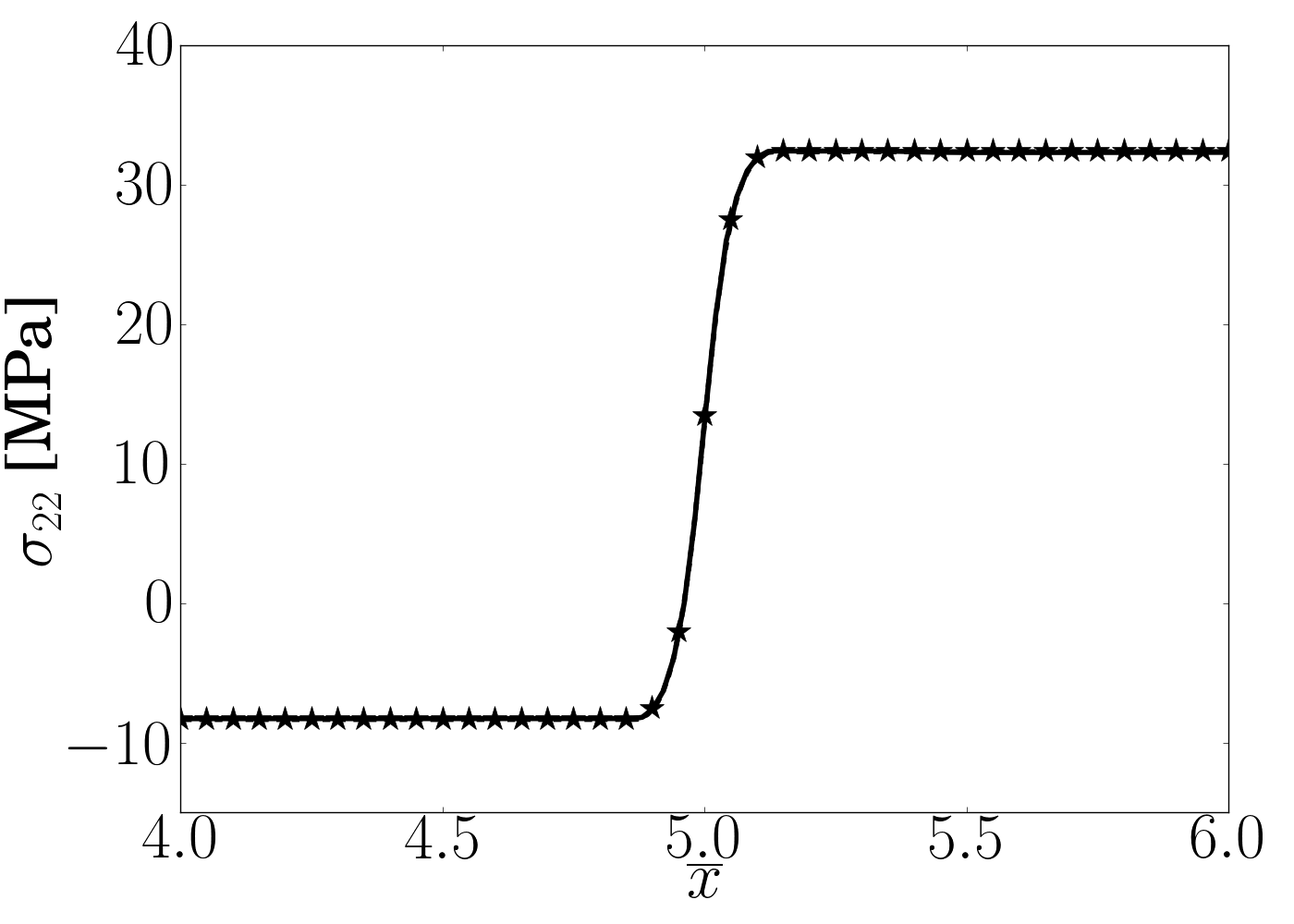}
\includegraphics[height=5.5cm]{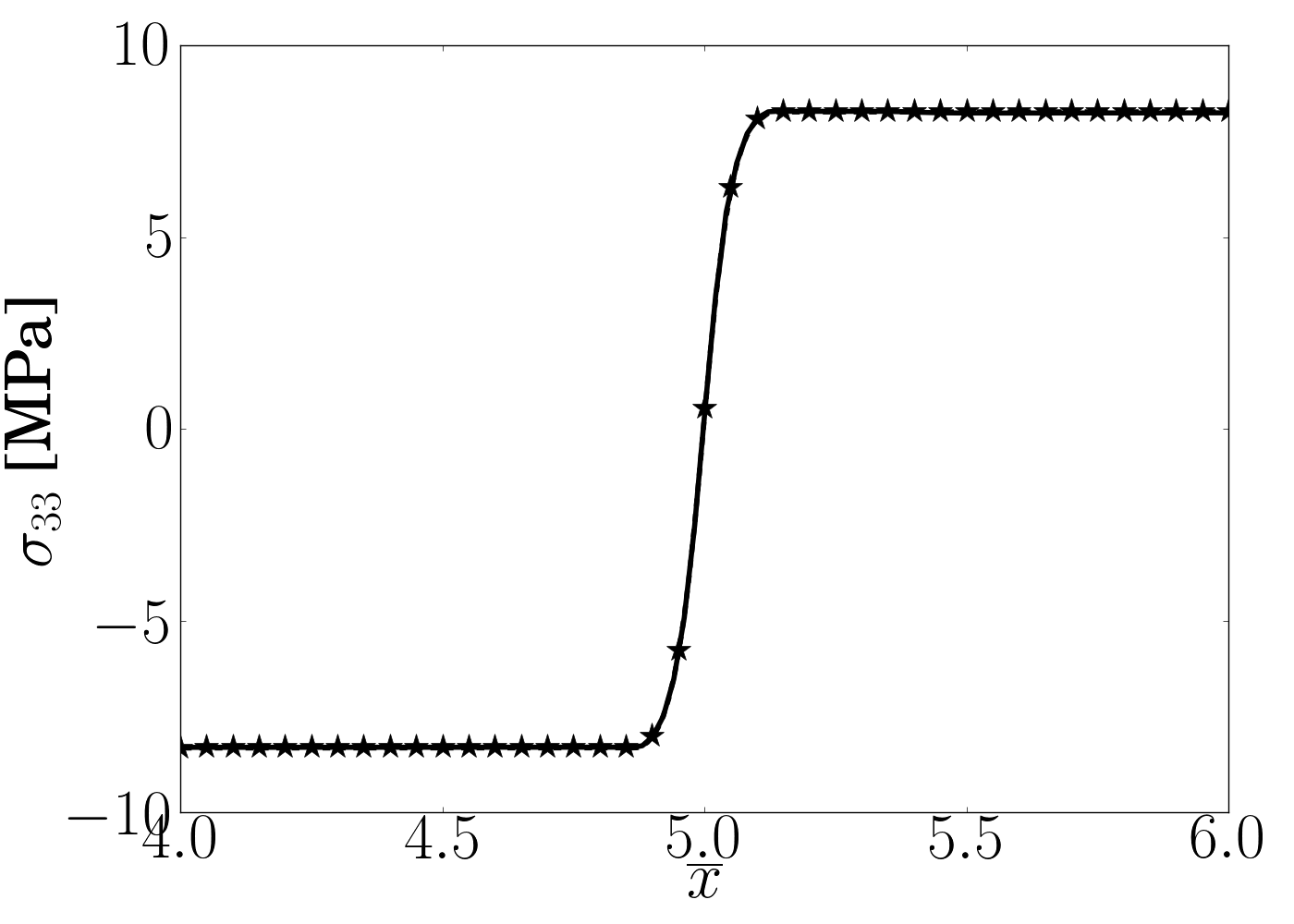}
\caption{\label{bigrain_shear_simple} Calculated stress profiles for simple shear loading and anisotropic elasticity. Solid lines show the result from finite element calculations and dashed lines with markers show the analytic solution.  Left: $\sigma^s_{22}$ and right: $\sigma^s_{33}$ with 1000 finite elements.}
\end{figure}

The parameters for cubic elastic anisotropy for pure copper are taken to be $C_{11}=160$ GPa, $C_{12}=110$ GPa and $C_{44}=75$ GPa.
Due to the misorientation between the two grains, the stress components
are not uniform and not limited to shear in each phase.
Figure \ref{bigrain_shear_simple} shows the computed profiles for the stress components $\sigma^s_{22}$ (left) and $\sigma^s_{33}$ (right) at simple shear. A finite element discretization along direction 1 of 1000 (bottom) elements is used. The results calculated by finite elements (solid lines) are compared to the analytic solutions for the corresponding rotation profiles (dashed lines with markers). The stress components $\sigma^s_{11}$ as well as the component $\sigma^s_{12}$ of the symmetric stress tensor are uniform and the computed values are $\sigma^s_{11} = -12.1$ MPa and $\sigma^s_{12}=70.8$ MPa which corresponds to the analytic values for the given material parameters.

Figure \ref{bigrain_shear_simple_wtheta} (right) shows the rotation due to the shear deformation superposed on the relative rotation of the Cosserat directors. Due to the anisotropy and the periodicity conditions, the resulting reorientation due to the imposed deformation is not the same in the differently oriented grains, with the grain at $15^\circ$ initial orientation relative to the reference frame experiencing a larger reorientation. This can be seen in the figure \ref{bigrain_shear_simple_wtheta} (right) where the maximum (red squares) and minimum (red stars) values of the reorientation are superimposed on the relative rotation of the Cosserat directors of the respective grains, where the solid line represents the grain at $15^\circ$ initial orientation and dashed lines represent the grain at $0^\circ$ initial orientation. The values of the relative rotations have been magnified by a factor 100. This example demonstrates that the relative rotation of the Cosserat directors indeed follow the lattice reorientation during deformation, as expected.

\subsection{Grain boundary migration}

\begin{figure}[t!]
\centering
\includegraphics[height=5.5cm]{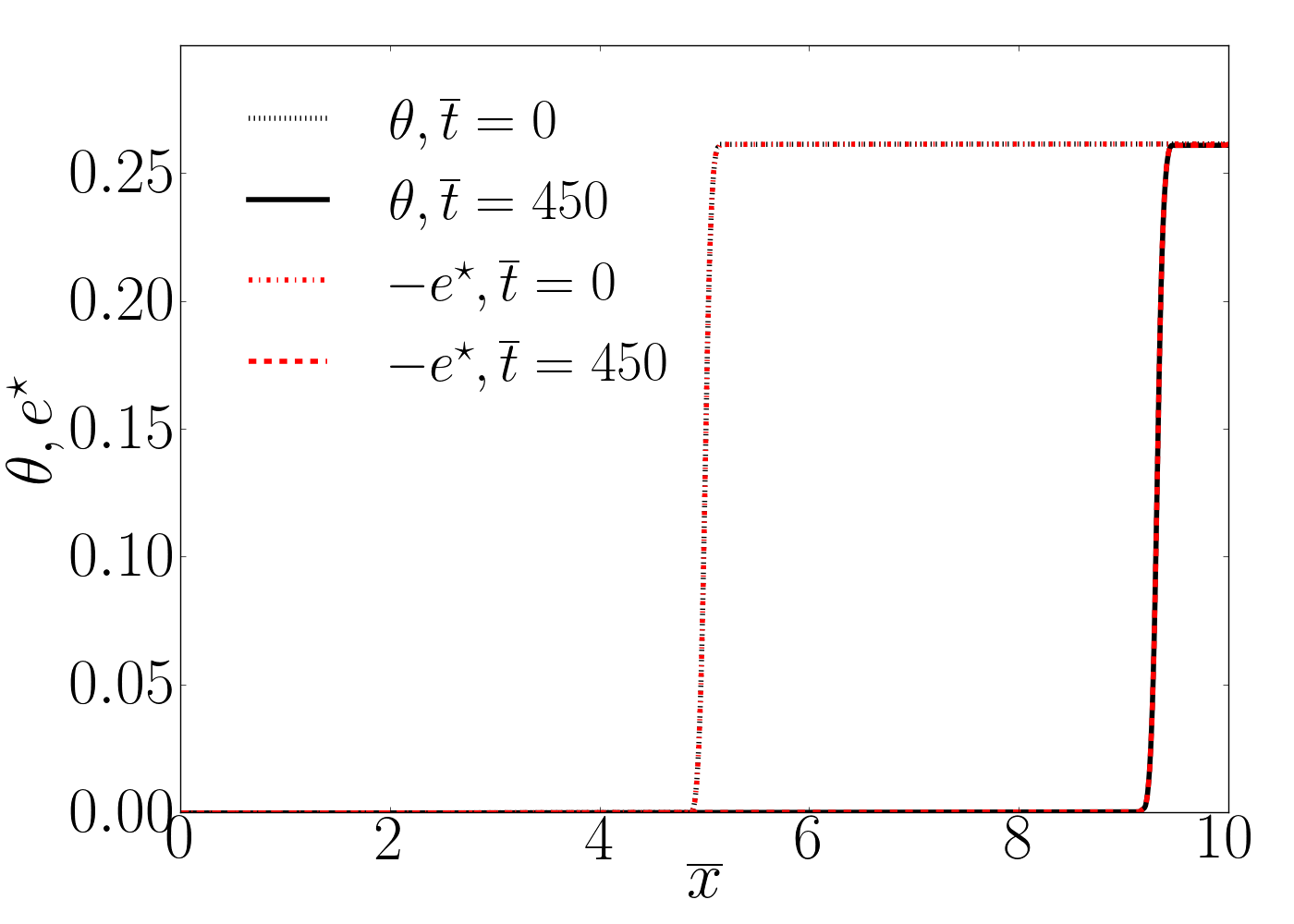}
\includegraphics[height=5.5cm]{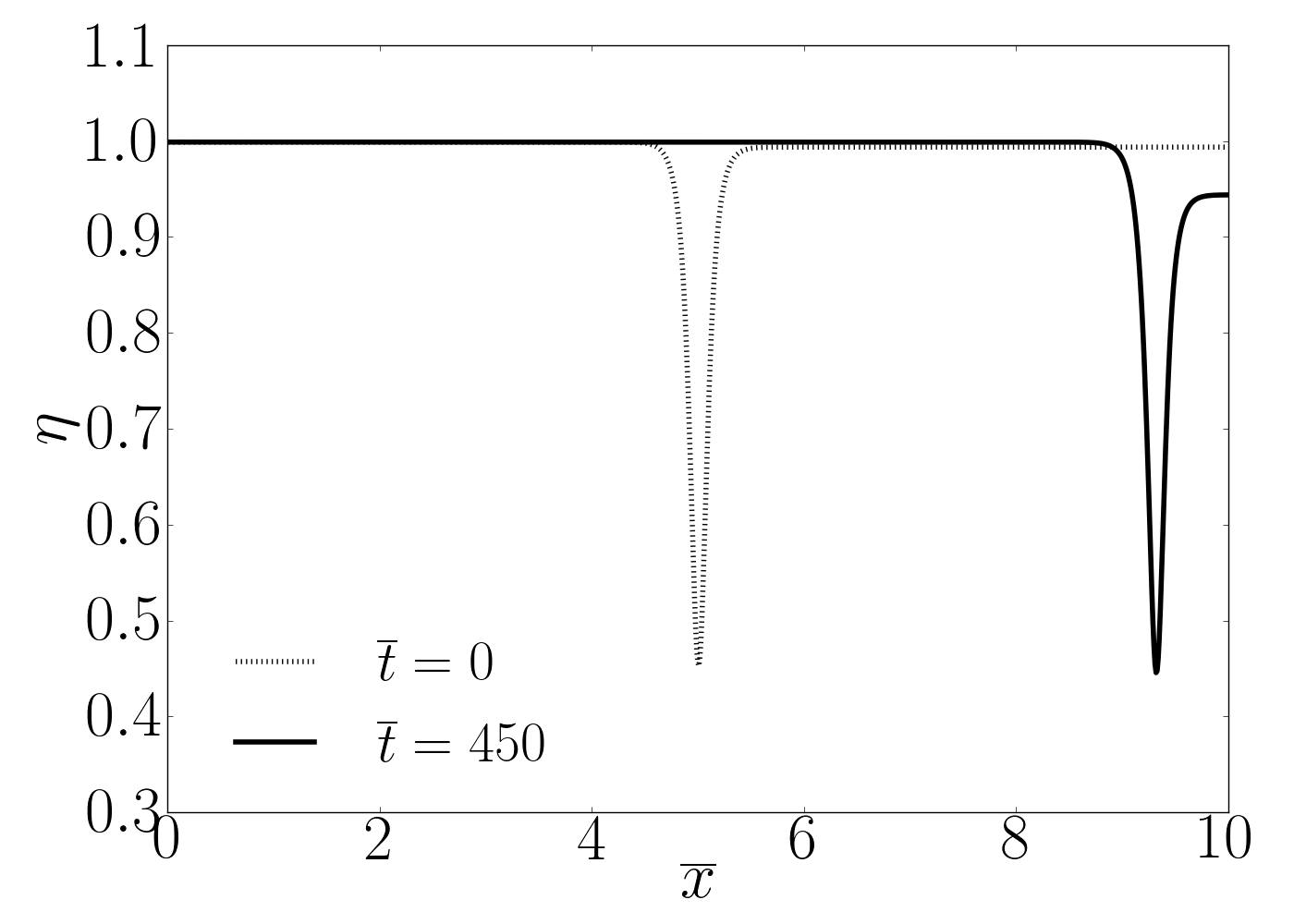}
\includegraphics[height=5.5cm]{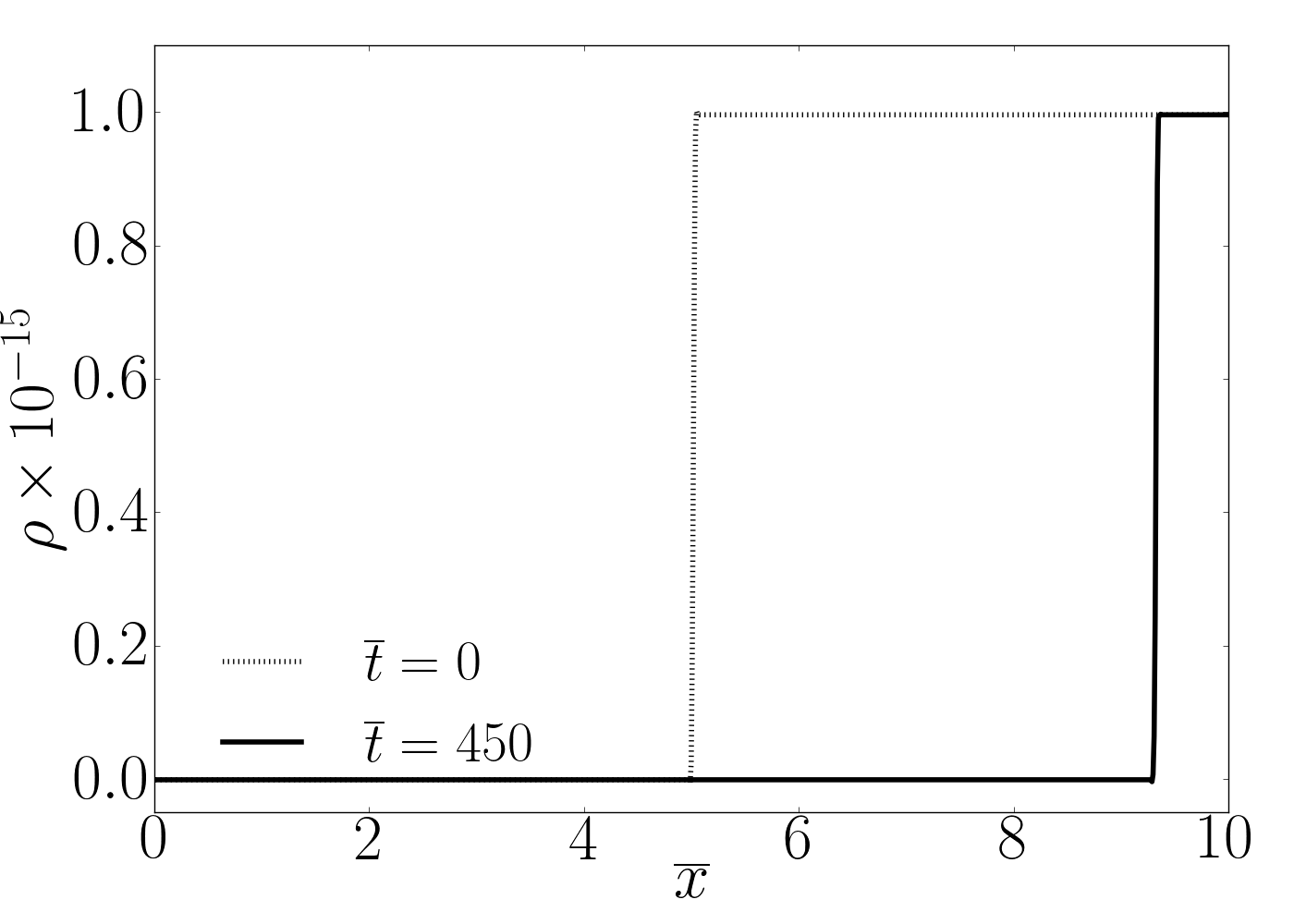}
\includegraphics[height=5.5cm]{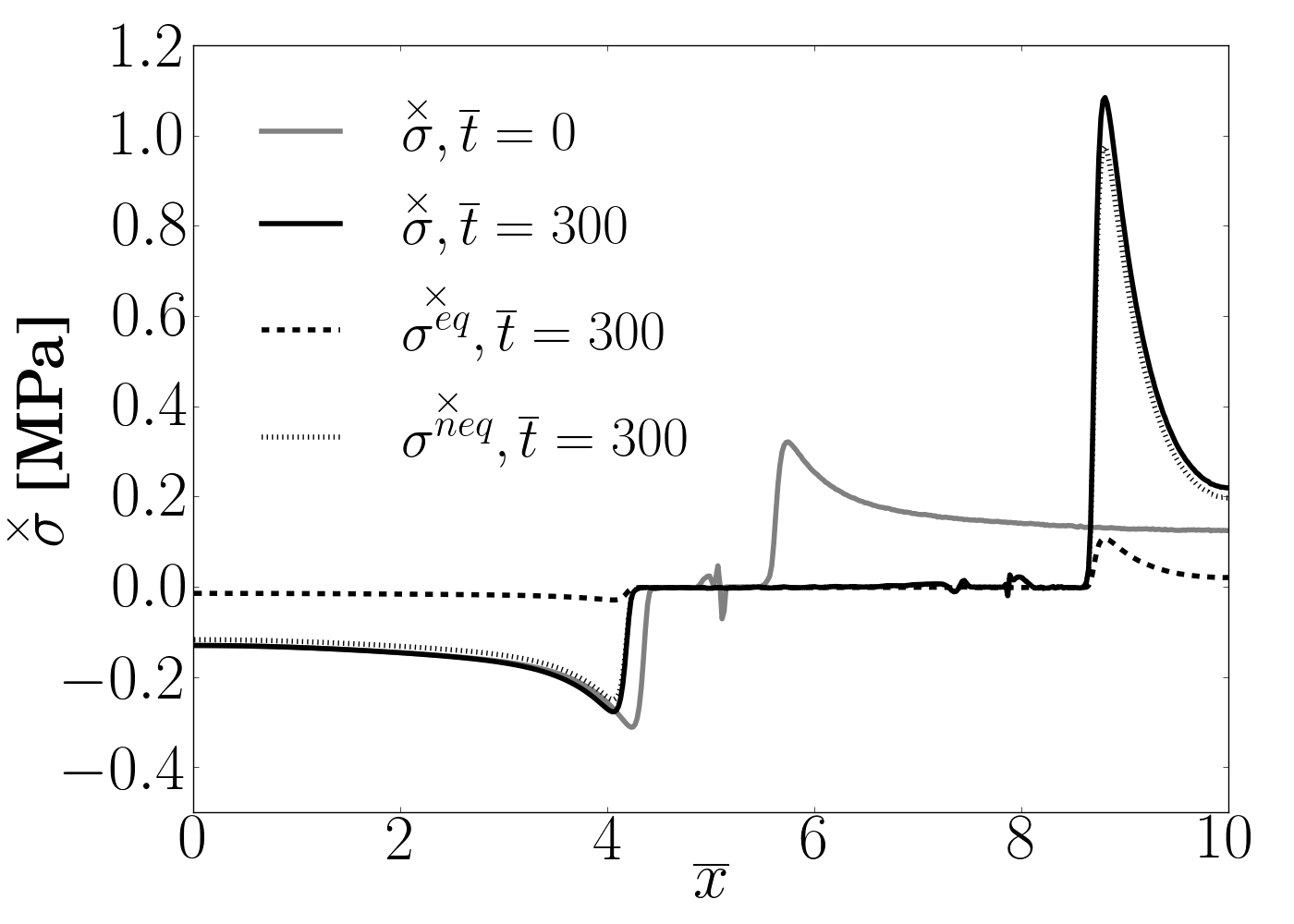}
\caption{\label{Estored} Tracking of the moving grain boundary. Top left and right shows the profiles of $\theta$ and $\eta$, respectively, at $\overline{t}=0$ and $\overline{t}=450$. Bottom left shows $\rho$ at $\overline{t}=0$ and $\overline{t}=450$.}
\end{figure}
\begin{figure}
\includegraphics[height=5.5cm]{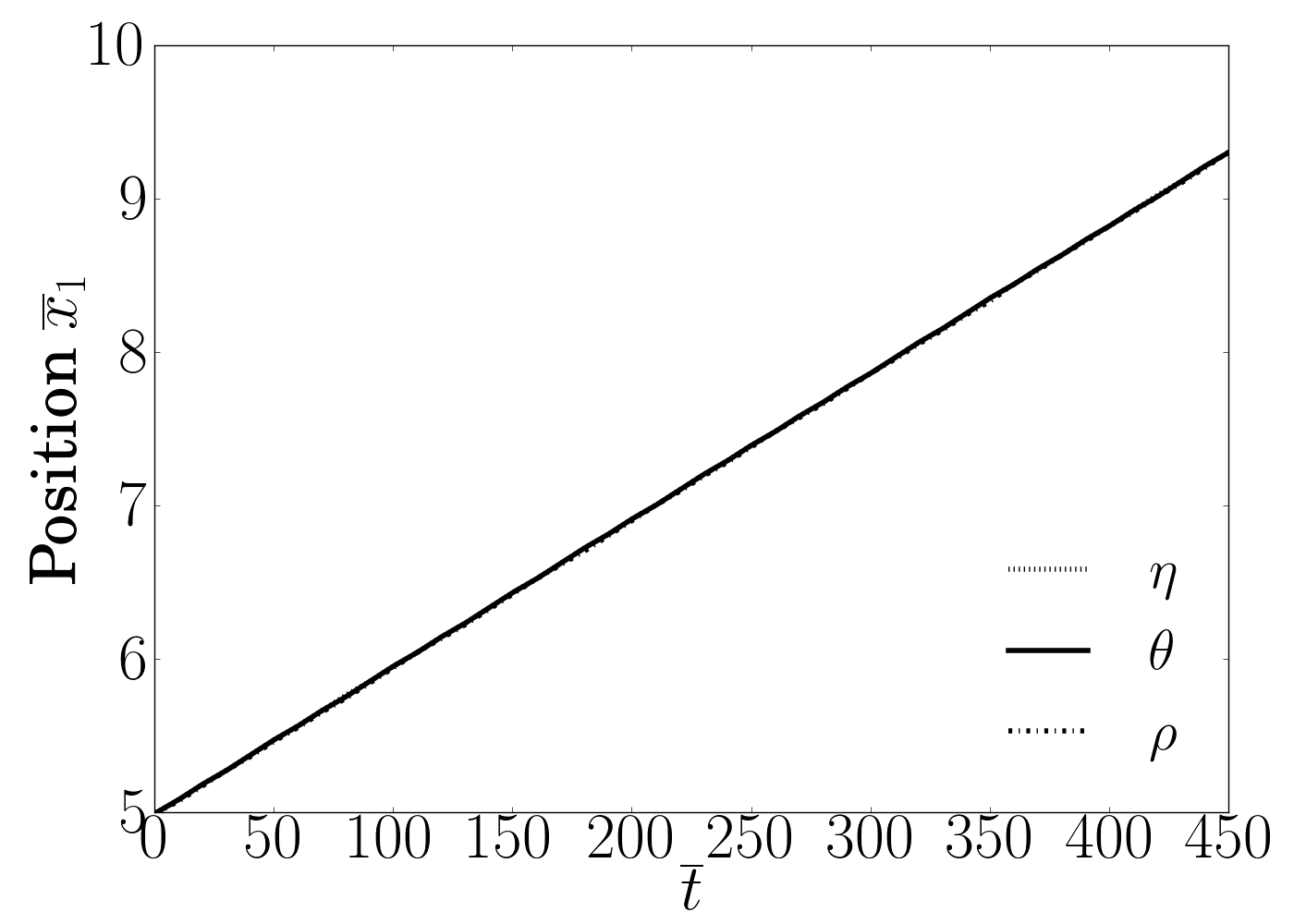}
\includegraphics[height=5.5cm]{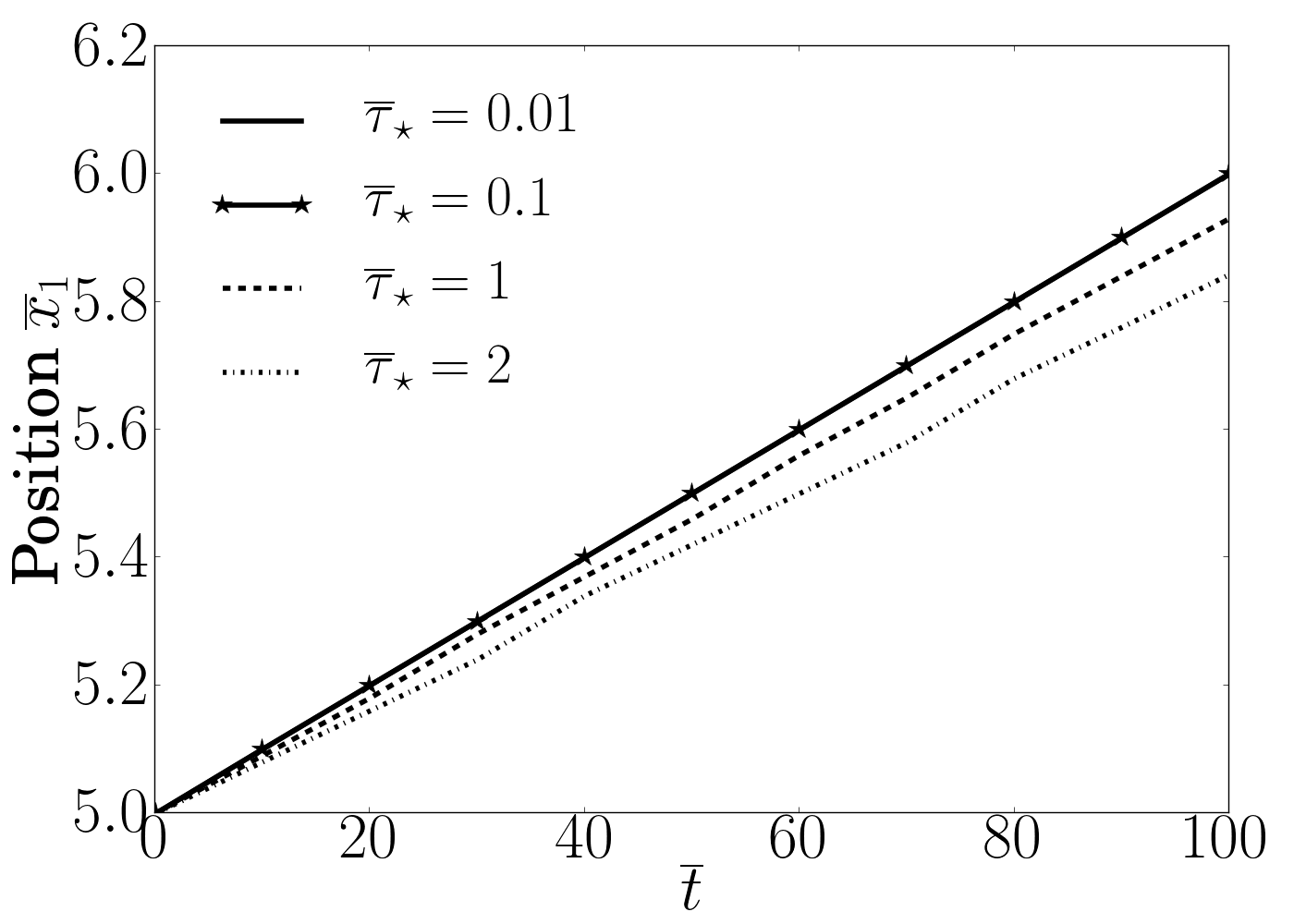}
\caption{\label{fronts_vel} {Position of the grain boundary center over time (left) for the respective profiles of $\eta$, $\theta$ and $\rho$. The influence of the parameter $\overline{\tau}_\star$ on the velocity of the grain boundary migration is also shown (right). Note that the solid line ($\overline{\tau}_\star = 0.01$) and the solid line with star markers ($\overline{\tau}_\star = 0.1$) coincide.}}
\end{figure}
\begin{figure}[t!]
\centering
\includegraphics[height=5.5cm]{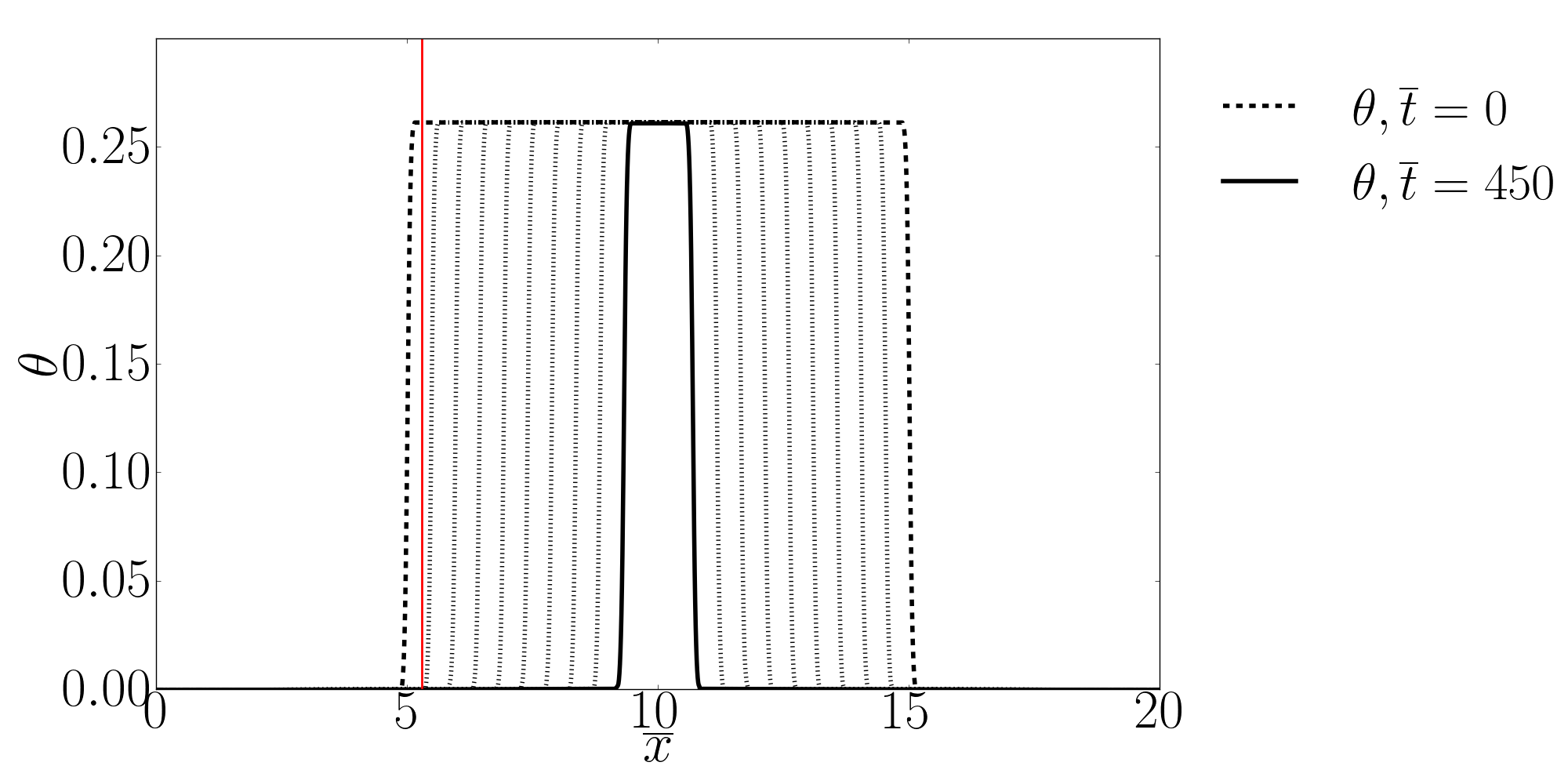}
\includegraphics[height=5.5cm]{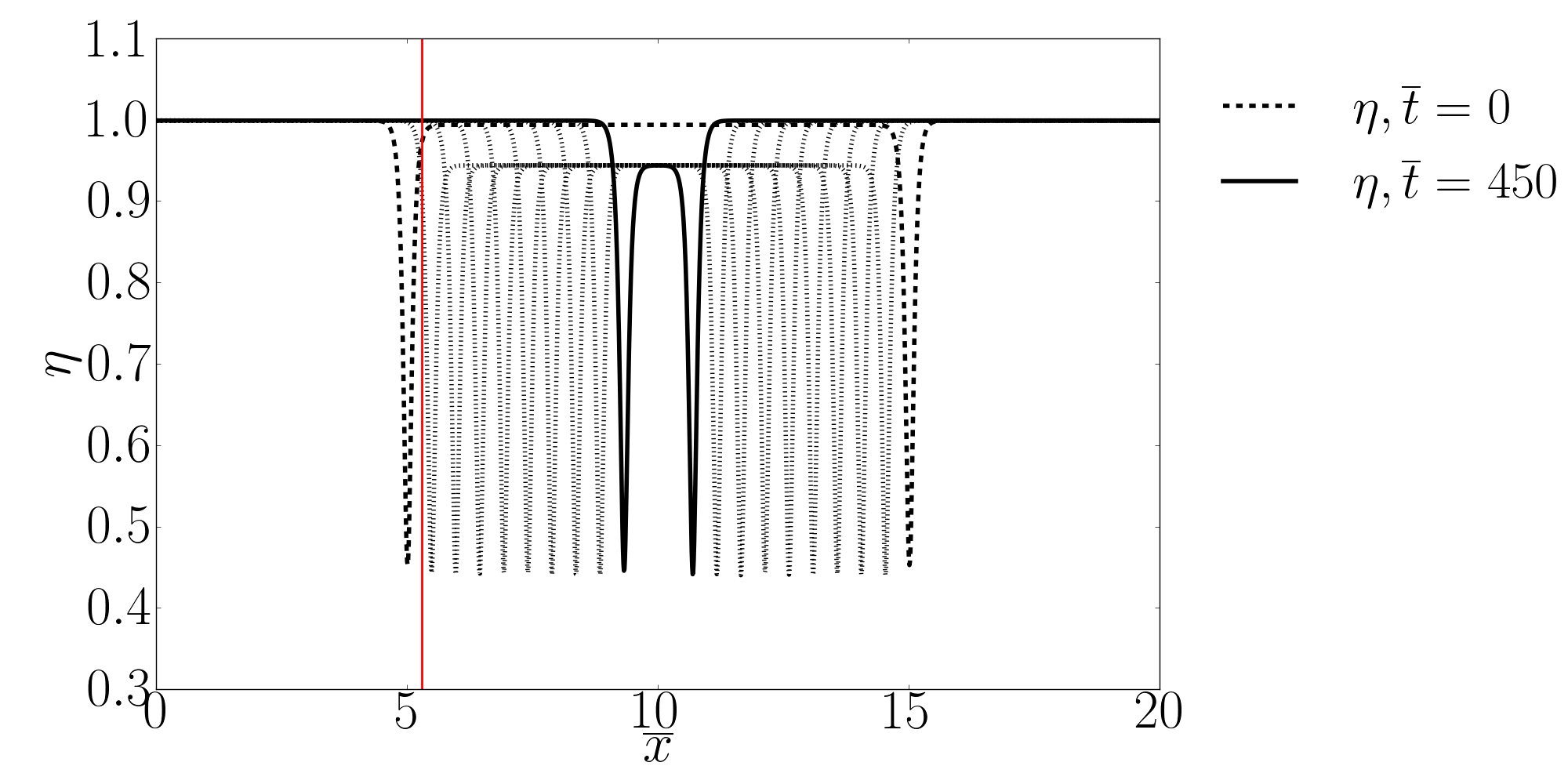}
\caption{\label{Estored_evol} Evolution in time of $\theta$ and $\eta$ at discrete time steps $\Delta \overline{t}$ = 50. The dislocation free grain grows and will eventually annihilate the grain of high stored dislocation density. The red line shows the position of the fixed node where results are extracted over time in figure \ref{Estored_node}.}
\end{figure}

A gradient in the stored energy due to non--homogeneous plastic deformation is an important driving force for grain boundary migration. In the proposed model, {the energy due to accumulation of dislocations (SSD) is included in the term $\psi_\rho(\eta,r^\alpha)$ in the free energy density (\ref{energy_isotropic}). 
This contribution to
stored energy is due to statistically stored dislocations (SSD) according to equations (\ref{dislo_energy}) and (\ref{rrho}).} Assuming for simplicity that only one slip system is active, the energy contribution due to statistically stored dislocations becomes
\begin{equation}
\psi_\rho(\eta,r) = \frac{1}{2}\eta \, \lambda \, \mu \, b^2 \rho \,.
\end{equation}
The Burgers vector for copper is $b=0.2556$ nm. The shear modulus can be calculated from the elastic properties given in \ref{app2} and $\lambda=0.3$ is used. Assuming isotropic elasticity and a dislocation density of $\rho = 10^{15}$ m$^2$ for cold worked copper \citep{humphreys04}, the stored energy is then around $0.45\,\eta$ MJ/m$^3$. With $f_0=8.2$ MJ/m$^3$ it follows that the dimensionless stored energy is 
\begin{equation}
\overline{\psi}_\rho(\eta,r) = \frac{\psi_\rho(\eta,r)}{f_0} = 0.055\,\eta \,.
\end{equation}
\begin{figure}[t!]
\centering
\includegraphics[height=5.5cm]{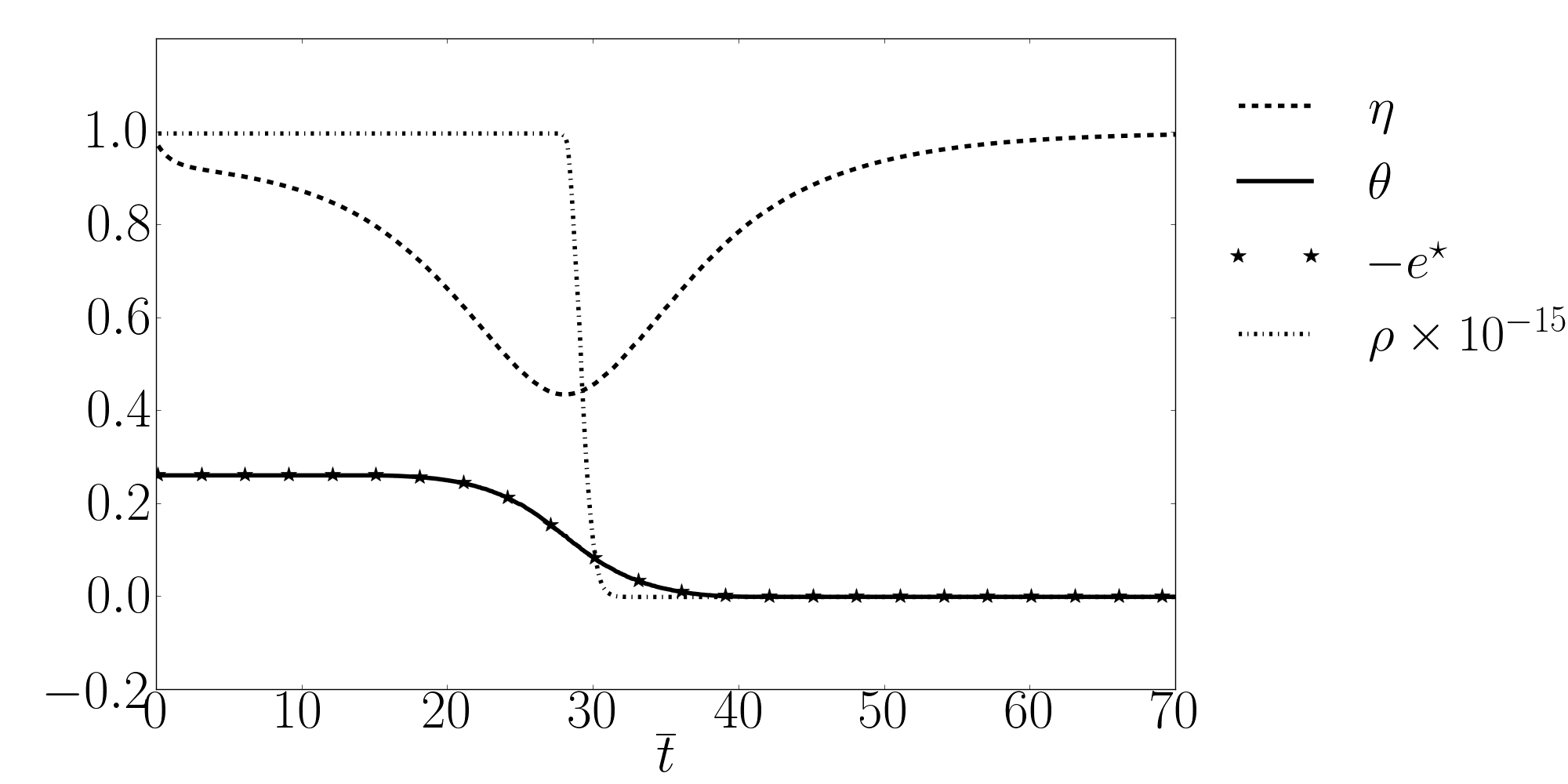}
\caption{\label{Estored_node} {Evolution} in time of $\eta$ and $\theta$ together with $\rho$ at a fixed point swept by the moving GB.}
\end{figure}
In order to study the influence of stored energy on the behavior of the model, 
the $15^\circ$ oriented grain of the unit cell is assigned an initial density of statistically stored dislocations of $\rho_0 = 10^{15}$~m$^{-2}$ whereas the $0^\circ$ grain is assumed to be dislocation free (see figure \ref{Estored}, bottom left). It is expected that the grain free of dislocations 
will expand through migration of the grain boundaries and that static recovery 
will take place behind the moving front according to equation (\ref{KocksMecking}). The initial values of $\vec \varTheta$, $\vx e\,^\star$ and $\eta$ are those found in the first example and used throughout in this section. {In the simulations, $\mu_c/f_0 = 1000$, $\overline{\tau}_\theta=1$, $\overline{\mu}_P=0.5$. All other material parameters, along with those governing the recovery, are given in table \ref{material_parameters_table} in \ref{app2}.}

Figure \ref{Estored} shows the profiles of $\eta$, $\theta$ and $\rho$ at $\overline{t}=0$ and $\overline{t}=450$, initially and just before the dislocation free grain entirely has replaced the grain of high dislocation content. When the dislocation content is non-zero in a grain, this will be reflected in the value of the order parameter which now stabilizes at a value $\eta < 1$,
as can be seen in figure \ref{Estored} (top right). 
This value of the order parameter corresponds to a uniform static equilibrium 
solution of the
equation (\ref{evol_eta}) and {can be expressed in terms of the involved material 
parameters\footnote{{At equilibrium, $\eta$ inside the grain is given by $\eta = 1 - \psi_{\rho,\eta}/f_0$.}},
 including the function $f_{,\eta}$ as calculated by \cite{Abrivard2012a}.}
During grain boundary migration, the eigen-rotation should follow the moving front in order to maintain the correct reference orientation of the growing grain. This is ensured by the evolution law (\ref{evol_estar_isotropic}) with the mobility parameter specified in (\ref{estar_evol_1D}). Figure \ref{Estored} (top left) shows the profiles of $\theta$ and the only non-zero component of the eigen-rotation, $e^\star = \overset{\times}{e}\,^\star_3$ (in red) at $\overline{t}=0$ and $\overline{t}=450$. For the displacement-free case, vanishing skew-symmetric stress $\vx \sigma\,^{eq}$ requires that $\vx e\,^\star = -\vec \varTheta$ which is clearly fulfilled in this case. Furthermore, the eigen-rotation follows the migrating grain boundary as expected, ensuring that the growing grain keeps its stress-free reference orientation of $0^\circ$. 
{Figure \ref{Estored} (bottom right) shows the profile of the 
skew-symmetric stress $\overset{\times}{\sigma}\,_3$ at $\overline{t}=0$ 
(gray line) and $\overline{t}=300$ (black line), respectively. 
A peak of the skew--symmetric stress is observed in the grain boundary.
It results from the crystal orientation changes in the grain boundary
and is relaxed after the passing of the grain boundary due to the
change of lattice reference represented by the eigenrotation variable.
It can be noticed that the skew--symmetric stress inside the grains is not 
strictly zero in contrast to 
what woud be expected in the initial state at $\overline{t}=0$. 
This is a consequence of the KWC model which is used to find the initial conditions on the orientation. Due to the
periodic boundary conditions, in the pure KWC model the grains would rotate 
which is prevented by a large value of the mobility function inside the grains. In the coupled
model, the rotation is instead prevented by the Cosserat penalty term.
The driving force for the evolution of eigenrotation is $\overset{\times}{\sigma}\,^{eq}_3$ which is plotted as a dashed line at $\overline{t}=300$. 
The non-equilibrium stress $\overset{\times}{\sigma}\,^{neq}_3$ (dotted line) 
is non-vanishing inside the grains. It is clear that this would lead to bulk grain 
rotation in the case of $\mu_c = 0$, i.e. in the original KWC model.
The skew symmetric stress is relaxed to zero after the passage of the grain
boundary due to the lattice reorientation via the evolution of the eigenrotation.}

{Figure \ref{fronts_vel} (left)} tracks the migrating grain boundary over time for, respectively, the profile of $\eta$, $\theta$ or $\rho$. From the figure it is seen that the grain boundary velocity is constant and that the relative position of the grain boundary remains the same for all variables. The dimensionless velocity\footnote{If the true grain boundary velocity $v$ at a given temperature and driving force (corresponding to the energy due to the given dislocation content) is known then the characteristic time $\tau_0=t/\overline{t}$ can be calculated as $\tau_0=\Lambda \overline{v}/v$.} of the interface is $\overline{v} = 9.6 \cdot 10^{-3}$. {Figure \ref{fronts_vel} (right) shows the influence of the parameter $\overline{\tau}_\star$ on the velocity of the grain boundary. This is the viscosity parameter--non-dimensionalized according to equation (\ref{tau_nodim})--for the evolution of the eigenrotation. As can be expected, in order not to impact the velocity of the grain boundary it is required that $\overline{\tau}_\star$ is smaller than the other viscosity type parameters which are $\overline{\tau}_\eta = \overline{\tau}_\theta = 1$ in this case. On the other hand, if $\overline{\tau}_\star$ is sufficiently small, it does not influence the grain boundary velocity as evidenced by the fact that the result is the same for $\overline{\tau}_\star = 0.1$ and $\overline{\tau}_\star = 0.01$.}

Figure \ref{Estored_evol} shows the profiles of $\theta$ and $\eta$ over time at fixed times from $\overline{t}=0$ to $\overline{t}=450$ with $\Delta \overline{t} = 50$, illustrating again how the dislocation free grains will grow and eventually annihilate grains of high stored dislocation density. In figure \ref{Estored_node} the evolution of the variables at a fixed point, swept by the moving interface, is represented. The point is at a distance of 0.3 $\mu$m to the right of the grain boundary inside the $15^\circ$ grain, shown by a red line in figure \ref{Estored_evol}. Figure \ref{Estored_node} shows how the values of $\eta$ (dashed line) and $\theta$ (solid line) change at the node as the GB passes it. The evolution of the eigen-rotation (stars) follows the evolution of $\theta$. Figure \ref{Estored_node} also shows the evolution of the dislocation density at the node (dotted line). The parameter $C_D$ in equation (\ref{KocksMecking}) has been chosen to be $C_D=100$. This value ensures that full static recovery takes place, as evidenced by the fact that the dislocation content is zero when the grain boundary has passed. The static recovery becomes active at the time when $\dot{\eta}$ shifts from negative to positive. 
 
\section{Conclusions}

The novel features of the proposed theory are the following:
\begin{enumerate}
\item A general 3D anisotropic constitutive framework
was proposed that intimately couples grain boundary motion and mechanics.
The generalization of the 2D
KWC approach leads to a full 3D Cosserat framework involving the microrotation
tensor and the torsion-curvature tensor $\ten \kappa$ identified as the
lattice curvature tensor. The KWC orientation field equation was interpreted
as a balance equation for couple stresses. 
\item An essential feature of the model
is the concept of relative rotation representing the difference
between lattice rotation and Cosserat microrotation. This difference
vanishes in the bulk of the grain but is generally non-zero inside the
grain boundary. In previous model formulations, the rotation phase field variable and
the material or lattice rotation as computed from crystal plasticity simulations were separate variables, whose difference
was left uncontrolled.  
\item An eigen-rotation variable with its relaxation equation is
introduced to control the magnitude of skew-symmetric stresses inside the
grain boundaries. The skew-symmetric part of the stress tensor in the Cosserat
theory is the driving force for material reorientation inside grain boundaries.
The eigen-rotation is necessary in particular to represent
stress-free
initial or relaxed grain boundaries. The corresponding relaxation time is
assumed to be much smaller than grain boundary migration characteristic
times.
\item The energetic contribution of the lattice curvature tensor has been
recognized as an approximate evaluation of the stored energy due to the dislocation density tensor. This explains the introduction of the norm of the
lattice curvature tensor in the free energy density, as initially proposed in the KWC model for different reasons, 
in addition to a quadratic contribution
also used in strain gradient plasticity \citep{wulfinghoffForest14}.
\item The model contains two contributions of dislocations to the stored
energy and its evolution: stored energy due to so-called geometrically
necessary dislocations related here to the lattice curvature tensor,
and stored energy due to statistically stored dislocations described by the
corresponding densities $\rho^\alpha$. The Cosserat mechanics phase field
contribution controls the evolution of lattice curvature whereas an
evolution law was proposed for the static recovery of stored dislocations
during grain boundary migration.
\item In the uncoupled KWC model (spurious) grain rotations had to be controlled by careful construction of the mobility $\tau_\theta$. It is now naturally limited by the Cosserat coupling
modulus, $\mu_c$. This is a significant improvement of the model since the grain boundary migration rate can now be identified independently of the
overall rotation rate. 
\end{enumerate}

Simple examples were provided highlighting the consistency of the phase field
mechanical coupling. In the proposed theory, lattice rotation can have several
origins: rigid body motion, lattice stretching due to applied strains and
grain boundary migration. The sweeping of the dislocated crystal by the grain
boundary leads to the (possibly partial) vanishing of dislocations 
controlled by a static recovery term coupled with the phase field in the
dislocation density evolution equation.

Several features of grain boundary behavior are not accounted for in 
the present model:
the detailed defect structure of GB that can be found in more sophisticated
models by \citep{berbenni13}, the interaction between dislocations and 
grain boundary
(dislocation absorption, annihilation, emission as addressed in 
gradient crystal plasticity contributions \cite{wulfinghoff13IJP,geersGB15}).

The next steps of the development of the model are the following.
First examples of lattice rotation induced by plastic slip and its
interaction with GB migration will be explored. Comparison with recent
experimental results by \cite{bacroix} will be performed.
The generalization to the finite deformation and finite rotation framework
will then be exposed. {The infinitesimal framework 
used in this presentation allows for a direct comparison between the proposed framework and the KWC model but the next step is to extend it to the
finite deformation framework.} Finally, 
the anisotropic GB energy, i.e. an  energy landscape depending on GB 
orientation and grain misorientation, must be incorporated explicitly 
by appropriate potentials to be developed.

\section*{Acknowledgements} This project has received funding from the European Research Council (ERC) under the European Union's Horizon 2020 research and innovation program (grant agreement n$^\circ$ 707392 MIGRATE).

\section*{References}
\bibliography{bibCOSSKWC} 

\newpage
\appendix

\section{Model equations in 2D}\label{app1}

Restricting the calculations to two dimensions and the axis of rotation to be the $x_3$-axis gives $\vec \varTheta =  [\begin{array}{ccc} 0 & 0 & \theta_3 \end{array}]^T$ with $\theta_3$ the angle of rotation. Since there are no out-of-plane rotations it also follows that $\vx \omega = [\begin{array}{ccc} 0 & 0 & \omega_{21} \end{array}]^T $ and $\vx e = [\begin{array}{ccc} 0 & 0 & \overset{\times}{e}_3 \end{array}]^T$. Furthermore, there is only one non-zero term in the skew-symmetric stress $\vx \sigma = [\begin{array}{ccc} 0 & 0 & \overset{\times}{\sigma}_3 \end{array}]^T$.

This allows for considerable simplification of the model equations. Dropping the subscripts on the quantities above so that $\theta = \theta_3$, $\omega = \omega_{21}$, $\overset{\times}{e}  = \overset{\times}{e}_3 $ and $\overset{\times}{\sigma}  = \overset{\times}{\sigma}_3 $, the principle of virtual power (ignoring external loads) reduces to
\begin{equation}
p^{(i)} 
= -\pi_\eta \virt \eta + \vec \xi_\eta \cdot \nabla \virt \eta
-2\,\overset{\times}{\sigma} \virt \theta + \vec m_\theta \cdot \nabla \virt \theta
+ 
\ten \sigma :
\virt{\vec u}\otimes \nabla 
\,.
\label{pvp_internal_2D}
\end{equation}
where $\vec m_\theta = [\begin{array}{ccc} m_{31} & m_{32} & m_{33} \end{array}]^T$ contains the non-zero contributions to $\ten m $ of equation (\ref{pvp_internal_coupled}). The weak format of the governing equations (which are the starting point for the finite element implementation) is then given by
\begin{align}
& 
\int_\omega \left[\,
\vec \xi_\eta \cdot \nabla \virt \eta - \pi_\eta\,\virt \eta
\,\right] \, {\rm d} V = \int_{\partial \omega} \virt \eta \,\vec \xi_\eta \cdot \vec n \, {\rm d} S \,, \\
& 
\int_\omega \left[\,
\vec m_\theta \cdot \nabla \virt \theta - 2\,\overset{\times}{\sigma}\,\virt \theta
\,\right] \, {\rm d} V = \int_{\partial \omega} \virt \theta \,\vec m_\theta \cdot \vec n \, {\rm d} S \,, \\
&
\int_\omega \left[\,
\ten \sigma^{\rm sym} : \virt{\ten \varepsilon} \cbox{+ 2\,\overset{\times}{\sigma} \, \omega}
\,\right] \, {\rm d} V =  \int_{\partial \omega} \virt{\vec u} \cdot \ten \sigma \cdot \vec n \, {\rm d} S \,.
\end{align}
The term highlighted in {gray} in the last equation is a coupling term due to the Cosserat formulation that was not present in e.g. the formulation used in \cite{Abrivard2012a}. 

The simple choice $g(\eta)=h(\eta)=\eta^2$ is made in line with \citep{KWC2000}, together with $\mu_c = $ constant. For the function $f(\eta)$ it is assumed to have a single minimum at $\eta=1$ and is therefore taken to be $f(\eta)=\tfrac{1}{2}[\,1-\eta\,]^2$. With these choices, the free energy function (\ref{energy_isotropic}) becomes
\begin{equation}
\begin{aligned}
\psi(\eta,\nabla \eta, \ten e^e, \ten \kappa,r^\alpha) 
= \, &
f_0 
\left[ 
\frac{1}{2}[\,1-\eta\,]^2 + \frac{a^2}{2}|\nabla \eta|^2 + 
s\,\eta^2\,|\nabla \theta| + 
\frac{\varepsilon^2}{2}\eta^2\,|\nabla \theta|^2 
\right ] 
\\
+ \, & \frac{1}{2}\ten \varepsilon^e : \TEN E^s :
\ten \varepsilon^e + 2\,\mu_c\,[\overset{\times}{e}\,^e]^2 +\eta\,\hat{\psi}^\alpha(\rho^\alpha)
\,,
\end{aligned}
\label{energy_isotropic_2D}
\end{equation}
where
\begin{equation}
\hat{\psi}^\alpha(\rho^\alpha) =  \sum_{\alpha = 1}^N \frac \lambda 2
\mu r^{\alpha 2} \,.
\end{equation}
The constitutive equations are given by
\begin{align}
\vec \xi_\eta = \, &  f_0\,  a^2 \nabla \eta  \\
\vec m_\theta = \, & f_0\left[
\,s\,\eta^2\frac{\nabla \theta}{|\nabla \theta|}
 +\varepsilon^2\,\eta^2\,\nabla \theta\,
\right]  \\
\pi_\eta = \, &  
 - f_0 \left[\,-[\,1-\eta\,] + 
2\,s\,\eta\,|\nabla \theta| +
\varepsilon^2\,\eta\,|\nabla \theta|^2  
\,\right] 
-  \hat{\psi}^\alpha(\rho^\alpha)  + \pi_\eta^{neq}\\
\overset{\times}{\sigma}\,^e = \, & 2\,\mu_c\, \overset{\times}{e}\,^e \\ 
\ten \sigma^{\rm sym} = \, & 
\TEN E^s : \ten \varepsilon^e 
\end{align} 
together with the evolution equations
\begin{align}
\pi_\eta^{neq} = \, & -\tau_\eta\,\dot{\eta}  \,, \label{eta_evol_1D}\\
\overset{\times}{\sigma}\,^{neq} = \, & \,\frac{1}{2}P(\nabla \theta)\,\hat{\tau}_\theta\,\eta^2\,[\omega^e - \dot{\theta}] \,, \label{theta_evol_1D}\\
\dot{\overset{\times}{e}}\,^\star = \, & \,\hat{\tau}\,_\star^{-1}\tanh(C_\star^2|\nabla \eta|^2)\,\overset{\times}{\sigma}\,^{eq} \,, \label{estar_evol_1D}
\end{align}
where specific choices have been made for the viscosity type parameters $\tau_\eta(\eta,\nabla \eta, \nabla \theta,T)$, $\tau_\theta(\eta,\nabla \eta, \nabla \theta,T)$ and $\tau_\star(\eta,\nabla \eta, \nabla \theta,T)$ which govern the evolution of the model toward equilibrium. In particular, by allowing them to depend on $\eta$, $\nabla \eta$ and $\nabla \theta$ it is possible to distinguish between behavior in the grain and in the grain boundary. In general the mobility functions are strongly temperature dependent although, for brevity, this will not be explored in this work and it will be assumed that all simulations are done at constant temperature. Temperature dependence can easily be introduced in the formulation by e.g. an Arrhenius type law as in \cite{Abrivard2012a}. In accordance with \cite{KWC2003}, it is assumed that $\tau_\eta$ is constant and that 
\begin{equation}
\tau_\theta(\eta,\nabla \eta, \nabla \theta,T) = \frac{1}{2} P(\nabla \theta)\,\hat{\tau}_\theta\,\eta^2
\label{tau_phi}
\end{equation}
with
\begin{equation}
P(\nabla \theta) =1 - \left[\,1-\frac{\mu_P}{\varepsilon}
\,\right]\,\exp(-\beta_P\,\varepsilon\,|\nabla \theta|) \,.
\label{Pmob_dim}
\end{equation}
The mobility function $P$ differentiates between the behavior in the bulk, where it can be made arbitrarily large through the choice of the parameter $\mu_P$, and the behavior in the grain boundary where it approaches $P=1$ for large values of the parameter $\beta_P$. This provides a means of separating the time scales for grain boundary migration and grain rotation. It is necessary that $P$ be dimensionless so the unit of $\mu_P$ is [m] while $\beta_P$ has no unit. For the evolution of the eigen-rotation the viscosity type parameter is taken as
\begin{equation}
\tau_\star = \hat{\tau}\,_\star\tanh^{-1}(C_\star^2|\nabla \eta|^2) \,, \label{tau_star}
\end{equation}
where the format has been chosen so that the evolution is restricted to the grain boundary. The parameter $C_\star$ has unit [m].

The balance equations together with the above results now give
\begin{align}
& \tau_\eta\,\dot{\eta} = f_0\,a^2 \Delta \eta - f_0 \left[\,-[\,1-\eta\,] + 
2\,s\,\eta \,|\nabla \theta| +
\varepsilon^2\,\eta\,|\nabla \theta|^2  
\,\right] 
-  \hat{\psi}^\alpha(\rho^\alpha)  \,,\\
& P(\nabla \theta)\,\tau_\theta\,\eta^2\,[\cbox{\omega^e} - \dot{\theta}] = - \nabla \cdot f_0\left[
\,s\,\eta^2\frac{\nabla \theta}{|\nabla \theta|}
 +\varepsilon^2\,\eta^2\,\nabla \theta\,
\right] \cbox{- 4\,\mu_c \overset{\times}{e}\,^e} \,. \label{2Devol_theta}
\end{align}
The terms highlighted in {gray} are the terms which are new compared to the model by \cite{KWC2000,KWC2003} and \citep{Abrivard2012a}. {Equation (\ref{2Devol_theta}) is of singular diffusive type and requires regularization in the numerical treatment \citep{KobayashiGiga1999}. A regularization scheme proposed in \cite{KWC2003} is adopted. In this scheme, the linear term $|\nabla \theta|$ is replaced by the function $A_\gamma(|\nabla \theta|)$, where
\begin{equation}
A_\gamma(\xi) = \left \lbrace
\begin{aligned}
& \frac{\gamma}{2}\,\xi^2  \qquad  \,  \rm{for} \,\, 0 \leq \xi \leq 1/\gamma  \,, 
\\
& \xi - \frac{1}{2\gamma} \qquad  \rm{for} \,\, \xi > 1/\gamma \,,
\end{aligned}
\right.
\end{equation}
where the constant $\gamma$ is a large positive number}. The function $A(||\ten  \kappa||)$ in equation (\ref{KocksMecking}) is chosen to be 
\begin{equation}
A(|\nabla \theta|) = \tanh(C_A^2\,|\nabla \theta|^2)\,,
\label{Afunction}
\end{equation}
where the parameter $C_A$ has unit [m].

\section{Model parameters}\label{app2}

Model parameters are chosen to be in the reasonable range for a test material, in this case pure copper, although a rigorous parameter fitting to experimental data has not been performed at this stage of general model development. Some parameters, such as elastic constants, can easily be found from reference literature. For the Cosserat parameter, $\mu_c$, it is required that it is large enough to penalize differences between the rate of lattice rotation and the rate of Cosserat rotation in the bulk of the grain. The parameters $f_0$, $a$, $s$ and $\varepsilon$ together with the mobility parameters can be identified using the results of \cite{LobkovskyWarren2001} who analyzed the KWC model in the sharp interface limit using formal asymptotics. They found expressions for the grain boundary energy and mobility as functions of the model parameters and the misorientation between the grains. 

In order to non-dimensionalize the energy, a suitable length scale $\Lambda$ of unit [m] is introduced to obtain the dimensionless coordinates $(\overline{x}, \overline{y}, \overline{z})=\frac{1}{\Lambda}\,(x,y,z)$. The dimensionless differential operator $\overline{\nabla}$ is then given by $\overline{\nabla} = \Lambda \, \nabla$. Furthermore, $\overline{\ten \kappa}=\vec \varTheta \otimes \overline{\nabla}= \Lambda \, \ten \kappa$. By division with $f_0$, the dimensionless free energy function is then found to be
\begin{equation}
\begin{aligned}
\overline{\psi} = \frac{\psi}{f_0}
= \, &
f(\eta) + \frac{\overline{a}^2}{2}|\overline{\nabla} \eta|^2 + 
\overline{s}\,g(\eta)|\overline{\nabla} \theta| + 
\frac{\overline{\varepsilon}^2}{2}h(\eta)|\overline{\nabla} \theta|^2
\\
+ \, & \frac{1}{2}\ten \varepsilon^e : \overline{\TEN E}^s :
\ten \varepsilon^e + 2\,\overline{\mu}_c(\eta)\,[\overset{\times}{e}\,^e]^2 + \eta\ \overline{\psi}^\alpha(\rho^\alpha)
\,,
\end{aligned}
\end{equation}
where $\overline{a} = a/\Lambda$,  $\overline{s} = s/\Lambda$ and  $\overline{\varepsilon} = \varepsilon/\Lambda$ are now dimensionless and
\begin{equation}
\overline{\TEN E}^s = \frac{1}{f_0}\TEN E^s \,, \qquad 
\overline{\mu}_c = \frac{\mu_c}{f_0} \,, \qquad
 \overline{\psi}^\alpha = \frac{\hat{\psi}^\alpha}{f_0} \,.
\end{equation}
The total energy is non-dimensionalized by observing that ${\rm d} \overline{\Omega} = \Lambda^3 {\rm d} \Omega$, so that 
\begin{equation}
\overline{\mathcal{F}} = \frac{\mathcal{F}}{\Lambda^3\,f_0} = \int_{\overline{\Omega}}
\overline{\psi} \, {\rm d} \overline{\Omega} 
\end{equation}
The parameters $\tau_\eta$, $\hat{\tau}_\theta$ and $\hat{\tau}_\star$ are made dimensionless by
\begin{equation}
\overline{\tau}_\eta = \frac{\tau_\eta}{f_0\,\tau_0} \,, \qquad
\overline{\tau}_\theta = \frac{\hat{\tau}_\theta}{f_0\,\tau_0} \,, \qquad
\overline{\tau}_\star = \frac{\hat{\tau}_\star}{f_0\,\tau_0} \,,
\label{tau_nodim}
\end{equation} 
where $\tau_0$ is a suitable time scale and $\overline{t} = t/\tau_0$. The model parameters can now be fitted to real data by appropriate choices of $f_0$, $\Lambda$ and $\tau_0$. The mobility function $P$ of equation (\ref{Pmob_dim}) is dimensionless by construction. By defining $\overline{\mu}_P = \mu_P/\Lambda$, it can be written in terms of the dimensionless parameters as
\begin{equation}
P(\nabla \theta) =1 - \left[\,1-\frac{\overline{\mu}_P}{\overline{\varepsilon}}
\,\right]\,\exp(-\beta_P\,\overline{\varepsilon}\,|\overline{\nabla} \theta|) \,.
\label{Pmob_nodim}
\end{equation}
The parameter $C_A$ in the function $A(|\nabla \theta|)$ of equations (\ref{KocksMecking}) and (\ref{Afunction}) is non-dimensionalized by $\overline{C}_A = C_A/\Lambda$ so that
\begin{equation}
A(|\nabla \theta|) = \tanh(C_A^2\,|\nabla \theta|^2) = \tanh(\overline{C}_A^2\,|\overline{\nabla} \theta|^2) \,.
\end{equation}
Likewise, $C_\star$ in equation (\ref{tau_star}) is non-dimensionalized by  $\overline{C}_\star = C_\star/\Lambda$.

For full details on the asymptotic analysis, see \cite{LobkovskyWarren2001}. The analysis is performed starting from the scaled dimensionless energy functional $\widetilde{\mathcal{F}} = \overline{\mathcal{F}}/\overline{\varepsilon}$. The grain boundary has two regions. In the inner region, $\nabla \theta \neq 0$. The profiles of $\eta^0$ and $\theta^0$, where superscript $0$ indicates the linearized function at lowest order, can be found from integration of 
\begin{equation}
\frac{\partial \eta^0}{\partial z} = \left\lbrace
\begin{aligned}
\, & \frac{1}{\tilde{a}}\sqrt{2\,f^0 - \frac{\tilde{s}^2}{h^0}
(g^0_{max}-g^0)} \,, \qquad & \eta^0_{min} \leq \eta^0 \leq \eta^0_{max} \,, \\
\, & \frac{1}{\tilde{a}} \sqrt{2\,f^0}\,, \qquad & \eta^0_{max} < \eta^0 \leq 1 \,,
\end{aligned}
\right. 
\end{equation}
where the notation $g^0_{max} = g^0(\eta^0_{max})$ is used, and
\begin{equation}
\frac{\partial \theta^0}{\partial z} = 
\frac{\tilde{s}\,[g^0_{max}-g^0]}{h^0} \,, \qquad 0 \leq z \leq \delta z\,.
\end{equation}
The scaled dimensionless parameters are given by $\overline{a} = \tilde{a}\,\overline{\varepsilon}$ and $\overline{s} = \tilde{s}\,\overline{\varepsilon}$ and the stretched coordinate by $z=x/\overline{\varepsilon}$, with the inner region bounded by $|z|\leq \delta z$. The values of $\eta^0_{min}$ (at the center of the grain boundary $z=0$) and $\eta^0_{max}$ (at the limit of the inner region of the grain boundary) can be found from the matching condition at $\delta z$
\begin{equation}
\frac{\tilde{s}}{2}\frac{\,[g^0_{max}-g_{min}^0]^2}{h_{min}^0} - f_{min}^0 = 0 \,,
\end{equation}
together with
\begin{equation}
\frac{\Delta \theta}{2} = \int_0^{\delta z}\frac{\partial \theta^0}{\partial z}\,{\rm d} z = 
\int_{\eta^0_{min}}^{\eta^0_{max}}\frac{\partial \theta^0}{\partial z}
\frac{\partial z}{\partial \eta^0}\,{\rm d} \eta^0 \,.
\end{equation}

The expression for the non-dimensional grain boundary energy in the sharp interface limit is then found to be 
\begin{equation}
\tilde{\gamma}_{gb} 
= 
2\,\tilde{a}^2\int_{\eta^0_{min}}^{1}\frac{\partial \eta^0}{\partial z} \, {\rm d} \eta^0
+ \,\tilde{s}\,g^0_{max} \Delta \theta
\,.
\label{gb_energy}
\end{equation}
The dimensionless mobility can be calculated from
\begin{equation}
\frac{1}{\widetilde{\mathcal{M}}} = \int_{-\infty}^{+\infty} Q\,\tilde{\tau}_\eta \left(\frac{\partial \eta^0}{\partial z}\right)^2
{\rm d} z
+ \int_{-\infty}^{0} P\,\tilde{\tau}_\theta\,(\eta^0)^2\left(\frac{\partial \theta^0}{\partial z}\right)^2  {\rm d} z
+ \int_0^{+\infty} P\,\tilde{\tau}_\theta\,(\eta^0)^2\left(\frac{\partial \theta^0}{\partial z}\right)^2  {\rm d} z \,.
\label{inv_mobil}
\end{equation}

Using equations (\ref{gb_energy}) and (\ref{inv_mobil}) the parameters $f_0$ and $\tau_0$ can be calibrated to fit experimental data. The true grain boundary energy can be calculated as
\begin{equation}
\gamma_{gb} = f_0\,\Lambda\,\overline{\varepsilon}\,\tilde{\gamma}_{gb} \,.
\end{equation}
which gives the parameter $f_0$ as
\begin{equation}
f_0 = \frac{\gamma_{gb}}{\Lambda\,\overline{\varepsilon}\,\tilde{\gamma}_{gb}} \,.
\label{f0_calib}
\end{equation}
The true mobility can be calculated as
\begin{equation}
\mathcal{M} = \frac{\overline{\varepsilon}\,\Lambda}{\tau_0\,f_0}\widetilde{\mathcal{M}}\,.
\end{equation}
If the true mobility and grain boundary energy are known from experiments, a suitable time scale can then be determined by fixing all other parameters and using
\begin{equation}
\tau_0  = \frac{\overline{\varepsilon}\,\Lambda}{f_0}\frac{\widetilde{\mathcal{M}}}{\mathcal{M}}\,.
\label{calib_tau0}
\end{equation}

\begin{figure}[t!]
\centering
\includegraphics[width=0.48\linewidth]{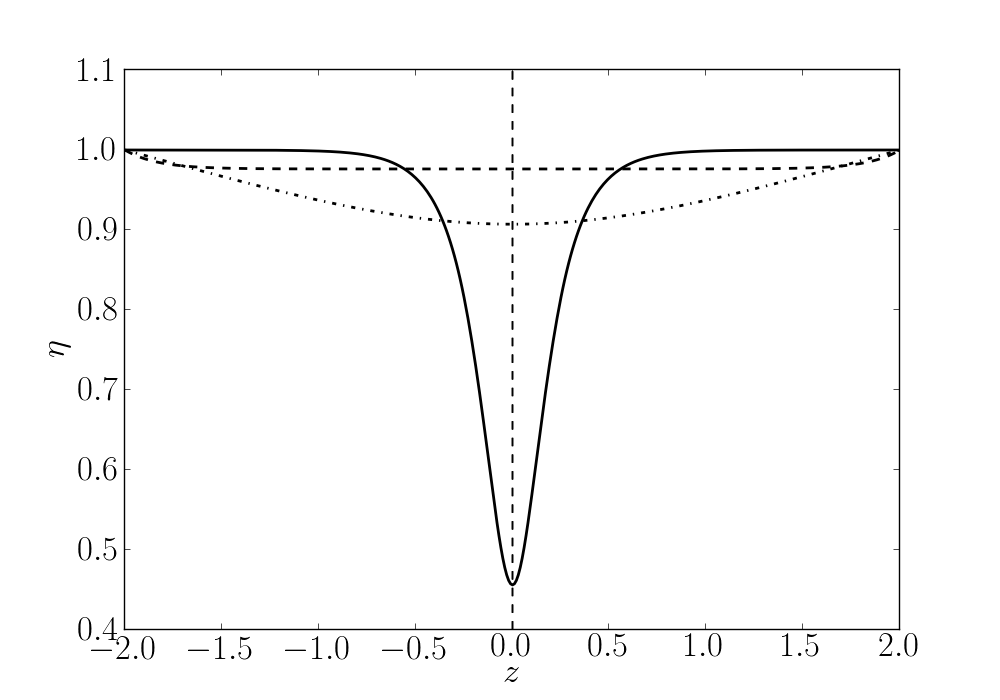}
\includegraphics[width=0.48\linewidth]{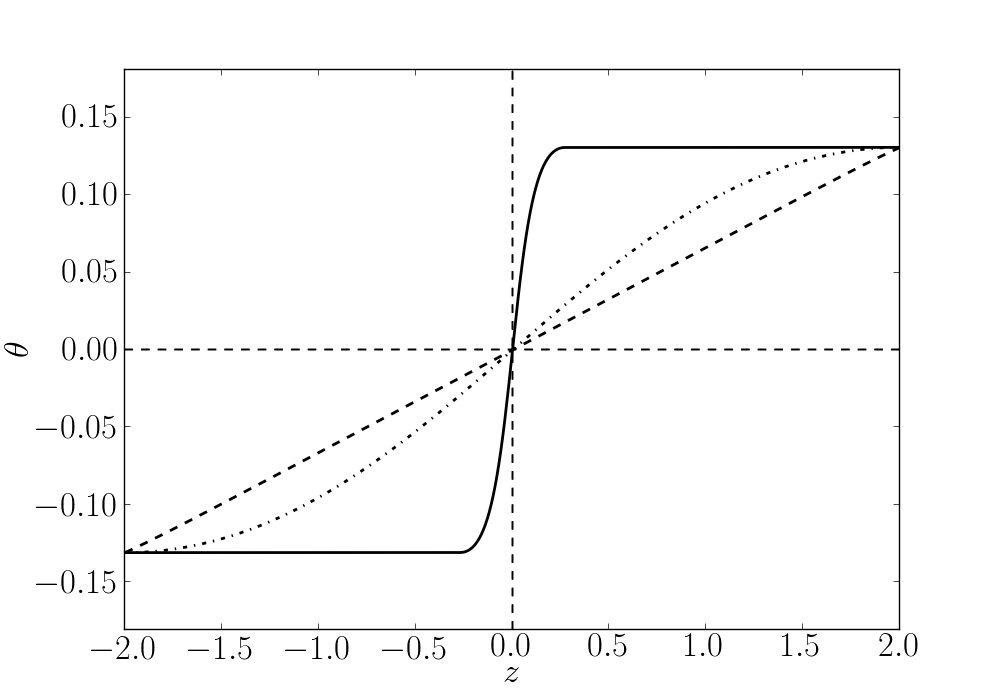}
\caption{\label{as_calib} Profiles for $\eta^0$ (left) and $\theta^0$ (right, in rad) for different choices of $\tilde{a}$ and $\tilde{s}$. Solid lines are for $\tilde{a}=0.15$ and $\tilde{s}=0.5$, the values chosen for calculations. Dashed lines are for $\tilde{a}=0.15$ and $\tilde{s}=0.15$ and dashed-dotted lines are for $\tilde{a}=0.5$ and $\tilde{s}=0.25$.}
\end{figure}

The scaled parameters $\tilde{a}$ and $\tilde{s}$ are found by choosing $\eta^0_{min}$ and $\eta^0_{max}$ for $\Delta \theta=15$ degrees. This is so as to have grain boundaries that are distinct. If these values are not chosen with some care there is a risk that the profile for $\eta$ is too shallow ($\eta_{min}$ close to 1) for small to moderate misorientations. This is demonstrated in figure \ref{as_calib}, where the profiles of $\eta^0$ and $\theta^0$ for $\tilde{a}=0.15$ and $\tilde{s}=0.5$, the values chosen for calculations, are shown as solid lines. The dashed lines are the profiles for $\tilde{a}=0.15$ and $\tilde{s}=0.15$ and dashed-dotted lines are for $\tilde{a}=0.5$ and $\tilde{s}=0.25$, demonstrating the importance to take advantage of the asymptotic analysis when choosing the parameters. 

\begin{figure}[thb!]
\centering
\includegraphics[width=0.5\linewidth]{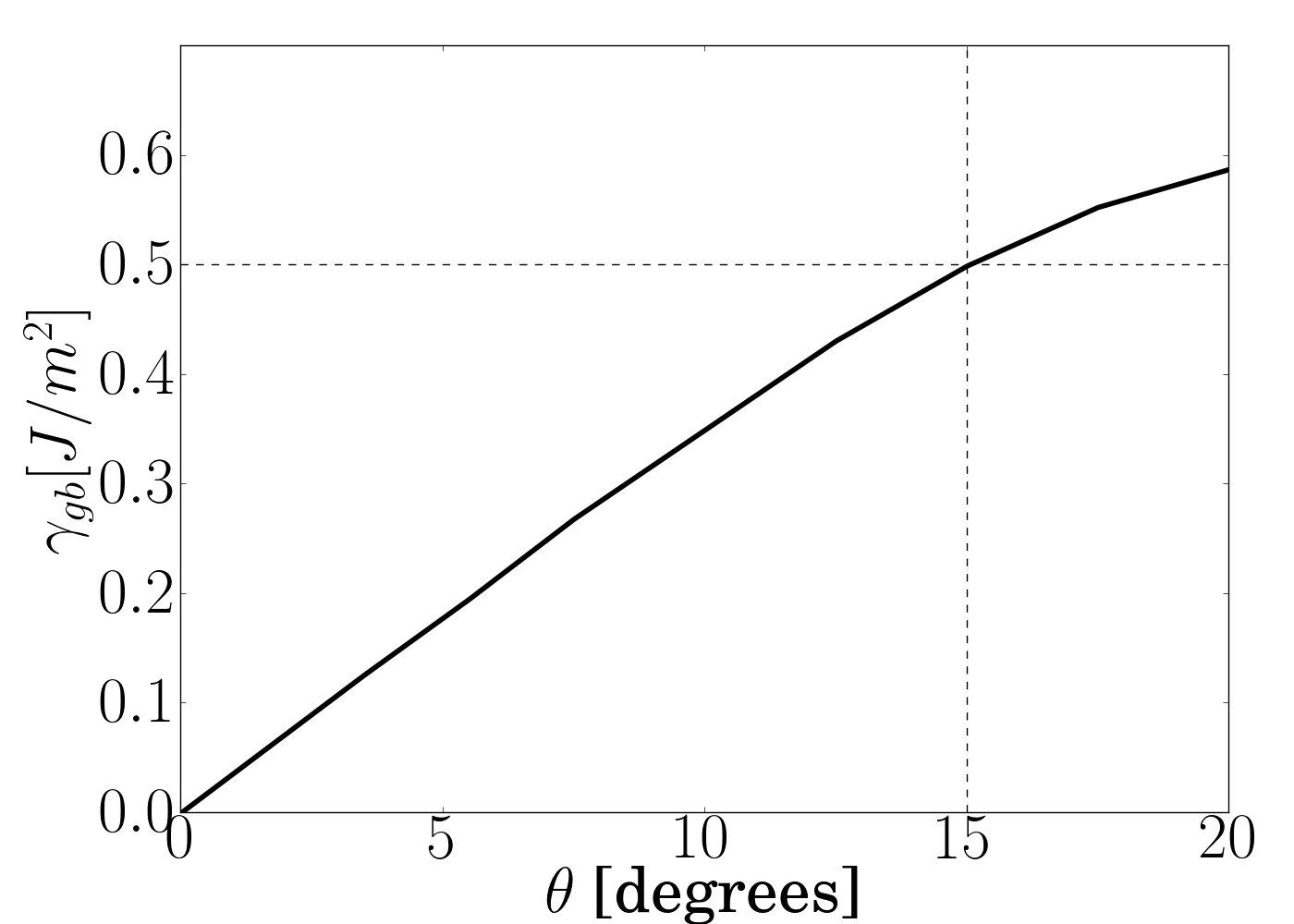}
\caption{\label{ggb_calib} Grain boundary energies for misorientaions up to $20^\circ$ and $f_0=8.2$ MPa, calibrated to an approximate grain boundary energy of 0.5 J/m$^2$ at $15^\circ$ misorientation.}
\end{figure}

A suitable length scale for the simulations is $\Lambda = 1$ $\mu$m. In the simulations of two grains it is assumed that each grain is 10 $\Lambda$, i.e. $10$ $\mu$m in length. A true grain boundary is only some nm wide, which indicates that an appropriate choice of $\varepsilon$ (which together with $a$ governs the width of the grain boundary) is in the nm range. However, in the phase field model the grain boundaries are diffuse and thus wider than the physical grain boundaries. The choice of $\varepsilon$ is also restricted by the finite element discretization as the grain boundary needs to be resolved by at least a few elements. In the simulations, $\overline{\varepsilon} = 0.5$ is used, which in turns gives $\overline{a} = 0.075$ and $\overline{s}=0.25$. With $\delta z \approx 0.2$ the choice $\overline{\varepsilon}=0.5$ gives a diffuse grain boundary width of around 200 nm.

For simplicity, the choice $g(\eta)=h(\eta)=\eta^2$ was made in \ref{energy_isotropic_2D}. \cite{Abrivard2012a} used a higher-order polynomial function for $g(\eta)$ to be able to fit the grain boundary energy to a Read-Shockley curve. This option will not be pursued here. Instead, the calibration is done by taking the average grain boundary energy at 15 degree misorientation to be 0.5 J/m$^2$. This value is only approximate but is in a reasonable range for copper (see e.g. \cite{Tschopp2015}). Using equation (\ref{f0_calib}), this gives $f_0 = 8.2$ MJ/m$^3$ with the dimensionless grain boundary energy calculated using equation (\ref{gb_energy}). The calculated grain boundary energy for misorientations from $0^\circ$ up to $20^\circ$ is shown in figure \ref{ggb_calib}.
\begin{table}[thb!]
\centering
\begin{tabular}{c|c|c}
Name  & Unit & Value\\
\hline
$f_0$  & [J/m$^3$] = [Pa] & 8.2 MPa \\
$a$  & [m] & 0.075 $\mu$m \\
$s$  & [m]  & 0.25 $\mu$m \\
$\varepsilon$ & [m]  & 0.5 $\mu$m \\
$C_D$  & &  100 \\
$C_A$  & [m] & $\sqrt{10}$ $\mu$m \\
$C_\star$  & [m] & $\sqrt{10}$ $\mu$m \\
$\tau_\eta$  & [J$\cdot$s/m$^3$] = [Pa\,s]& 1 $f_0\,\tau_0$ \\
$\hat{\tau}_\theta$  & [Js/m$^3$] = [Pa\,s] & 1 $f_0\,\tau_0$\\
$\hat{\tau}_\star$  & [J$\cdot$s/m$^3$] = [Pa\,s] &  0.1 $f_0\,\tau_0$\\
$\lambda$ & & 0.3 \\
$b$ & [m] & 0.2556 nm \\
\hline
\multicolumn{3}{c}{Cubic anisotropy} \\
\hline
$C_{11}$  & [J/m$^3$] &  160 GPa \\
$C_{12}$  & [J/m$^3$] &  110 GPa \\
$C_{44}$  & [J/m$^3$] &  75 GPa \\
\end{tabular}
\caption{\label{material_parameters_table} List of model parameters used to produce the numerical examples, unless otherwise stated.}
\end{table}

The full set of parameters along with their units is presented in table \ref{material_parameters_table}. The time scale parameter can easily be fitted to a true mobility value using equation (\ref{calib_tau0}) together with (\ref{inv_mobil}).

\newpage

\section{Analytic solution of the stresses during shear loading}\label{app3}

Periodic boundary conditions are assumed so that loading is applied by imposing
\begin{equation}
\begin{aligned}
\vec u = \ten B \cdot \vc x + \vec p 
\end{aligned}
\end{equation}
where $\vec p$ is the periodic fluctuation vector taking the same value in opposing points of the boundary and the components of the tensor $\ten B$ are given by
\begin{equation}
\ten B = \left[\,
\begin{array}{ccc}
0 & B_{12} & 0 \\
B_{21} & 0 & 0 \\
0 & 0 & 0 
\end{array} \,
\right] \,,
\end{equation} 
for shear loading. Specifically, for simple shear with $B_{12}=0$ and the problem at hand, the components of the displacement vector are given by
\begin{equation}
\begin{aligned}
u_1 = \, & p_1(x_1)\,, \\
u_2 = \, & B_{21} \, x_1 + p_2(x_1) \,, \\
u_3 = \, & 0  \,.
\end{aligned}
\end{equation}
The only non-zero strains (in the small deformation setting) are then given by
\begin{equation}
\varepsilon_{11} =  \frac{\partial u_1}{\partial x_1} = p_{1,1} \,, \qquad
\varepsilon_{12} = \varepsilon_{21} = \frac{1}{2}\left[\, \frac{\partial u_1}{\partial x_2} + 
\frac{\partial u_2}{\partial x_1} \,\right] = \frac{1}{2} \left[\,B_{21} + p_{2,1}\,\right] \,,
\end{equation}
whereas
\begin{equation}
\varepsilon_{22} = 0 
\end{equation}
due to the imposed boundary conditions and all other strain components are zero due to the plane strain conditions. Furthermore, the following holds 
\begin{equation}
\left\langle \varepsilon_{11} \right\rangle = 0 \,, \qquad \left\langle \varepsilon_{12} \right\rangle = \frac{1}{2}B_{21} \,.
\end{equation}
Cubic anisotropy is assumed so that the elastic stiffness tensor has three independent parameters $C_{11}$, $C_{12}$ and $C_{44}$. Likewise, the compliance tensor has only three independent parameters $S_{11}$, $S_{12}$ and $S_{44}$ which can be related to the modulus of elasticity through
\begin{equation}
S_{11} = \frac{C_{11} + C_{12}}{(C_{11} + 2\,C_{12})\,(C_{11} - C_{12})} \quad 
S_{12} =  \frac{-C_{12}}{(C_{11} + 2\,C_{12})\,(C_{11} - C_{12})}  \quad
S_{44} =  \frac{1}{S_{44}} \,.
\end{equation}
Strain and stress are rotated between the global frame and the local frame of the lattice through transformation matrices constructed using the non-constant rotation field parameter $\theta$. The constitutive relation between stress and strain is applied in the local frame. Finally, an expression for the strains in the global frame is obtained as
\begin{equation}
\begin{aligned}
\varepsilon_{11} = \, & A \, \sigma_{11} + B \, \sigma_{22} - C \, \sigma_{12} \,, \\
\varepsilon_{22} = \, & B \, \sigma_{11} + A \, \sigma_{22} + C \, \sigma_{12} \,, \\
\varepsilon_{12} = \, & -\frac{1}{2} C \left[\, \, \sigma_{11} - A \, \sigma_{22} \,\right]
 + D \, \sigma_{12}  \,, \\
\varepsilon_{33} = \, &  S_{12}\,\left[\,\sigma_{11}+\sigma_{22}\,\right] + S_{11}\,\sigma_{33} \,,
\end{aligned}
\end{equation}
where the coefficients are given by
\begin{equation}
\begin{aligned}
A = \, & S_{11} + S_{D} + S_{A}\,\sin^2(2\theta) \,, \\
B = \, & S_{12} + S_{D} - S_{A}\,\sin^2(2\theta) \,, \\
C = \, & S_{A}\,\sin(4\theta) \,, \\
D = \, & S_B + S_{A}\,\cos(4\theta) \,, \\
S_A =  \, & \frac{1}{2}\left[\,-S_{11} + S_{12} + \frac{1}{2}S_{44}
\,\right] \,, \\
S_B =  \, & \frac{1}{2}\left[\,S_{11} - S_{12} + \frac{1}{2}S_{44}
\,\right] \,, \\
S_D = \, & -\frac{S_{12}^2}{S_{11}} \,,
\end{aligned}
\end{equation}
and it was used that $\varepsilon_{33}=0$ (plane strain) to express $\sigma_{33}$ in terms of $\sigma_{11}$ and $\sigma_{22}$ in the equations for $\varepsilon_{11}$ and $\varepsilon_{22}$. 

\begin{figure}[thb!]
\centering
\includegraphics[height=5.5cm]{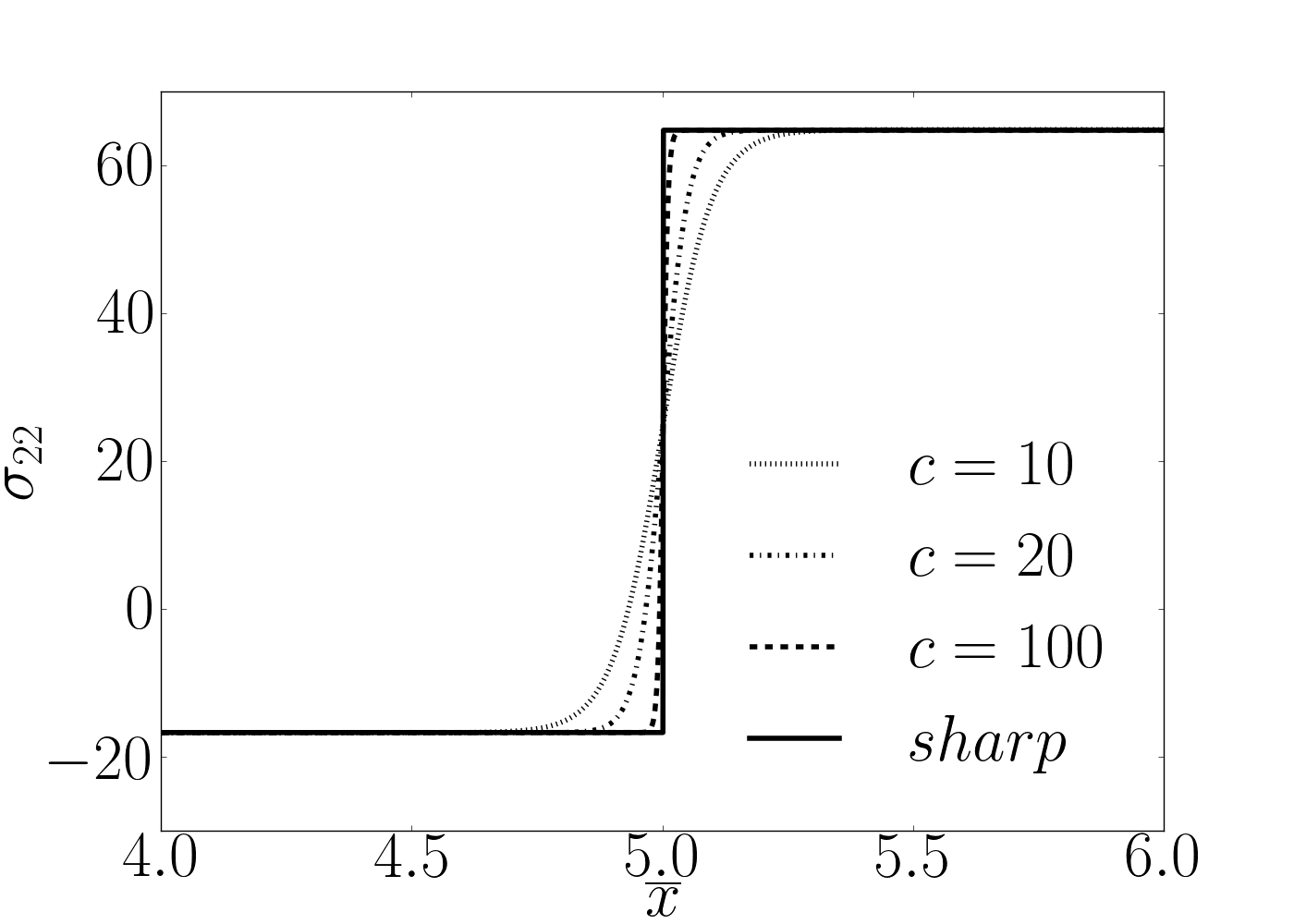}
\includegraphics[height=5.5cm]{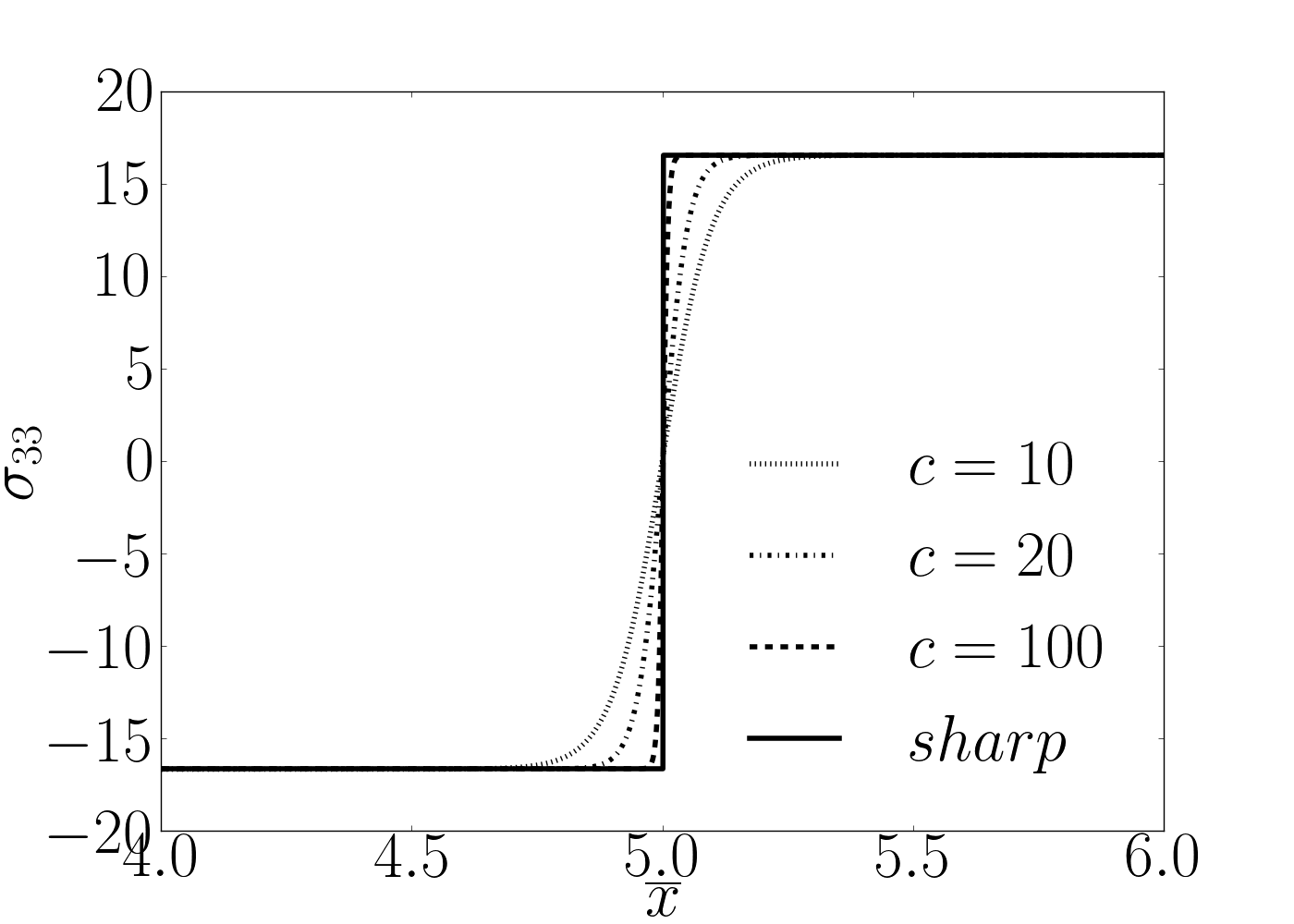}
\caption{\label{bigrain_shear_sharp} Analytically resolved stress profiles for pure shear loading and anisotropic elasticity. The solid line is the sharp interface solution and the other lines represent $\tanh$ formats of the orientation with increasingly high prefactor.}
\end{figure}

It is now possible to construct a system of equations to solve for the stresses. First, as already noted, due to the imposed loading (pure strain) the out-of-plane strain is zero
\begin{equation}
\varepsilon_{33} =  S_{12}\,\left[\,\sigma_{11}+\sigma_{22}\,\right] + S_{11}\,\sigma_{33} = 0
\quad \Rightarrow \quad \sigma_{33} = - \frac{S_{12}}{S_{11}}\left[\,\sigma_{11}+\sigma_{22}\,\right] \,.
\end{equation}
For the stresses, the following holds for the averages 
\begin{equation}
\left\langle \sigma_{33} \right\rangle = 0\,, \qquad
\left\langle \sigma_{11} \right\rangle = \sigma_{11} \,, \qquad
\left\langle \sigma_{12} \right\rangle = \sigma_{12} \,.
\end{equation}
This together with $\varepsilon_{33}=0$ gives the first equation
\begin{equation}
\left\langle \sigma_{33} \right\rangle = 0 \quad \Rightarrow \quad \sigma_{11} = - \left\langle \sigma_{22} \right\rangle \,.
\end{equation}
The second equation comes from the fact that $\varepsilon_{22}=0$, i.e.
\begin{equation}
B \, \sigma_{11} + A \, \sigma_{22} + C \, \sigma_{12} = 0 \,,
\end{equation}
and the third equation comes from the condition
\begin{equation}
\left\langle \varepsilon_{12} \right\rangle = \frac{1}{2}B_{21} \,, 
\end{equation}
so that finally a system of equation in three unknowns $\sigma_{11}$, $\sigma_{22}$ and $\sigma_{12}$ can be written as
\begin{equation}
\left\lbrace
\begin{aligned}
\, & B \, \sigma_{11} + A \, \sigma_{22} + C \, \sigma_{12} = 0 \,,  & (1) \\
\, & \left\langle \,
-\frac{1}{2} C \left[\, \, \sigma_{11} - A \, \sigma_{22} \,\right]
 + D \, \sigma_{12}
\, \right\rangle = \frac{1}{2}B_{21} \,, & (2) \\
\, & \sigma_{11} = - \left\langle \sigma_{22} \right\rangle  \,. & (3)
\end{aligned}
\right.
\end{equation}
Solving the system of equations and recalling the result previously obtained for $\sigma_{33}$ results in
\begin{equation}
\begin{aligned}
\sigma_{12} = \, & \frac{1}{2\,F} B_{21} \,, \\
\sigma_{11} = \, & E \, \sigma_{12} \,, \\
\sigma_{22} = \, & -\frac{B}{A}\sigma_{11} - -\frac{C}{A}\sigma_{12} \,, \\
\sigma_{33} = \, & - \frac{S_{12}}{S_{11}}\left[\,\sigma_{11}+\sigma_{22}\,\right] \,,
\end{aligned}
\label{stress_system}
\end{equation}
with the coefficients $E$ and $F$ given by
\begin{equation}
\begin{aligned}
E = \, & \frac{\left\langle C/A \right\rangle}{1 - \left\langle B/A \right\rangle} \,, \\
F = \, & \left\langle\,
-\frac{C}{2}\left[\,E\,\left[\,1+\frac{B}{A}\,\right] + \frac{C}{A}\,\right] + D
\, \right\rangle \,.
\end{aligned}
\label{coeff_system}
\end{equation} 
Note that, in general, $A$, $B$, $C$ and $D$ all depend on the angle $\theta$ and, in the case where the angle is not constant, must be integrated to obtain the averages. The influence of an increasingly sharp grain boundary profile is shown in figure \ref{bigrain_shear_sharp}. The orientation field is here represented by a $\tanh$ function of the following format
\begin{equation}
\theta = \frac{\pi}{24}\,\tanh(c\,[\,\overline{x}+\overline{x}_{shift}\,]+1)
\end{equation}
where $\overline{x}_{shift}$ shifts the profiles center from origo and the parameter $c$ determines the width of the region over which $\theta$ varies significantly. The larger the value of $c$, the sharper the profile. 

\end{document}